\documentclass[11pt, notitlepage, tightenlines,reprint, aps, prd, superscriptaddress, floatfix]{revtex4-2}
\usepackage[utf8]{inputenc}
\usepackage{rotating}
\usepackage{amsfonts}
\usepackage{amssymb,amsmath}
\usepackage{mathtools}
\usepackage[usenames, dvipsnames]{color}
\usepackage{graphicx}
\usepackage{subfigure}
\graphicspath{ {image/} }

\usepackage{multirow}
\usepackage{xcolor} 
\usepackage{soul}
\usepackage{ulem}
\newcommand*\diff{\mathop{}\!\mathrm{d}}
\usepackage{braket}
\usepackage{bm}
\usepackage{txfonts}
\usepackage{mathrsfs} 
\usepackage{slashed}
\usepackage{float}
\usepackage{simplewick}

\newcommand{\GeV}{{{\,}\textrm{GeV}}}
\newcommand{\keV}{{{\,}\textrm{keV}}}
\newcommand{\TeV}{{{\,}\textrm{TeV}}}

\colorlet{darkgreen}{green!50!black}
\colorlet{brightyellow}{yellow!75!red}
\colorlet{orange}{red!50!yellow}
\colorlet{darkgray}{gray!50!black}
\colorlet{darkred}{red!50!black}

\definecolor{lcolor}{rgb}{0.5,0,0}
\definecolor{citcolor}{rgb}{0,0.3,0.0}
\usepackage[breaklinks,colorlinks,urlcolor=blue,citecolor=citcolor,linkcolor=lcolor]{hyperref}

\begin{document}
 \title{Light-front wavefunctions of mesons by design}
 \author{Meijian Li}
 \email{mliy@jyu.fi}
\affiliation{Department of Physics, P.O. Box 35, FI-40014 University of Jyv\"{a}skyl\"{a},
Finland}
\affiliation{
Helsinki Institute of Physics, P.O. Box 64, FI-00014 University of Helsinki,
Finland
}
\affiliation{
Instituto Galego de F\'\i sica de Altas Enerx\'\i as (IGFAE), Universidade de Santiago de Compostela, E-15782 Galicia, Spain
}

\author{Yang Li}
\email{leeyoung1987@ustc.edu.cn}
\affiliation{Department of Modern Physics, University of Science and Technology of China, Hefei 230026, China}

\author{Guangyao Chen}
\email{ychen3@ju.edu}
\affiliation{Department of Physics, Jacksonville University, Jacksonville, FL, 32211, USA}
\affiliation{Department of Physics and Astronomy, Iowa State University, Ames, IA, 50011, USA}

\author{Tuomas Lappi}
\email{tuomas.v.v.lappi@jyu.fi}
\affiliation{Department of Physics, P.O. Box 35, FI-40014 University of Jyv\"{a}skyl\"{a},
Finland}
\affiliation{
Helsinki Institute of Physics, P.O. Box 64, FI-00014 University of Helsinki,
Finland
}

\author{James P. Vary}
\email{jvary@iastate.edu}
\affiliation{Department of Physics and Astronomy, Iowa State University, Ames, IA, 50011, USA}

\begin{abstract}
We develop a mechanism to build the light-front wavefunctions (LFWFs) of meson bound states on a small-sized basis function representation. 
Unlike in a standard Hamiltonian formalism, the Hamiltonian in this method is implicit, and the information of the system is carried directly by the functional form and adjustable parameters of the LFWFs.
In this work, we model the LFWFs for four charmonium states, $\eta_c$, $J/\psi$, $\psi'$, and $\psi(3770)$ as superpositions of orthonormal basis functions.
We choose the basis functions as eigenfunctions of an effective Hamiltonian, which has a longitudinal confining potential in addition to the transverse confining potential from light-front holographic QCD. 
We determine the basis function parameters and superposition coefficients by employing both guidance from the nonrelativistic description of the meson states and the experimental measurements of the meson decay widths.
With the obtained wavefunctions, we study the features of those meson states, including charge radii and parton distribution functions. 
We use the $J/\psi$ LFWF to calculate the meson production in diffractive deep inelastic scattering and ultra-peripheral heavy-ion collisions, and the $\eta_c$ LFWF to calculate its diphoton transition form factor. 
Both results show good agreement with experiments.
The obtained LFWFs have simple-functional forms and can be readily used to predict additional experimental observables.
\end{abstract}
\maketitle

\normalem 
\section{Introduction}
Understanding and describing hadrons, the bound states of quantum chromodynamics (QCD), is crucial to increase our comprehension of the strong interaction and the constitution of matter.
The charmonium sector has attracted extensive experimental investigations, including the mass spectrum, transitions between excited and low-lying states, and photoproduction of the vector mesons in heavy-ion collisions~\cite{Zholents:1980qu, Brambilla:2010cs, Kowalski:2006hc, Baltz:2007kq, Armesto:2014sma, ALICE:2021gpt}.
Theoretical efforts also contribute from an array of complementary perspectives, both Euclidean and Minkowskian formalisms.       
Euclidean formulations of quantum field theories such as Dyson-Schwinger equations~\cite{Roberts:1994dr, Blaschke:2000zm} and lattice gauge theory~\cite{Gattringer:2010zz, Dudek:2007wv} offer methods of performing a first-principles computation of the charmonium spectrum and other observables.  
On the other hand, Hamiltonian methods formulated in Minkowski spacetime also provide a detailed description of the meson's internal structure and dynamics through wavefunctions. 
The wavefunctions play a central role in describing the bound states and computing the physical observables in Hamiltonian formalism.

In a standard Hamiltonian formalism, the wavefunctions are solved from the Schr\"{o}dinger(-like) equations where the Hamiltonian governs the physics of the system.
Since the discovery of the $J/\psi$ resonance in 1974~\cite{Aubert:1974js}, various potential models were developed to describe the heavy quarkonium system, including the Buchmüller-Tye potential~\cite{Buchmuller:1980su}, power-law potential~\cite{Martin:1980jx}, logarithmic potential~\cite{Quigg:1977dd}, and the Cornell potential~\cite{Eichten:1978tg}.
These phenomenological models from the early years were inspired by various aspects of QCD,
and they successfully described the spectrum, especially the $\psi$ family and the $\Upsilon$ family.

Later on, nonrelativistic QCD (NRQCD) was developed as an effective field theory by incorporating standard quantum field theory techniques such as dimensional regularization~\cite{Bodwin:1994jh, Pineda:2011dg, Brambilla:2004jw}. 
It captures the nonrelativistic nature of the heavy system, and the relativistic corrections can be incorporated systematically, although calculations show that the relativistic corrections may be large for selected charmonium observables~\cite{Feng:2015uha}.

One shared feature of these studies is that the meson wavefunctions are given in their rest frames. 
Hadrons in high-energy processes, however, are correctly described by wavefunctions on the light front. A Lorentz transformation is required to apply rest-frame wavefunctions to modern deep inelastic scattering (DIS) experiments. In principle, the light-front wavefunctions can be generated directly from the light-front Hamiltonian approach~\cite{Hiller:2016itl,Brodsky:1997de,Miller:2000kv}, which combines light-front quantization and affords boost-invariant light-front wavefunctions (LFWFs).
Light-front holography (LFH) exploits the AdS/CFT correspondence between string states in anti-de Sitter (AdS) space and conformal field theories (CFT) in physical space-time to obtain a semiclassical first approximation to QCD~\cite{deTeramond:2008ht,Brodsky:2020ajy}. 
It generates effective potentials for valence quarks of hadron bound states, and the resulting LFWFs are relativistic and analytically tractable.
Basis light-front quantization (BLFQ)~\cite{Vary:2009gt} has been used to improve LFH by incorporating a longitudinal confinement and a realistic one-gluon exchange interaction~\cite{Li:2015zda,Li:2017mlw,Li:2021jqb}. 

A different path to obtain the meson LFWF is to model it directly or determine it from other formalisms. 
In such an approach, the LFWF is not solved from an eigenvalue equation and does not require one to assume a specific form for the Hamiltonian. However, the functional form can be inspired by a phenomenological Hamiltonian when modeling the LFWF directly.
The vector meson wavefunction is modeled as predominantly a quark-antiquark state.
The Dosch, Gousset, Kulzinger, and Pirner (DGKP) model ~\cite{Dosch:1996ss}, and the widely used boosted Gaussian~\cite{Kowalski:2006hc,Nemchik:1994fp,Nemchik:1996cw} are in this category. 
In these parametrizations, the helicity and polarization structure is the same as in the photon perturbatively calculated in QCD.
One main advantage of such modeled wavefunctions is their simplicity.
They have become an important element for calculating meson production cross sections, e.g., exclusive processes at the electron-ion collider (EIC)~\cite{Accardi:2012qut}.
There are also works in determining the LFWF from other formalisms.
LFWFs determined from the Dyson-Schwinger and Bethe-Salpeter approach embed information from higher Fock states, which is achieved by projecting the covariant Bethe-Salpeter wavefunctions onto the light front~\cite{Shi:2018zqd, Mezrag:2016hnp,dePaula:2020qna, Shi:2021taf}. 
Insights from NRQCD could also be carried through LFWFs by boosting the meson wavefunctions from the rest frame to the light-front frame and employing Melosh rotations on the spin structure~\cite{Krelina:2019egg, Lappi:2020ufv}.

As a complementary study to the existing modeled LFWFs, we propose a method of designing the LFWFs of meson bound states with a simple-functional form.
A unique feature of the formalism in this work is its close relation to the light-front Hamiltonians through the choice of basis functions.
More specifically, for designing the charmonium LFWFs, we choose eigenfunctions of a generalized light-front holographic confining potential as the basis functions, which were first introduced in a study on heavy quarkonia by the BLFQ approach~\cite{Li:2015zda, Li:2017mlw}. 
Consequently, this work is closely related to the light-front Hamiltonians through the choice of basis functions, and shares several advantages of the LFH framework and the BLFQ framework. 
Additionally, this choice of functional forms provides us with some guidance on interpreting the meson's internal structure. 
While the works on and based on LFH give connections and predictions across the entire mass spectrum of hadrons~\cite{Brodsky:2006uqa,Karch:2006pv,deTeramond:2005su,Dosch:2015bca,Dosch:2016zdv,Brodsky:2014yha,Brodsky:2020ajy,Li:2015zda,Li:2017mlw,Tang:2018myz,Jia:2018ary,Tang:2019gvn,Qian:2020utg,Li:2021jqb}, our proposed approach has more flexibility in adjusting the wavefunctions and facilitates a better agreement with targeted experimental observables such as the dilepton and diphoton decay widths studied in this work. 

The implicit Hamiltonian that describes the designed system is thereby understood as an effective Hamiltonian which is an extension of the LFH/BLFQ Hamiltonian, and its information is carried directly by the functional form and the adjustable parameters in the wavefunctions. 
We determine the parameters and basis coefficients by adopting guidance from the nonrelativistic description of the meson states and the experimental measurements of the meson's decay widths. 

In designing the charmonia LFWFs, our major considerations and motivations are the following:
\begin{itemize}
    \item \emph{Approximation to QCD}. The designed LFWFs inherit an approximation to QCD from LFH through the basis functions.
    \item \emph{Symmetries}. The designed states are invariant under kinematical symmetries, including mirror parity $\hat m_{\mathsf P}$ and charge conjugation $\mathsf C$, and approximately under rotational symmetries, including the total spin $J$ and parity $\mathsf P$.
    In addition, when calculating Lorentz invariants, rotational symmetry is respected by using different current components and different polarized states. 
    \item \emph{Matching the NR limit}. 
   In the NR limit, the LFWFs reduce to the solutions of the Schrödinger equation with a spherically symmetric harmonic oscillator potential, and the three-dimensional rotational symmetry is restored.
    \item \emph{Decay widths}. The LFWFs give the correct diphoton decay width for the pseudoscalar and the dilepton decay width for the vector meson. 
    These constraints are most sensitive to the behavior of the wavefunctions at short distances, which is also the most important region for perturbative scattering processes.
\end{itemize}
We call our approach ``by design'', because we are choosing by hand to apply exactly the constraints that we consider to be the most important for phenomenological applications of the wavefunctions.
It is an alternative phenomenological approach to obtain LFWFs that is not limited by a particular choice of Hamiltonian. 
This involves a certain amount of judgement as to how many basis functions to include and which constraints to impose. 
We make these choices in such a way that we can exactly satisfy all the constraints that we are using. 
Alternative schemes are also possible, for example, having the number of observables exceed the number of parameters, and one then needs to choose weights for different observables.
One primary advantage of our approach is that the resulting LFWFs are analytically tractable and can be used to calculate a wide variety of physical observables.

In this paper, we calculate the charge radii and parton distribution functions of those meson states with the obtained wavefunctions. We estimate the masses of those designed states by evaluating their expectation with existing Hamiltonians, and they are in a reasonable range compared with experimental values. We use the wavefunction of $J/\psi$ to calculate meson production in heavy-ion collisions and compare it with other model calculations. 
We use the wavefunction of $\eta_c$ to calculate the diphoton transition form factor.
We found that all observables calculated are in reasonable agreement with experimental measurements.

The layout of this paper is as follows. We first introduce the formalism of the basis function representation in Sec.~\ref{sec:basis}. We then construct the LFWFs for $J/\psi$, $\eta_c$, $\psi'$, and $\psi(3770)$ in Sec.~\ref{sec:LFWF}. 
With the obtained wavefunctions, we study selected key features of those meson states in Sec.~\ref{sec:observables} and calculate the $J/\psi$  production in high-energy scattering and the $\eta_c$ diphoton decay.
We conclude the work in Sec.~\ref{sec:summary}.

\section{Basis function representation}\label{sec:basis}
In this section, we introduce our formalism of designing the meson LFWFs in a basis function representation and the basis functions we use in this work.

\subsection{Light-front wavefunctions in a basis space}\label{sec:basis_method}
Consider a meson state $h$ consisting of a quark and an antiquark, with momentum $(P^+, \vec P_\perp)$, and expand its wavefunction on an orthonormal basis $\{\beta_1, \beta_2, \ldots, \beta_{N_\beta}\}$,
\begin{equation}
  \psi_h(\vec k_\perp, x) = \sum_{i=1}^{N_\beta} C_{h,i} \beta_i(\vec k_\perp, x)\;,
\end{equation}
where $C_{h,i}$ are the basis coefficients for $h$ and $N_\beta$ is the number of basis states.
Here we are writing the wavefunction in a relative coordinate, where $x=p^+_q/P^+$ is the longitudinal momentum fraction of the quark and $\vec k_\perp = \vec k_{q\perp} - x \vec P_\perp$ is the relative transverse momentum. 

The wavefunctions should satisfy the orthonormalization relation
\begin{equation}\label{eq:norm}
\sum_{i=1}^{N_\beta} C_{h,i}C^*_{h',i} = \delta_{h,h'}\;.
\end{equation}
Physical quantities and observables ($O$) such as decay widths and charge radius are functions ($f_O$) of the basis coefficients,
\begin{equation}\label{eq:Oh}
  O_h = f_O (C_{h,i})\;.
\end{equation}
The constraints Eqs.~\eqref{eq:norm} and \eqref{eq:Oh} form a system of equations, and the unknowns are the basis coefficients $C_{h,i}$ and could also include parameters in the basis functions. 
The procedure of designing LFWFs is, in essence, solving such a system of equations. 

In the continuum limit of $N_\beta\to \infty$, both the number of constraints and the number of unknowns are infinite.
For the purpose of designing LFWFs with a simple-functional form, it is favorable to use a small number of basis states. At the same time, the constraints are chosen from observables that are most relevant to the physics one wants the wavefunctions to describe.  
A solution can be determined uniquely when the number of equations is equal to the number of unknowns. It could also be true that the system is overdetermined if there are more equations than unknowns, in which case one could obtain a solution by using optimization algorithms to minimize the deviation from the constraint.
If there are more unknowns than equations, such that the system is underdetermined, one might obtain multiple solutions, with additional criteria needed to choose the preferred one. 
The approach we choose in this paper is to work in a relatively small basis and to impose exactly enough constraints to obtain a unique solution that is a physically reasonable description of the lowest states $J/\psi, \psi', \psi(3770)$ and $\eta_c$. In the future, this work could be extended to a larger basis and to include additional constraints. 

\subsection{A generalized holographic basis representation}\label{sec:LFWFbasis}
In this paper, we are working with a highly truncated basis space, and our goal is to describe the charmonium bound states in a simple-functional form. 
It is, therefore, crucial to choose a basis function that has been successfully adopted for solving light-front Hamiltonian problems.
For this purpose, we take the basis introduced in BLFQ studies on heavy quarkonia~\cite{Li:2015zda,Li:2017mlw}.
These basis functions are the eigenfunctions of a generalized holographic confining potential, 
\begin{align}\label{eq:H0}
  \begin{split}
  H_0 = &
  \frac{\vec k_\perp^2 + m_q^2}{x}
  + 
  \frac{\vec k_\perp^2 + m_{\bar q}^2}{1-x}\\
  &+\kappa^4 x(1-x) r_\perp^2
  -\frac{\kappa^4}{(m_q + m_{\bar q})^2}\partial_x(x(1-x)\partial_x)
  \;,
  \end{split}
\end{align}
where $m_q$ ($m_{\bar q}$) is the mass of the quark (antiquark), and for charmonium we have $m_q=m_{\bar q}=m_f$.
The first two terms are the light-front kinetic energy of the constituent quark and antiquark, the third term is a transverse holographic confining potential, and the last term is a longitudinal confining potential.
The confining strength is characterized by the parameter $\kappa$.
This Hamiltonian, although with independent strength parameters for the transverse and longitudinal confining terms, is referred to as $\mathrm{BLFQ_0}$ in a recent work on light mesons~\cite{Li:2021jqb}. 

This basis, obtained from the eigenfunctions of the Hamiltonian~\eqref{eq:H0}, consists of a two-dimensional (2D) harmonic oscillator (HO) function $\phi_{nm}$ in the transverse direction and a modified Jacobi polynomial $\chi_l$ in the longitudinal direction. 
The transverse basis function in the momentum space reads
\begin{align}
  \begin{split}
  \phi_{nm}(\vec k_\perp)=&\kappa^{-1}\sqrt{\frac{4\pi n!}{(n+|m|)!}}{\bigg(\frac{k_\perp}{\kappa}\bigg)}^{|m|}
  \exp(-k_\perp^2/(2\kappa^2))\\
  &L_n^{|m|}({k_\perp}^2/\kappa^2)\exp(im\theta_k)
  \;,    
  \end{split}
\end{align}
where $k_\perp=|\vec k_\perp |$, $\theta_k=\arg (k^x+ik^y)$, $n=0,1,2,\ldots$ is the principal number, and $m=0,\pm1,\pm2,\ldots$ is the orbital number. 
The transverse confining strength $\kappa$ functions as the basis parameter here.
The orthonormality relation is
\begin{align}
  \int\frac{\diff^2\vec k_\perp}{{(2\pi)}^2}\phi^*_{n' m'}(\vec k_\perp)  \phi_{nm}(\vec k_\perp)=\delta_{nn'}\delta_{mm'}
  \; .
\end{align}
Note that the HO functions are a natural generalization of the Gaussian type wavefunctions widely adopted in the literature~\cite{Kowalski:2006hc,Dosch:1996ss}. 
The longitudinal basis function is 
\begin{align}
  \begin{split}
  \chi_l(x)=&\sqrt{4\pi(2l+\alpha+\beta+1)}\sqrt{\frac{\Gamma(l+1)\Gamma(l+\alpha+\beta+1)}{\Gamma(l+\alpha+1)\Gamma(l+\beta+1)}}\\
  &x^{\beta/2}{(1-x)}^{\alpha/2}P_l^{(\alpha,\beta)}(2x-1)
  \; ,    
  \end{split}
\end{align}
where $P_l^{(\alpha,\beta)}$ is the Jacobi polynomial with $l=0,1,2,\ldots$. 
The dimensionless basis parameters $\alpha$ and $\beta$ are related to the longitudinal confining strength and the fermion mass in $H_0$ as $\alpha =4m_q^2/\kappa$ and $\beta = 4m_{\bar q}^2/\kappa$. For charmonia, $\alpha=\beta$.
The orthonormality relation is
\begin{align}
  \frac{1}{4\pi}\int_0^1\diff x \chi_l(x)\chi_{l'}(x)=\delta_{l l'}
  \;.
\end{align}

Each basis state is characterized by five quantum numbers,  $\{n,m,l,s,\bar s\}$, where $s$ ($\bar s$) is the light-front helicity of the quark (antiquark). 
The basis is constructed to conserve the magnetic projection of the total angular momentum: $m_j=m+s+\bar s$, and $m$ is interpreted as the orbital angular momentum projection. 

\begin{figure*}[ht]
  \centering
  \subfigure[\ $\psi_{nml}(k_\perp,\theta_k=0, x=0.5)$ ]
  {\includegraphics[width=0.35\textwidth]{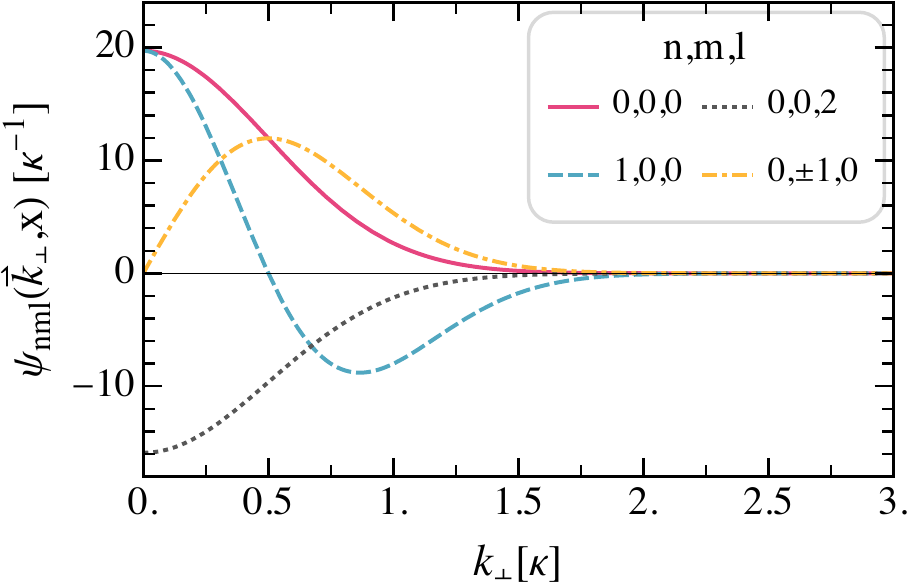}
  } 
  \subfigure[\ $\psi_{nml}(\vec k_\perp=\vec 0_\perp,x)$ ]
  {\includegraphics[width=0.24\textwidth]{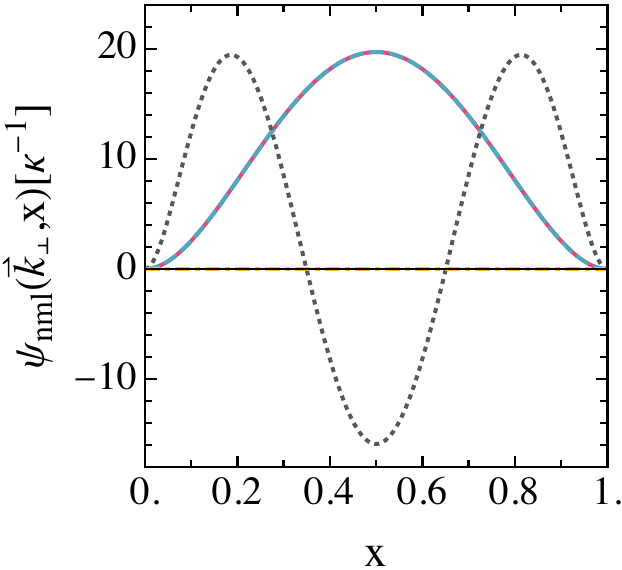}
  } 
  \caption{The basis function $\psi_{nml}(\vec k_\perp,x)$ [defined in Eq.~\eqref{eq:psi_nml}]: 
  (a) as a function of $k_\perp$ (in the unit of $\kappa$) at $\theta_k=0$ and $x=0.5$, and
  (b) as a function of $x$ at $\vec k_\perp=\vec 0_\perp$.
  Note that in (b), the curves of $\psi_{000}$ and $\psi_{100}$ are the same, and $\psi_{0,\pm 1,0}$ is zero.
  In the plots, we take $\alpha=\beta=4$, and the basis function is in the unit of $\kappa^{-1}$. 
  }
  \label{fig:psinml_k_x}
\end{figure*}

The LFWF of a meson state $h$ is written as an expansion on this basis function representation:
\begin{equation}\label{eq:LFWF_full}
  \psi^{(m_j)}_{h}(\vec k_\perp, x) = \sum_{n, m, l,s, \bar s} \psi_h^{(m_j)}(n, m, l, s, \bar s) \beta_{ n, m, l, s, \bar s}(\vec k_\perp,x)\;,
\end{equation}
where $\psi_h(n, m, l, s, \bar s) $ is the basis coefficient, and the basis function is defined as \(\beta_{ n, m, l, s, \bar s} (\vec k_\perp,x)\equiv \psi_{nml}(\vec k_\perp,x)\sigma_{s  \bar s}\). The spatial dependence is in $\psi_{nml}(\vec k_\perp,x)$, which combines the transverse and the longitudinal basis functions as
\begin{align}\label{eq:psi_nml}
  \psi_{nml}(\vec k_\perp,x)\equiv  \phi_{nm}(\vec k_\perp/\sqrt{x(1-x)}) \chi_l(x)
  \;.
\end{align}
The parameters in the transverse basis function and those in the longitudinal basis function are connected by $\alpha =\beta = 4m_f^2/\kappa$, so we have two adjustable parameters for the basis function, $\kappa$ and $m_f$.
Figure~\ref{fig:psinml_k_x} presents some special cases of the  basis function $\psi_{nml}(\vec k_\perp,x)$ as functions of $k_\perp$ and $x$.

The helicity configuration of the quark and the antiquark is denoted by $\sigma_{s  \bar s}$, and we use $\uparrow (\downarrow)$ to indicate $1/2 (-1/2)$, the value of $s$ and $\bar s$. We also define $\sigma_{\pm}\equiv 1/\sqrt{2}(\sigma_{\uparrow\downarrow}\pm\sigma_{\downarrow\uparrow})$.
It is convenient to write out the wavefunction for a specific spin component, 
\begin{equation}\label{eq:LFWF}
  \psi^{(m_j)}_{s\bar s/h}(\vec k_\perp, x) = \sum_{n, m, l} \psi_h^{(m_j)}(n, m, l, s, \bar s) \, \psi_{nml}(\vec k_\perp,x)\;.
\end{equation}

The orthonormalization of the LFWF in Eq.~\eqref{eq:LFWF} reads, 
\begin{align}\label{eqn:normalization}
  \begin{split}
 \sum_{s, \bar s}  
 &\int_0^1\frac{\diff x}{2x(1-x)} 
\int \frac{\diff^2  k_\perp}{(2\pi)^3} 
\psi^{(m_j')*}_{s \bar s/h'}(\vec k_\perp, x)
\psi^{(m_j)}_{s \bar s/h}(\vec k_\perp, x) \\
 &= \delta_{hh'}\delta_{m_j,m_j'}.    
  \end{split}
\end{align}

The LFWF in the transverse coordinate space is obtained by performing a 2-dimensional Fourier transformation,
\begin{equation}\label{eq:FT_kt_rt}
  \tilde\psi_{s\bar s} (\vec r_\perp, x) \equiv \frac{1}{\sqrt{x(1-x)}}\int \frac{\diff^2 k_\perp}{(2\pi)^2}  e^{i \vec k_\perp \cdot
\vec r_\perp} \psi_{s \bar s} (\vec k_\perp, x).
\end{equation} 
The corresponding orthonormalization is,
\begin{align}\label{eqn:normalization_r}
  \begin{split}
&\sum_{s, \bar s} \int_0^1\frac{\diff x}{4\pi} \int \diff^2 r_\perp \, \tilde\psi^{(m_j')*}_{s \bar s/h'}(\vec r_\perp, x)
\tilde\psi^{(m_j)}_{s \bar s/h}(\vec r_\perp, x) \\
&= \delta_{hh'}\delta_{m_j,m_j'}.
\end{split}
\end{align}
The basis function representation becomes 
\begin{align}
  \begin{split}
  \tilde \psi^{(m_j)}_{ss'/h}(\vec r_\perp, x) 
  = &\sqrt{x(1-x)} 
  \sum_{n, m, l} 
  \psi^{(m_j)}_h(n, m, l, s, s') \\
 & \tilde \phi_{nm}(\sqrt{x(1-x)} \vec r_\perp) 
  \chi_l(x)\;,
\end{split}
\end{align}
and the transverse basis function in the coordinate space reads
\begin{align}
  \begin{split}
  \tilde \phi_{nm}(\vec r_\perp) = 
  &\kappa \sqrt{\frac{n!}{\pi(n+|m|)!}} 
  (\kappa r_\perp)^{|m|} 
  \exp\big(-\kappa^2 r^2_\perp/2\big)\\
 &L_n^{|m|}(\kappa^2 r_\perp^2) 
 \exp\big[ i m\theta_r + i \pi (n+|m|/2) \big]\;,  
  \end{split}
 \end{align}
 where $r_\perp=|\vec r_\perp |$ and $\theta_r=\arg (r^x +i r^y)$. 
 
The parity and charge conjugation can be evaluated from this basis representation of the LFWFs~\cite{Li:2015zda}.
We exploit the mirror parity $\hat m_{\mathsf P} = \hat R_x(\pi) \hat P$ that is related to $\mathsf P$ as,
$ \hat m_{\mathsf P}\ket{\psi_h^{(m_j)}} = (-1)^j\mathsf P
\ket{\psi_h^{(-m_j)}}
$. It follows that,
\begin{align}\label{eq:LFP}
  \begin{split}
    (-1)^j\mathsf P =&  \langle \psi_h^{(-m_j)}|\hat m_P | \psi_h^{(m_j)}\rangle \\
  =& \sum_{n,m,l,s,\bar s} (-1)^m \psi_h^{(-m_j)*}(n, -m, l, -s, -\bar s)\\
  &
 \psi_h^{(m_j)}(n, m, l, s, \bar s)
 \;.
  \end{split}
\end{align}
We categorize the basis states $\psi_{nml}\sigma_{s  \bar s}$  by mirror parity in Table~\ref{tab:mp_basis}. 
\begin{table}[tb!]
  \caption{\label{tab:mp_basis} 
  Basis states $\psi_{nml}\sigma_{s  \bar s}$ categorized by the mirror parity $m_{\mathsf P}$ according to Eq.~\eqref{eq:LFP}.
  }
  \centering
  \begin{ruledtabular}
  \begin{tabular}{llll }
    $m_j$ &$m$& ~~~$m_\mathsf P=1$~~~ ~~~& ~~~~~~$m_\mathsf P=-1$~~~~~~\\
    \hline
    \multirow{2}{*}{$0$}
    & $0$ & $\psi_{n,0,l}\sigma_+$ & $\psi_{n,0,l}\sigma_-$ \\  
   &$\pm 1$& $\frac{1}{\sqrt{2}}(\psi_{n,-1,l}\sigma_{\uparrow\uparrow}-\psi_{n,1,l}\sigma_{\downarrow\downarrow})$
     & $\frac{1}{\sqrt{2}}(\psi_{n,-1,l}\sigma_{\uparrow\uparrow}+\psi_{n,1,l}\sigma_{\downarrow\downarrow})$\\ 
     \hline
     \multirow{3}{*}{$1,-1$}
     & $0$ & $\psi_{n,0,l}\sigma_{\uparrow\uparrow}$, $\psi_{n,0,l}\sigma_{\downarrow\downarrow}$ 
     & $\psi_{n,0,l}\sigma_{\uparrow\uparrow}$, $-\psi_{n,0,l}\sigma_{\downarrow\downarrow}$  \\  
    &$\pm 1$& $\psi_{n,1,l}\sigma_{\pm}$, $\mp \psi_{n,-1,l}\sigma_{\pm}$ 
    & $\psi_{n,1,l}\sigma_{\pm}$, $\pm\psi_{n,-1,l}\sigma_{\pm}$ 
    \\  
    & $\pm 2$ & $\psi_{n,2,l}\sigma_{\downarrow\downarrow}$, $\psi_{n,-2,l}\sigma_{\uparrow\uparrow}$ 
    & $\psi_{n,2,l}\sigma_{\downarrow\downarrow}$, $-\psi_{n,-2,l}\sigma_{\uparrow\uparrow}$  
  \end{tabular}
  \end{ruledtabular}
\end{table}

The charge conjugation on the basis reads
\begin{align}\label{eq:C}
  \begin{split}
 \mathsf C =& \langle\psi_h^{(m_j)}|\hat C | \psi_h^{(m_j)}\rangle\\
  =& \sum_{n,m,l,s,\bar s} (-1)^{m+l+1} \psi_h^{(m_j)*}(n, m, l, \bar s, s)\\
 &\psi_h^{(m_j)}(n, m, l, s, \bar s)
 \;. 
\end{split}
\end{align}
The basis states $\psi_{nml}\sigma_{s  \bar s}$ are categorized by charge conjugation in Table~\ref{tab:C_basis}. 
\begin{table}[tb!]
  \caption{\label{tab:C_basis} 
    Basis states $\psi_{nml}\sigma_{s  \bar s}$ categorized by the charge conjugation $\mathsf C$ according to Eq.~\eqref{eq:C}.
  }
  \centering
  \begin{ruledtabular}
  \begin{tabular}{llll }
   $m+l$ ~~~& ~~~~~~$\mathsf C=1$~~~ ~~~& ~~~~~~$\mathsf C=-1$~~~~\\
    \hline
   even ~~~
    & $\psi_{n,m,l}\sigma_-$  & $\psi_{n,m,l}\sigma_+$,
    $\psi_{n,m,l}\sigma_{\uparrow\uparrow}$,
    $\psi_{n,m,l}\sigma_{\downarrow\downarrow}$\\  
    odd ~~~& $\psi_{n,m,l}\sigma_+$,
    $\psi_{n,m,l}\sigma_{\uparrow\uparrow}$,
    $\psi_{n,m,l}\sigma_{\downarrow\downarrow}$
    & $\psi_{n,m,l}\sigma_-$
  \end{tabular}
  \end{ruledtabular}
\end{table}
We use those eigenstates of mirror parity and charge conjugation as building blocks to construct meson states. 

\subsubsection{Light-front spectroscopic state}
We have seen that the spatial part of the basis function, $\psi_{nml}$, is the eigenfunction of $H_0$ in Eq~\eqref{eq:H0}.
In the nonrelativistic limit $m_f\gg \kappa$, the potential in $H_0$ reduces to the 3D harmonic oscillator potential. 
Consequently, the basis functions $\psi_{nml}$ are also related to the 3D harmonic oscillators (HOs). 
Such a relation is helpful in designing the LFWFs of charmonium, which has been studied extensively in the NRQCD framework.
We shall first use this nonrelativistic limit to construct approximate orbital angular momentum states (denoted by $\psi_{\text{LF}-W}$) as combinations of our basis functions. 
We will then combine these spatial wavefunctions with the helicity structure to construct what we call \emph{light-front spectroscopic states} $\psi_{\text{LF}-n\ell,j\mathsf P \mathsf C}$. 
These states are exact eigenstates of mirror parity and charge conjugation, which are good symmetries on the light front. 
Although they do not exactly correspond to orbital angular momentum $\ell$ eigenstates, they are close enough to be clearly identified with states of specific $n$ (the principle number), $j$ (the total angular momentum), $\ell$ and $s$ (the parton spin).

We first construct the light-front spatial state $\psi_{\text{LF}-W}$ that has an approximate orbital angular momentum, using the basis function $\psi_{nml}$. 
The ``W" part contains the information of the orbital angular momentum $\ell$ and its projection $m(=-\ell,-\ell+1,\ldots ,\ell)$, and the principal number $n$.
We take the spectroscopic notation, in which $\ell=0,1,2$ corresponds to the S, P, and D, respectively, and $n=1,2,\ldots$ labels the energy level in the ascending order. 
The notation $\text{LF}-W$ distinguishes the light-front spatial state from the spectroscopic $W$ state of the 3D HO.
In the nonrelativistic limit, the former should reduce to the latter with the corresponding $n,\ell, m$ numbers.
Building $\psi_{\text{LF}-W}$, therefore, suggests a transformation between the 3D HO and the light-front functional basis $\psi_{nml}$.
Such a transformation is non-trivial and not exact since angular momentum is dynamical on the light front.
However, considering the similarity of our basis function and the 3D HO in the cylindrical coordinates, we could use the transformation of 3D HO from the spherical coordinates to the cylindrical coordinates as a guidance.
The transformation coefficients are calculated in Appendix.~\ref{app:3DHO}.

The lowest basis state in the chosen basis function representation is the ground state $\psi_{0,0,0}$. 
We expect it to provide a close approximation to the 1S state,
\begin{align}\label{eq:LF_1S}
  \psi_{\text{LF}-1S} = 
    \psi_{0,0,0}  
  \;.
\end{align}
The 1P state has three components differing by magnetic projections, and their light-front basis correspondences are 
\begin{subequations}\label{eq:LF_1P}
  \begin{align}\label{eq:LF_1P0}
    \psi_{\text{LF}-1P0} = 
      \psi_{ 0,0,1}  
    \;,
  \end{align}
  \begin{align}\label{eq:LF_1Ppm1}
    \psi_{\text{LF}-1P\pm 1} = 
      -\psi_{0,\pm 1,0}  
    \;.
  \end{align}
\end{subequations}
In the 3D HOs, the radial excited state and the angular excited state are distinct from each other, whereas in the light-front basis, the radial and angular excitations mix automatically. 
The first radial excited state, the 2S state, can be approximated as a linear superposition of $\psi_{1,0,0}$ and $\psi_{ 0,0,2}$. 
We take their relative coefficients from the cylindrical representation, which leads to
\begin{align}\label{eq:LF_2S}
    \psi_{\text{LF}-2S} = 
    \sqrt{\frac{2}{3}}\psi_{ 1,0,0}  
    -\sqrt{\frac{1}{3}}\psi_{ 0,0,2}  
    \;.
\end{align}
The 1D state has five components differentiated by magnetic projections, and their light-front basis correspondences are 
\begin{subequations}
  \begin{align}\label{eq:LF_1D0}
    \psi_{\text{LF}-1D0} = 
      \sqrt{\frac{1}{3}}
      \psi_{ 1,0,0}  
    +    \sqrt{\frac{2}{3}}
    \psi_{ 0,0,2}  
    \;,
  \end{align}
  \begin{align}\label{eq:LF_1Dpm1}
    \psi_{\text{LF}-1D\pm 1} = 
      -\psi_{0,\pm 1,1}  
    \;,
  \end{align}
  \begin{align}\label{eq:LF_1Dpm2}
    \psi_{\text{LF}-1D\pm 2} = 
      \psi_{0,\pm 2,0}  
    \;.
  \end{align}
\end{subequations}
These light-front spatial states $\psi_{\text{LF}-W}$ are orthogonal to each other.
In the nonrelativistic limit, they reduce to the spherical harmonics with their corresponding angular numbers (see the derivation in Appendix~\ref{app:NR}).
We assume that these $\psi_{\text{LF}-W}$ states are the spatial components of the physical heavy quarkonia LFWFs.

We now combine the spatially dependent part with the quark-antiquark helicity structure into the form of $\psi_{\text{LF}-W}\sigma_{s \bar{s}}$. 
Each of these states inherits the principal quantum number $n$, the approximate orbital number $\ell$, and its projection $m$ from the spatial part, and the parton spin projection $m_s(=s+\bar s)$ from the helicity part.
Consequently, it acquires the $m_j(=m+m_s)$ number as well.
We know from Sec.~\ref{sec:LFWFbasis} that we can further combine these components into eigenstates of mirror parity $\hat{m}_{\mathsf P}$ and charge conjugation $\mathsf C$ (see the eigenstates listed in Tables~\ref{tab:mp_basis} and \ref{tab:C_basis}).
Then by specifying the total spin $j$, we can construct the \emph{light-front spectroscopic state} $\psi_{\text{LF}-n\ell,j\mathsf P \mathsf C}$, which is identified by $j^{\mathsf P \mathsf C}$ values.
The total spin is a vector summation of the orbital angular momentum and the parton spin, $\vec j=\vec \ell + \vec s$. 
Note that for the light-front basis state, the value of orbital angular momentum $\ell$ is approximate, and the parton spin number $s$ is $0$ ($1$) for the singlet $\sigma_-$ (triplet $\{ \sigma_{\uparrow\uparrow},\sigma_+,\sigma_{
\downarrow\downarrow}\}$) configuration.
In the following, we find the light-front spectroscopic states for meson states with $j^{\mathsf P \mathsf C}=0^{-+}$ and $1^{--}$.

A pseudoscalar state with quantum number $j^{\mathsf P \mathsf C}=0^{-+}$ (recall that $j$ is only approximated in the valence sector) could be a LF-nS state or LF-nP state.
The LF-nS component is
\begin{align}
  \psi_{\text{LF}-nS, 0-+}=\psi_{\text{LF}-nS}\sigma_-
  \;.
\end{align}
The LF-nP component is
\begin{align}
  \psi_{\text{LF}-nP,0-+}=
  \psi_{\text{LF}-nP-1}\sigma_{\uparrow\uparrow}
  +\psi_{\text{LF}-nP1}\sigma_{\downarrow\downarrow}
  \;.
\end{align}
The vector state with $j^{\mathsf P \mathsf C}=1^{--}$ could be a LF-S state or LF-D state.
Each LF-nS $m_j$ state has one component:
\begin{align}
  \begin{split}
    \psi_{\text{LF}-nS, 1--}^{(m_j=0)}=&\psi_{\text{LF}-nS}\sigma_+\;,\\
    \psi_{\text{LF}-nS, 1--}^{(m_j=1)}=&\psi_{\text{LF}-nS}\sigma_{\uparrow\uparrow}\;,\\
    \psi_{\text{LF}-nS, 1--}^{(m_j=-1)}=&\psi_{\text{LF}-nS}\sigma_{\downarrow\downarrow}
    \;.
  \end{split}
\end{align}
Each LF-1D $m_j$ state has multiple components with different helicity structures, and the relative magnitude and phase among those components are not determined. Here we take the Clebsch-Gordan (CG) coefficients such that in the nonrelativistic limit, the light-front wavefunction components reduce to the spectroscopic state with $j=1, \ell=2, s=1$,
\begin{align}
  \begin{split}
    \psi_{\text{LF}-1D, 1--}^{(m_j=0)}=&  -\sqrt{\frac{3}{10}}
    \psi_{\text{LF}-1D-1}
    \sigma_{\uparrow \uparrow}
    -\sqrt{\frac{2}{5}}
  \psi_{\text{LF}-1D0}
    \sigma_+\\
  &+
  \sqrt{\frac{3}{10}}
  \psi_{\text{LF}-1D1}
  \sigma_{\downarrow\downarrow}\;,\\
    \psi_{\text{LF}-1D, 1--}^{(m_j=1)}=& 
    \sqrt{\frac{3}{5}}
    \psi_{\text{LF}-1D2}
\sigma_{\downarrow\downarrow}
    -
    \sqrt{\frac{3}{10}}
    \psi_{\text{LF}-1D1}
\sigma_+\\
  &
    + \sqrt{\frac{1}{10}}
    \psi_{\text{LF}-1D0} 
    \sigma_{\uparrow\uparrow}
    \;,\\
    \psi_{\text{LF}-1D, 1--}^{(m_j=-1)}=&\sqrt{\frac{3}{5}} 
    \psi_{\text{LF}-1D-2}
    \sigma_{\uparrow\uparrow}
  +   \sqrt{\frac{3}{10}}
  \psi_{\text{LF}-1D-1} 
  \sigma_+\\
  &
    + \sqrt{\frac{1}{10}}
    \psi_{\text{LF}-1D0} 
    \sigma_{\downarrow\downarrow}
  \;.
  \end{split}
\end{align}
We will use these light-front spectroscopic states as components in designing the charmonium LFWFs. 

\subsubsection{The Hamiltonian is implicit}
Unlike the standard Hamiltonian approach, the heavy quarkonia LFWFs in this work are constructed directly on a chosen basis representation, without solving the eigenvalue equation of an explicit Hamiltonian. 
The advantage of doing so is that we are free from interpretating any modeled Hamiltonian as the physical Hamitonian responsible for the mass spectroscopy.
Instead, we use quantities that can be directly calculated from the wavefunction.
The basis space we use to design the wavefunctions is much smaller than in the computational method BLFQ.
Instead, our approach here is reminiscent of a mean-field theory where a many-body problem could be reduced into an effective one-body problem, and there is a simplified reference Hamiltonian with an effective interaction correction.
The basis states $\beta_{ n, m, l, s, \bar s} (\vec k_\perp,x)$ are a chosen subset of the eigenfunctions of a generalized holographic Hamiltonian $H_0$, as in Eq.~\eqref{eq:H0}. 
We could take $H_0$ as a reference Hamiltonian, and interpret the full implicit Hamiltonian as $H_0$ plus an effective interaction $\Delta H$,
\begin{align}\label{eq:H}
  H = H_0 + \Delta H
  \;.
\end{align}
The effective interaction $\Delta H$ should contain all the remaining interaction effects from truncated states, including interaction effects from higher Fock sectors. 

We could estimate the invariant mass for a designed state by calculating its expectation value of $H$, which gives a range instead of a definite value since $\Delta H$ is unknown.
The expectation value of $H_0$ for a basis state $\beta_{ n, m, l, s, \bar s}$ is
\begin{align}
  \begin{split}
M^2_{n,m,l}(\kappa, m_q, m_{\bar q}) = &
  (m_q + m_{\bar q})^2
  + 2 \kappa^2 (2n+|m|+l+\frac{3}{2})\\
 & +\frac{\kappa^4}{(m_q + m_{\bar q})^2}l(l+1)
  \;.
\end{split}
\end{align}

The expectation value of $H$ for a state by design is therefore 
\begin{align}\label{eq:tildeM2}
  \begin{split}
  \left(\tilde{M}^{(m_j)}_{h}\right)^2 
  =& 
  \sum_{n, m, l, s, \bar s} 
  \sum_{n', m', l', s', \bar s'}
  \psi_h^{(m_j)}(n, m, l, s, \bar s)\\
  &\times \psi_h^{(m_j)*}(n', m', l', s', \bar s')\\
  & 
\times \bigg[
  M^2_{n,m,l}\delta_{n,n'}\delta_{m,m'}\delta_{l,l'}
  \delta_{s,s'}\delta_{\bar s, \bar s'}\\
  & 
  +\bra{\beta_{ n', m', l', s', \bar s'}  }\Delta H \ket{\beta_{ n, m, l, s, \bar s} }
\bigg]
\;.    
  \end{split}
\end{align}

Here we give a tentative interpretation on $\Delta H$ by approximating it as an effective one-gluon exchange interaction $V_{\text{OGE}}$ from Ref.~\cite{Li:2015zda}. The one-gluon exchange interaction includes contributions from the $\ket{q\bar q g}$ sector. 
This corresponds to approximately interpreting the implicit Hamiltonian as the BLFQ Hamiltonian, which is $H_{BLFQ}=H_0+V_{\text{OGE}}$~\cite{Li:2015zda}. 
The effective one-gluon-exchange interaction, written explicitly, is
\begin{align}\label{eq:OGE}
  V_{\text{OGE}} =  - \frac{C_F 4\pi\alpha_s(q^2)}{q^2}\bar u_{s'}(k')\gamma_\mu u_s(k) \bar v_{\bar s}(\bar k) \gamma^\mu v_{\bar s'}(\bar k')
  \;,
\end{align}
  where $C_F=(N_c^2 -1)/(2N_c)=4/3$ is the color factor, and the energy denominator is the average 4-momentum squared carried by the exchanged gluon $q^2 = [-(k'-k)^2-(\bar k'-\bar k)^2]/2$ (see the details of the form and parameters in Ref.~\cite{Li:2017mlw}).

\section{Wavefunction and the decay width}\label{sec:LFWF}
\begin{figure}[tbh]
  \centering
  \includegraphics[width=.27\textwidth]{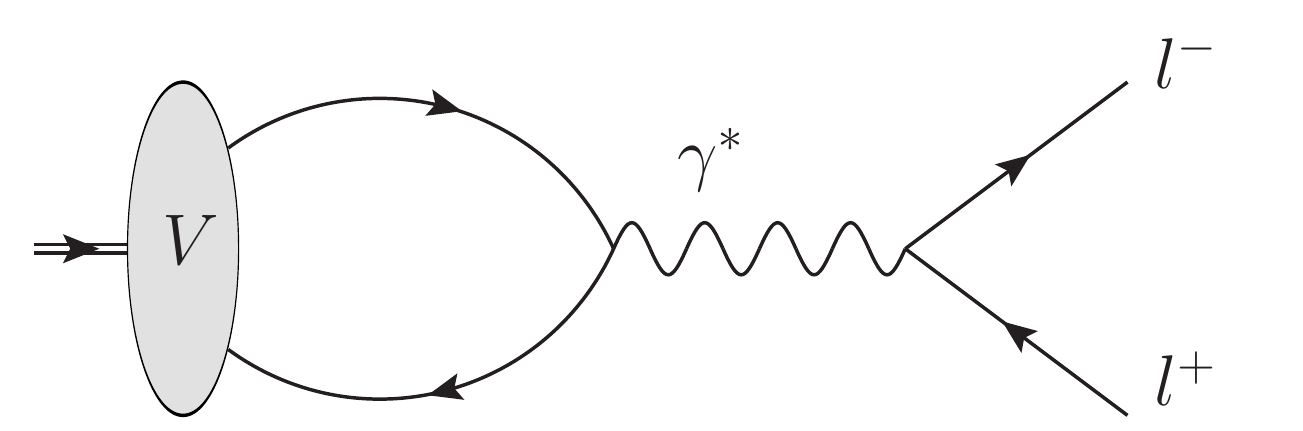}
  \caption{\label{fig:DecayCon_V} 
  The dilepton decay process of a vector meson $V$ (such as $J/\psi$, $\Upsilon$).}
\end{figure}
In this section, we design the LFWFs for the four charmonium states, $\eta_c, J/\psi$, $\psi'$, and $\psi(3770)$ in the basis function representation developed in Sec.~\ref{sec:LFWFbasis}. 
We determine the basis parameters and remaining coefficients in the wavefunctions by matching the calculated decay widths to the experimental values, in particular, the diphoton decay width for the pseudoscalar meson, and the dilepton decay width for the vector meson. 
\subsection{The vector meson dilepton decay and the decay constant}
The vector meson can decay into a dilepton pair via a virtual photon, see Fig. \ref{fig:DecayCon_V} for the process. 
The transition amplitude is factorized into the decay constant, which is defined with the electromagnetic current via the local vacuum-to-hadron matrix element
\begin{align}\label{eq:fv}
\bra{0} \bar{\Psi}(0)\gamma^\mu \Psi(0) \ket{\mathcal{V} (P,m_j)}
=m_{\mathcal{V}} e^\mu(P, m_j) f_{\mathcal{V}}
  \;.
\end{align}
Here $m_{\mathcal{V}}$ is the mass of the vector meson $\mathcal{V}$ and $f_{\mathcal{V}}$ the decay constant. 
It is related to the experimental decay width in the particle rest frame as~\cite{Feldmann:1997te,Becirevic:2013bsa}
\begin{align}\label{eq:exp_fV}
  \Gamma_{\mathcal{V}\to l^+ l^-}=\frac{4\pi}{3} \mathcal{Q}_f^2\alpha_{em}^2\frac{f_{\mathcal{V}}^2}{m_{\mathcal{V}}}
  \;.
\end{align}
Here, $\mathcal{Q}_f$ is the dimensionless fractional charge of the constituent quarks, $\mathcal{Q}_c=+2/3$ for the charm quark and $\mathcal{Q}_{b}=-1/3$ for the bottom quark.
The decay width sums over the contributions from all polarized states of the vector meson and the virtual photon, and the size of the individual contribution depends on the reference frame. 
On the contrary, the decay constant defined from Eq.~\eqref{eq:fv} is Lorentz invariant and should not depend on the polarization of the vector meson or the current component in the calculation. 
As a result, for the LFWFs obtained from the Hamiltonian formalism or model calculations, the decay constants extracted from different polarized states could provide a measure of the violations of rotational symmetry in the system, for example, in Ref. \cite{Li:2018uif}. 
For the formalism we are developing here, this provides a powerful constraint in designing the LFWF; that is, we require the decay constant calculated from different polarized states of the vector meson to be the same as the experimental value. 
In this way, we impose rotational invariance in the LFWF at the level of electromagnetic decay width. 

In the LFWF representation, the decay constant extracted from Eq.~\eqref{eq:fv} is an integral of the meson LFWF. The calculations using different current components in combination with different polarized states have been analyzed and discussed in Chapter. 2 of Ref.~\cite{Li:2019ijx}. 
In brief, for the $m_j=0$ state, using the $J^+$ and the $J^\perp$ current components are equivalent, and the result from the $J^-$ current only agrees with the two on the condition that the meson mass equals the invariant mass of the constituent quark and antiquark.  The latter is not guaranteed for the LFWF in the valence sector since the bound state mass also contains interactions besides the kinetic energy of the valence particles. We therefore adopt the $J^+$ or equivalently the $J^\perp$ current to calculate the decay constant from the $m_j=0$ state,
\begin{align}\label{eq:fV_mj0}
    \begin{split}
       f_{\mathcal{V}}|_{m_j=0}= &
      \sqrt{2N_c}
      \int_0^1\frac{\diff x}{\sqrt{x(1-x)}}
      \int\frac{\diff^2 k_\perp}{{(2\pi)}^3}
      \\
      & 
      \psi^{(m_j=0)}_{+/\mathcal{V}}(\vec k_\perp, x)
      \;,
    \end{split}
\end{align}
Recall that we adopt the notations for spin configurations as $\psi_{\pm}\equiv (\psi_{\uparrow\downarrow}\pm \psi_{\downarrow\uparrow})/\sqrt{2}$.
For the $m_j=\pm 1$ states, the $J^+$ channel is not available since $e^+(P, m_j=\pm 1)=0$, and using the $J^\perp$ and $J^-$ currents give the same result,
\begin{subequations}\label{eq:fV_mjpm1}
    \begin{align}\label{eq:fV_mj1}
    \begin{split}
    f_{\mathcal{V}}|_{m_j= 1}
    =&
    \frac{\sqrt{N_c}}{2m_{\mathcal{V}}}
    \int_0^1\frac{\diff x}{[x(1-x)]^{3/2}}
    \int\frac{\diff^2 k_\perp}{{(2\pi)}^3}\\
    &
    \bigg\{
      k^L
      [(1-2x)\psi^{(m_j=1)}_{+/\mathcal V}(\vec k_\perp, x)
      -
      \psi^{(m_j=1)}_{-/\mathcal V}(\vec k_\perp, x)]\\
      &
      -\sqrt{2} m_f\psi^{(m_j=1)}_{\uparrow\uparrow/\mathcal V}(\vec k_\perp, x)
    \bigg\}
   \;,
  \end{split}
\end{align}
    \begin{align}\label{eq:fV_mjm1}
      \begin{split}
      f_{\mathcal{V}}|_{m_j=-1}
      &=
      \frac{\sqrt{N_c}}{2m_{\mathcal{V}}}
      \int_0^1\frac{\diff x}{[x(1-x)]^{3/2}}
      \int\frac{\diff^2 k_\perp}{{(2\pi)}^3}\\
      &
      \bigg\{
        k^R
        [(1-2x)\psi^{(m_j=-1)}_{+/\mathcal V}(\vec k_\perp, x)
        +
        \psi^{(m_j=-1)}_{-/\mathcal V}(\vec k_\perp, x)]\\
        &
        +\sqrt{2} m_f\psi^{(m_j=-1)}_{\downarrow\downarrow/\mathcal V}(\vec k_\perp, x)
      \bigg\}
      \;.
      \end{split}
    \end{align}
\end{subequations}
Here, we use the notations $k^R=k^x+ik^y=k_\perp e^{i \theta_k}$ and $k^L=k^x-ik^y=k_\perp e^{-i \theta_k}$ with $k_\perp=|\vec k_\perp |$ 
(also see Eq.~\eqref{eq:RL} in Appendix.\ref{app:LF_cor} for the definitions of these quantities). 
Note that Eqs.~\eqref{eq:fV_mj1} and ~\eqref{eq:fV_mjm1} are equivalent up to an overall minus sign, which can be seen directly by noting that the two polarized states $m_j=\pm 1$ are related by the mirror parity, as in Eq. (17). But what matters in the decay width is the absolute value of the decay constant, so the two equations give the same constraint on the transverse polarized LFWF of the vector meson.
In the basis function representation introduced in Section.~\ref{sec:basis}, the integrals in Eqs.~\eqref{eq:fV_mj0} and \eqref{eq:fV_mjpm1} can be solved analytically, and those equations reduce to a summation over weighted basis coefficients. The expressions of the decay constant in the basis function representation can be found in Appendix.~\ref{app:fV_basis}.

In the nonrelativistic (NR) limit of $x\to 1/2+k_z/(2m_f)$, $m_f\to m_{\mathcal{V}}/2$ and $m_f \gg k_z$, only the dominant spin components, $\psi^{(m_j=0)}_{+/\mathcal{V}}$, $\psi^{(m_j=-1)}_{\downarrow\downarrow/\mathcal{V}}$ and $\psi^{(m_j=1)}_{\uparrow\uparrow/\mathcal{V}}$, survive as the NR wavefunction $\psi^{NR}_{\mathcal{V}}(\vec{k})$ [ $ \tilde\psi^{NR}_{\mathcal{V}}(\vec{r})$ in the coordinate space]. 
The two choices of extracting the decay constants, Eqs.~\eqref{eq:fV_mj0} and \eqref{eq:fV_mjpm1}, reduce to the same form, 
\begin{align}
    f^{NR}_{\mathcal{V}}|_{m_j=0}
    =
    f^{NR}_{\mathcal{V}}|_{m_j=\pm 1}
     =
     \frac{\sqrt{2N_c}}{m_f}
    \tilde\psi^{NR}_{\mathcal{V}}(\vec{r}=\vec 0)
    \;.
\end{align}

In designing the LFWF of the vector mesons, we impose the constraint that the decay constant calculated from different polarized states of the vector meson to be the same as the experimental value,
\begin{align}
    \left|f_{\mathcal{V}}|_{m_j=0}\right|
    =\left |f_{\mathcal{V}}|_{m_j= \pm 1}\right|
    = f_{\mathcal{V}, \text{experiment}}\;.
\end{align}
The significance of this condition is twofold. First, it matches the structure of the designed LFWF to the physical state, especially at short distances. Second, it enforces rotational symmetry in the vector meson state examined by an electromagnetic probe.

\subsection{The pseudoscalar meson diphoton decay}\label{sec:decay_Pgg}
\begin{figure}[tbh]
  \centering
  \includegraphics[width=.2\textwidth]{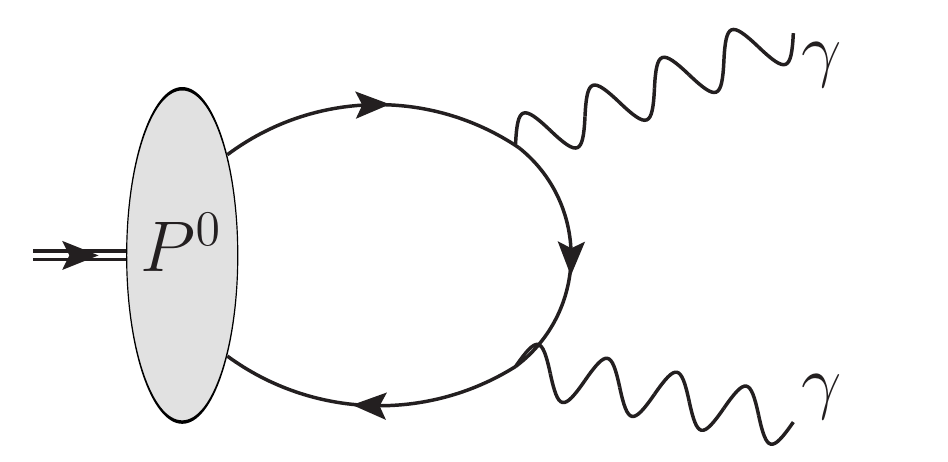}
  \caption{\label{fig:DecayCon_P_n}  
  The diphoton decay process of a neutral pseudoscalar meson $P^0$ (such as $\eta$, $\eta_c$ and $\eta_b$).}
\end{figure}

The simplest decay channel of a neutral pseudoscalar meson for our purposes is the decay into two photons, $\mathcal{P} (P)\to\gamma(q_1)+\gamma(q_2)$ (see Fig. \ref{fig:DecayCon_P_n} for the process). 
It involves only one transition form factor, $F_{\mathcal{P}\gamma}(Q_1^2, Q_2^2)$, where  $Q^2_i=-q_i^2\ge 0, i=1,2$, from which the decay width $\Gamma_{\mathcal{P}\to\gamma\gamma}$ can be extracted exactly at $Q_1^2=Q_2^2=0$, i.e., taking both photons on-shell. 
The width can also be approximately described by a decay constant.  
Similar to the dilepton decay for the vector meson, the diphoton decay provides a test of the structure of the pseudoscalar meson.

In designing the pseudoscalar meson LFWF, we use the decay width calculated through the transition form factor as a constraint. 
However, it is also helpful to check the decay constant, especially since this approximate route has a simpler form in the basis function representation.

\subsubsection{The diphoton decay width in terms of the pseudoscalar-photon transition form factor}
The transition form factor $F(Q_1^2, Q_2^2)$ for the $\mathcal{P} (P)\to\gamma^*(q_1)+\gamma^*(q_2)$ transition is defined from the Lorentz covariant decomposition of the electromagnetic transition matrix element~\cite{Babiarz:2019sfa, Lepage:1980fj},
\begin{align}\label{eq:FPg_def2}
  \begin{split}
    I^\mu_{\lambda_1}(P, q_1)\equiv & 
    \bra{\gamma^*(q_1,\lambda_1)} J^\mu(0) \ket{\mathcal{P}(P)}\\
    =&-i e^2 F_{\mathcal{P}\gamma}(Q_1^2, Q_2^2)\epsilon^{\mu\alpha\beta\sigma} {P}_\alpha {q_1}_\beta \epsilon^*_{\sigma,\lambda_1}(q_1)
    \;,
  \end{split}
\end{align}
where $P^\mu$ and ${q_1}^\mu$ are the four momenta of the incoming pseudoscalar meson and the outgoing photon, and ${q_2}^\mu= P^\mu - {q_1}^\mu$ is the momentum of the other photon. 
Here, $m_{\mathcal{P}}$ is the meson mass, and $e_{\sigma,\lambda_1}$ the polarization vector of the final photon $\gamma^*(q_1,\lambda_1)$, with $\lambda_1=0, \pm 1$ the magnetic projection. 
The transition amplitude $\mathcal{P} (P)\to\gamma^*(q_1)+\gamma^*(q_2)$ is obtained by contracting $I^\mu$ with the polarization vector of the other photon,
\begin{align}\label{eq:amplitude2}
  \mathcal{M}_{\lambda_1,\lambda_2}(P, q_1)= I^\mu_{\lambda_1}(P, q_1)
  \epsilon_{\mu,\lambda_2}^*(q_2)\;,
\end{align}
where $\epsilon_{\mu,\lambda_2}$ is the polarization vector of the final-state virtual photon $\gamma^*(q_2,\lambda_2)$ with its spin projection $\lambda_2=0, \pm 1$.
The two-photon decay width is calculated by averaging over the initial particle polarization (which is just 1 for the pseudoscalar) and summing over the final polarization ($\lambda_1$ and $\lambda_2$),
\begin{align}\label{eq:Pggwidth}
  \begin{split}
    \Gamma_{\mathcal{P}\to\gamma\gamma}
    =&\int\diff \Omega_q
    \frac{1}{32\pi^2}
    \frac{|\vec{q}|}{m_{\mathcal{P} }^2}
    \sum_{\lambda_1,\lambda_2}|\mathcal{M}_{\lambda_1,\lambda_2}|^2\\
    =&
    \frac{e^4 m_{\mathcal{P}}^3}{ 64\pi}
    {|F_{\mathcal{P}\gamma}(0,0)|}^2
    \;.
  \end{split}
\end{align}
The transition form factor defined by Eq.~\eqref{eq:FPg_def2} is Lorentz invariant and should not depend on the polarization of the photons or the current component in the calculation.
The hadron matrix element has the same structure as that in the electromagnetic transition between a vector meson and a pseudoscalar $\mathcal{P}\to\mathcal{V}+\gamma$, obtained by replacing the vector meson by a photon. 
The studies of the transition form factor in the LFWF representation have shown that there are only two combinations of the current component and the polarization of the vector meson that could unambiguously extract the transition form factor from the valence hadron matrix element: the plus current $J^+$ with the transverse polarized vector meson ($m_j=\pm 1$), and the transverse current $J^\perp$ with the longitudinally polarized vector meson ($m_j=0$)~\cite{Li:2018uif,Li:2019ijx,Li:2020wrn}. 
The transition form factors extracted from other choices are not invariant under the transverse boost $\vec P_\perp\to\vec P_\perp + P^+ \vec \beta_\perp$ ($\vec \beta_\perp$ an arbitrary velocity vector), indicating the necessity of higher Fock sector contributions to restore Lorentz invariance.
Here we take both choices and require the two derived diphoton decay widths to be the same as the experimental value. 
In this way, we impose rotational invariance in the LFWF at the level of the electromagnetic decay width, in the same spirit as what we do for the vector meson. 
Though the pseudoscalar meson has only one polarization state $m_j=0$, extracting the transition form factor using different current components and polarized states of the photon probes the pseudoscalar LFWF in different ways.

Taking the impulse approximation, in which the interaction of the external current with the meson is the summation of its coupling to the quark and the antiquark, the transition matrix elements in the LFWF representation read as an overlap of the pseudoscalar meson and the final photon $\gamma^*(q_1,\lambda_1)$. Here we take the photon wavefunction calculated from light-cone perturbation theory to the lowest order (see the explicit expression in Appendix~\ref{app:photon}).
The transition form factor at $Q_1^2=Q_2^2=0$ using the transverse current $J^+$ and the transverse polarized photon ($\lambda_1=\pm 1$) is
\begin{align}\label{eq:FPg_00_plus}
  \begin{split}
    F_{\mathcal{P}\gamma}(0,0)|_{J^+}
=&2 \mathcal{Q}^2_f\sqrt{N_c}
\int_0^\infty\frac{ k_\perp \diff k_\perp}{(2\pi)^2} 
    \int_{0}^1\frac{\diff x}{\sqrt{2x(1-x)}}\\
   &\frac{ 
   - m_f^2 \phi_{0/\mathcal P}(k_\perp, x)+\sqrt{2}m_f 
   k_\perp\phi_{1/\mathcal P}(k_\perp, x)
  }{[k_\perp^2+m_f^2]^2}\;.
  \end{split}
\end{align}
For simplicity, we have introduced two scalar functions that are related to the singlet and triplet helicity components as  $\phi_{0/\mathcal P}(k_\perp, x)=\psi_{-/\mathcal{P}}(\vec k_\perp, x)$ and
$\phi_{1/\mathcal P}(k_\perp, x)= \psi_{\uparrow\uparrow/\downarrow\downarrow/\mathcal{P}}(\vec k_\perp, x) e^{\mp i\theta_k}$.
Using the transverse current $J^\perp$ and the longitudinally polarized photon ($\lambda_1=0$), we get
\begin{align}\label{eq:FPg_00_perp}
  \begin{split}
    F_{\mathcal{P}\gamma}(0,0)|_{J^\perp}
    =&
    -2\mathcal{Q}^2_f \sqrt{N_c}
    \int_0^\infty\frac{ k_\perp \diff k_\perp}{(2\pi)^2}
     \int_{0}^1\frac{\diff x}{\sqrt{2x(1-x)}}\\
   &\frac{ \phi_{0/\mathcal P}(k_\perp, x)}{k_\perp^2+m_f^2}
    \;.
  \end{split}
\end{align}

According to the formalism of the covariant light-front dynamics (CLFD), the general helicity structure of a two-body pseudoscalar bound state takes the form 
\(
  \psi_{s\bar s/\mathcal P, CLFD}(\vec k_\perp, x) = \bar u_{s}(k_q) \phi_A(k_\perp, x) 
  \gamma_5v_{\bar s}(k_{\bar q})\),
  in which $\phi_A$ is a scalar function,
if we neglect the light-front orientation dependent term \cite{Carbonell:1998rj, Leitner:2010nx}. 
It follows that the two spin components are related as
\begin{align}\label{eq:P_WF_CLFD_phi0_phi1}
    -k_\perp\phi_{0/\mathcal P,CLFD}(k_\perp, x)=\sqrt{2}m_f \phi_{1/\mathcal P,CLFD}(k_\perp, x)\;.
\end{align}
With this condition, the transition form factors extracted using the two different currents, Eqs.~\eqref{eq:FPg_00_plus} and \eqref{eq:FPg_00_perp}, would be equivalent.

In the nonrelativistic (NR) limit of $x\to 1/2+k_z/(2m_f)$, $m_f\to m_{\mathcal{P}}/2$ and $m_f \gg k_z$, only the dominant spin components, $\psi^{(m_j=0)}_{-/\mathcal{P}}$ survive as the NR wavefunction $\psi^{NR}_{\mathcal{P}}(\vec{k})$ ( $ \tilde\psi^{NR}_{\mathcal{P}}(\vec{r})$ in the coordinate space). The two Equations.~\eqref{eq:FPg_00_plus} and \eqref{eq:FPg_00_perp}, reduce to the same form, 
\begin{align}
  F^{NR}_{\mathcal{P}\gamma}(0,0)|_{J^+} 
    =
    F^{NR}_{\mathcal{P}\gamma}(0,0)|_{J^\perp} 
     =
     \frac{8\sqrt{2N_c}\mathcal Q_f^2}{m_{\mathcal P}^3}
    \tilde\psi^{NR}_{\mathcal{P}}(\vec{r}=\vec 0)
    \;.
\end{align}

In designing the LFWF of the pseudoscalar meson, we impose the constraint that the transition form factors calculated from different currents are the same as the experimental value converted from the diphoton decay width according to Eq.~\eqref{eq:Pggwidth},
\begin{align}\label{eq:FPgg0_constraint}
    \left|F_{\mathcal{P}\gamma}(0,0)|_{J^+}\right|
    =\left |F_{\mathcal{P}\gamma}(0,0)|_{J^\perp}\right|
    = \left|F_{\mathcal{P}\gamma, \text{experiment}}(0,0)\right|\;.
\end{align}
In this constraint, the first equation enforces rotational symmetry in the pseudoscalar meson decay width (i.e. the integral over  $k_\perp,x$), but not at the level of the wavefunction [as in the  CLFD condition for the integrand in Eq.~\eqref{eq:P_WF_CLFD_phi0_phi1}].
The second equation then fixes coefficients in the LFWF based on the experimental measurement.
\subsubsection{The diphoton decay width in terms of the pseudoscalar decay constant}
The decay constant of the pseudoscalar $f_{\mathcal{P}}$ is defined with the axial current via the local vacuum-to-hadron matrix element,
\begin{align}\label{eq:fs}
\begin{split}
 & \bra{0} \bar{\Psi}(0)\gamma^\mu\gamma^5\Psi(0) \ket{\mathcal{P} (P)}
  =i P^\mu f_{\mathcal{P}} 
  \;.
\end{split}
\end{align}
In the LFWF representation, the pseudoscalar decay constant is an integral of the meson LFWF.
The calculation using the $J^+$ and the $J^\perp$ current components are equivalent, whereas the result extracted with the $J^-$ current has some nontrivial dependence on the meson's momentum (see Chapter. 2 of Ref.~\cite{Li:2019ijx} for more details). 
Here, we take the $J^+$ (or equivalently the $J^\perp$) current to calculate the decay constant of the pseudoscalar,
\begin{align}\label{eq:fP}
  \begin{split}
    f_{\mathcal{P}}
    =
    &\sqrt{2 N_c}
    \int_0^1\frac{\diff x}{\sqrt{x(1-x)}}
    \int\frac{\diff^2 k_\perp}{{(2\pi)}^3}
    \phi_{0/\mathcal P}(k_\perp, x)
    \;.
  \end{split}
\end{align}
The expressions of the decay constant in the basis function representation can be found in Appendix.~\ref{app:fV_basis}.

The decay constant of the pseudoscalar meson is related to the diphoton decay width through a single pole fit to the transition form factor~\cite{Feldmann:1997te, Becirevic:2013bsa}, 
$f_{\mathcal{P}}\approx F_{P\gamma}(0,0) m_{\mathcal P}^2/ (4 \mathcal Q^2_f)$ in the leading order approximation, thus
\begin{align}\label{eq:exp_fP}
  \Gamma_{\mathcal{P}\to \gamma\gamma}\approx 4\pi \mathcal{Q}_f^4\alpha_{em}^2\frac{f_{\mathcal{P}}^2}{m_{\mathcal{P}}}
  \;.
\end{align}
We could also see this approximation explicitly in the LFWF representation by comparing Eq.~\eqref{eq:fP} to Eq.~\eqref{eq:FPg_00_perp}. The left-hand side of Eq.~\eqref{eq:exp_fP} would reduce to the right-hand side in the nonrelativistic limit of $m_{\mathcal{P}}=2m_f$ and $m_f>> \braket{k_\perp}$.
The decay constant $f_{\eta_c}$ could also be extracted from the $B$ meson decay channels by assuming factorization, $\Gamma(B \to \eta_c K)/\Gamma(B\to J/\psi K)\propto (f_{\eta_c}/f_{J/\psi})^2$~\cite{CLEO:2000moj,VanRoyen:1967nq,Gourdin:1995zw,Gourdin:1994ac,Deshpande:1994mk,Ahmady:1994qf}.

As noted above, to design the pseudoscalar LFWF that gives the correct diphoton decay width, we use the more accurate equations from the transition form factor rather than the decay constant here. Still, calculating the decay constant as a simple analytical form provides an insightful approximation and a cross-check, especially for heavy systems.
\begin{figure*}[ht!]
  \centering
  \subfigure[\ $m_j = 0$]
  {\includegraphics[width=0.305\textwidth]{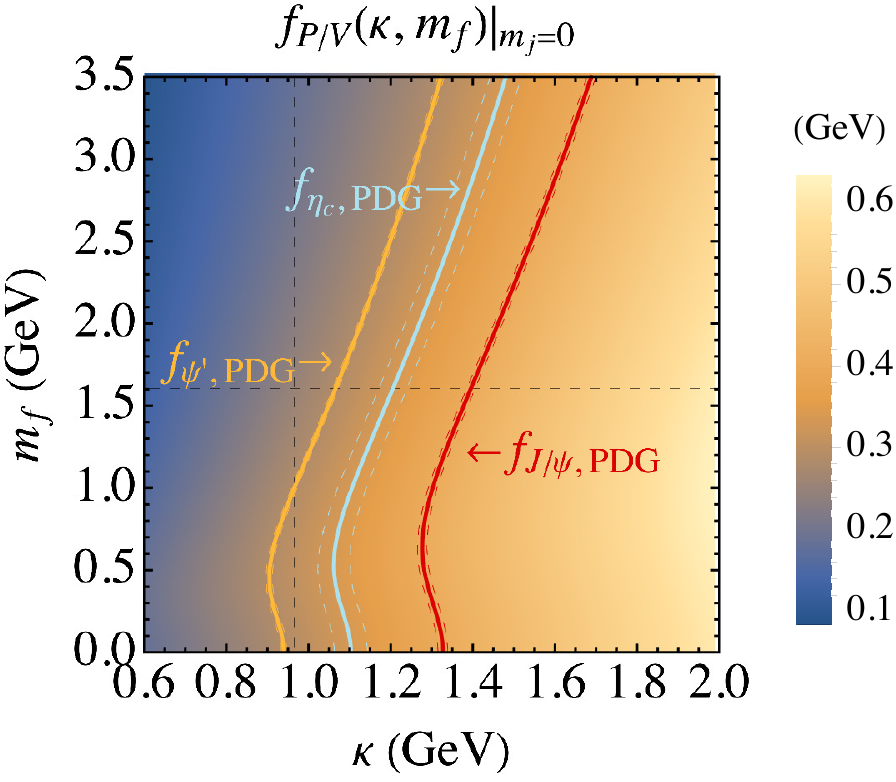}
  } 
  \subfigure[\ $m_j = \pm 1$]
  {\includegraphics[width=0.31\textwidth]{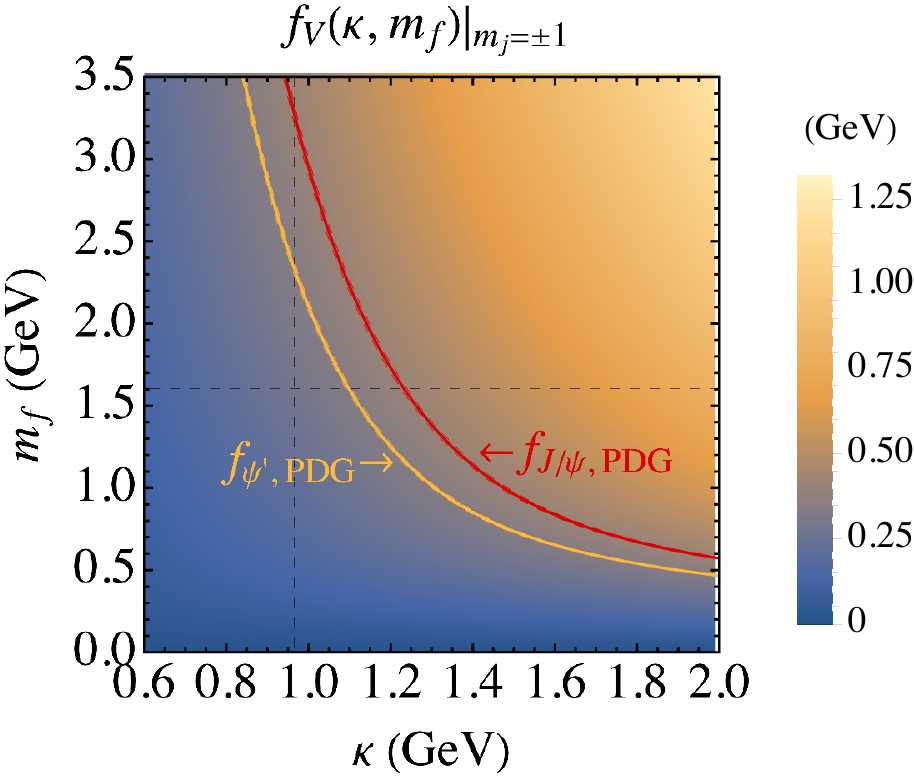}
  } 
  \subfigure[\ $J/\psi, m_j = 0, \pm 1$]
  {\includegraphics[width=0.24\textwidth]{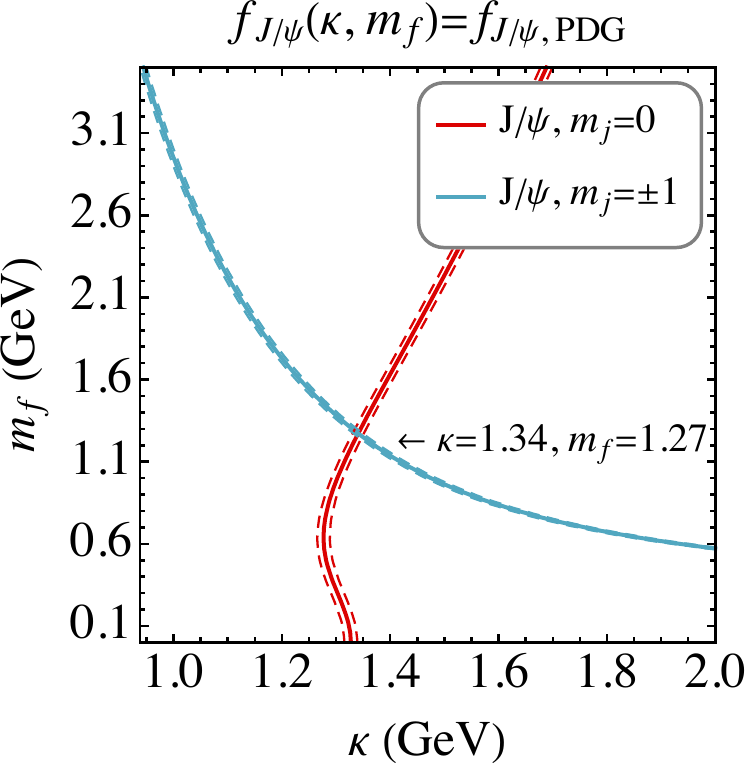}
  } 
  \caption{
    Decay constant of $\eta_c$, $J/\psi$ and $\psi'$ as single-basis states [see Eq.~\eqref{eq:Jpsi_WF_base1} and the text], at different values of $m_f$ and $\kappa$.
    Panel (a) is calculated from the $m_j=0$ state, and panel (b) is from the $m_j=\pm 1$ state.  
    In each panel, the curves are the contours where the calculated decay constants equate to the PDG values in Table.~\ref{tab:cc_mass_f}. 
    There are three lines for each state, the solid line in the middle is the mean value, and the two neighboring dashed lines are the lower and higher bounds by taking uncertainties into account.
    The values of $m_f$ and $\kappa$ from the BLFQ calculation~\cite{Li:2017mlw}, $\kappa_{BLFQ}=0.966 ~\GeV$ and $m_{f,BLFQ}= 1.603 ~\GeV$, are shown in thin dashed black lines in panels (a) and (b).
    In panel (c), decay constants of $J/\psi$  calculated from the $m_j=0$ and the $m_j=\pm 1$ states are plotted together. The two bands join at $\kappa = 1.34(1)~\GeV, m_f = 1.27(4) ~\GeV$.
  }
  \label{fig:C_kappa_mf_base_d}
\end{figure*}

We list the experimental values of the meson masses and the dilepton (diphoton) decay width for the vector (pseudoscalar) meson, of selected charmonium states, according to the Particle Data Group (PDG)~\cite{PDG2020} in Table. \ref{tab:cc_mass_f}.
The decay constant of the vector meson is converted from the associated dilepton decay width according to Eqs.~\eqref{eq:exp_fV} with $\alpha_{em}=1/134$. 
Note that we take the effective electromagnetic coupling at the meson mass scale.
The decay constant and transition form factor at zero four-momentum transfer of the pseudoscalar meson are converted from the corresponding diphoton decay width according to Eqs.~\eqref{eq:exp_fP}, and ~\eqref{eq:Pggwidth} with $\alpha_{em}=1/137$. Note that for this process, the coupling is taken at zero momentum transfer.
\begin{table}[ht]
  \caption{\label{tab:cc_mass_f} 
  Experimental values of selected charmonia states. 
  The values of meson masses and decay widths are taken from the  PDG~\cite{PDG2020}. 
  The values of decay constants and the transition from factor are calculated from the corresponding decay widths according to Eqs.~\eqref{eq:exp_fV}, ~\eqref{eq:exp_fP}, and ~\eqref{eq:Pggwidth} with $\alpha_{em}=1/134$ ($\alpha_{em}=1/137$) for the vector meson dilepton decay(the pseudoscalar meson diphoton decay).
  The uncertainties of the converted values are calculated by propagation from those of the meson masses and the decay widths. 
  The numbers in parentheses are the uncertainties, and they apply to the preceding significant digit or digits.
 }
  \centering
  \begin{ruledtabular}
  \begin{tabular}{llll}
      &  $m_{\mathcal{P}} (\GeV)$ & $\Gamma_{\mathcal{P}\to \gamma\gamma} (\keV)$  & $f_{\mathcal{P}} (\GeV)$, $F_{\mathcal{P}\gamma}(0) (\GeV^{-1})$ \\
  \hline
  $\eta_c$ & $2.9839(5)$ &$5.06(37)$ & $0.338(12)$,  $0.067(2)$
  \\
  \hline
  \hline
      &  $m_{\mathcal{V}} (\GeV)$ & $\Gamma_{\mathcal{V}\to l^+ l^-} (\keV)$ & $f_{\mathcal{V}} (\GeV)$\\
  \hline
  $J/\psi$ & $3.096900(6)$ & $5.53(10)$& $0.406(4)$ \\
  $\psi'$ & $3.68610(6)$ & $2.33(4)$ & $0.288(2)$ \\
  $\psi(3770)$ & $3.7737(4)$ & $0.262(18)$ & $0.0977(34)$ 
  \end{tabular}
  \end{ruledtabular}
\end{table}

\subsection[J/psi]{The LFWF of $J/\psi$ as a $1^{--}$ state}
\label{sec:single_basis}

In the charmonia system, $J/\psi$ is the ground vector state with $J^{PC}=1^{--}$. The dominant contribution to the $J/\psi$ state should be the ground light-front state, 1S state. For instance, the BLFQ calculations predict that the probability of finding the $J/\psi$ in a LF-1S state is larger than $99\%$~\cite{Li:2015zda, Li:2017mlw}. In consideration of that, we set the LFWF of $J/\psi$ to be a pure LF-1S state, $\psi_{\text{LF}-1S,1--}$, such that all three polarization states of $J/\psi$ have the same spatial dependence:
\begin{subequations}
  \label{eq:Jpsi_WF_base1}
  \begin{align}
    &\psi^{(m_j=0)}_{J/\psi} = 
   \psi_{\text{LF}-1S, 1--}^{(m_j=0)}
    \;,
    \label{eq:WF_base1_mj0}\\
    &
    \psi^{(m_j=1)}_{J/\psi} = 
   \psi_{\text{LF}-1S, 1--}^{(m_j=1)}
    \;,
    \label{eq:WF_base1_mj1}\\
    &
    \psi^{(m_j=-1)}_{J/\psi} = 
    \psi_{\text{LF}-1S, 1--}^{(m_j=-1)}
    \;.
    \label{eq:WF_base1_mjm1}
  \end{align}
\end{subequations}

\begin{figure}[htp!]
  \centering 
  \includegraphics[width=0.333\textwidth]{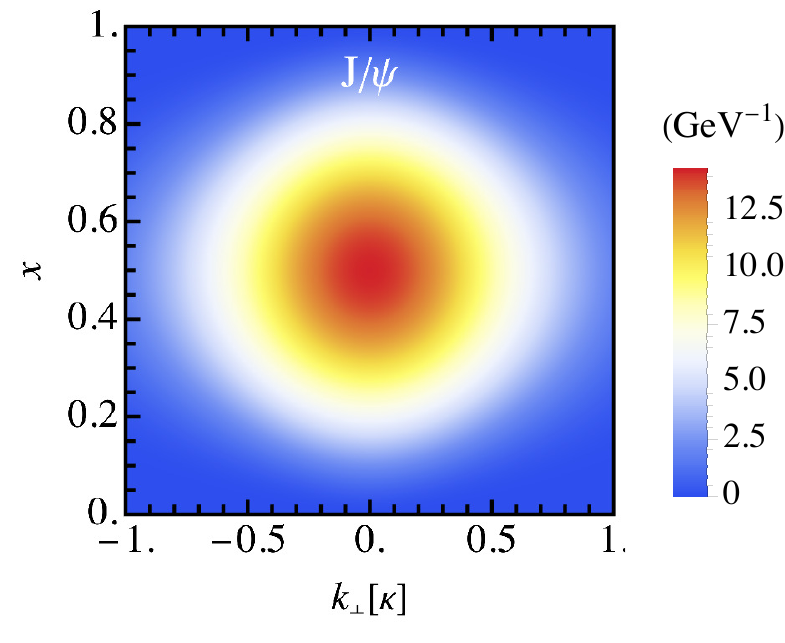} 
  \caption{Plot of the $J/\psi$ wavefunction according to Eq.~\eqref{eq:Jpsi_WF_base1} as a function of $x$ and $\vec k_\perp$ at $\theta_k=0,\pi$, with the basis parameter $\kappa =1.34~\GeV$. 
  All three polarized states of $J/\psi$ have the same spatial dependence, $\psi^{m_j=0}_{\uparrow\downarrow+\downarrow\uparrow/J/\psi} (\vec k_\perp, x) = \psi^{m_j=1}_{\uparrow\uparrow/J/\psi}(\vec k_\perp, x) = \psi^{m_j=-1}_{\downarrow\downarrow/J/\psi}(\vec k_\perp, x) $.
  The magnitude of the wavefunction is in $\GeV^{-1}$.}
  \label{fig:WF_Jpsi}
\end{figure}

With the wavefunction of Eqs.~\eqref{eq:Jpsi_WF_base1}, the decay constants of those states can be calculated as functions of $m_f$ and $\kappa$ according to Eqs.~\eqref{eq:fP_basis} to \eqref{eq:fV_mj1_basis}. 
The decay constant is plotted as a function of $m_f$ and $\kappa$ in Fig.~\ref{fig:C_kappa_mf_base_d}. Perhaps surprisingly, the lines of constant $f_{\mathcal{V}}$ in the $m_f, \kappa$-plane are very different for the transverse and longitudinal polarization states of the meson. 
Thus the point where the lines of experimentally measured $f_{J/\psi}$ intersect (i.e. where the decay width is rotational invariant and has the correct value as illustrated in Eq. (36)) gives us a very tight constraint on both $m_f$ and $\kappa$ simultaneously. The agreement of both decay constants happens at
$\kappa =\hat\kappa= 1.34(1) \GeV$ and $m_f =\hat m_f = 1.27(4)\GeV$. 
We, therefore, adopt these values of $m_f$ and $\kappa$ in our basis representation.
The LFWF for $J/\psi$ is now determined, and we present a plot of the spatial wavefunction in Fig.~\ref{fig:WF_Jpsi}.

Now that we have fixed the parameters in our basis functions, we can, as a cross-check, look at the mass of the state, using the two estimates $H_0$ and $H_{BLFQ}$ for the effective Hamiltonian. 
A calculation according to Eq.~\eqref{eq:tildeM2} yields
\begin{align*}
  \sqrt{\braket{H_0}}_{J/\psi^{(m_j=0)}} 
  = 3.44~\GeV
  ,
  \sqrt{\braket{H_{BLFQ}}}_{J/\psi^{(m_j=0)}} 
  = 2.93~\GeV\;,\\
  \sqrt{\braket{H_0}}_{J/\psi^{(m_j=\pm 1)}} 
  = 3.44~\GeV
  ,
  \sqrt{\braket{H_{BLFQ}}}_{J/\psi^{(m_j=\pm 1)}} 
  = 2.96~\GeV\;.
\end{align*}
Note that these are not the mass eigenvalues of the designed $J/\psi$ state, since we are not working with some specific Hamiltonian. These estimated masses are reasonably close to the experimental value in Table.~\ref{tab:cc_mass_f}.

Based on the nonrelativistic limit, we would expect the $\eta_c$ to be mostly a $\psi_{\text{LF}-1S,0-+}$ state with partons in the spin singlet state, i.e.  $\psi_{0,0,0}\sigma_-$. We would also expect the $\psi'$ to be predominantly a $\psi_{\text{LF}-2S,1--}$ state, and thus have a large $ \psi_{1,0,0}$ component [see Eq.~\eqref{eq:LF_2S}] with  partons in the spin triplet state, i.e.  $\psi_{1,0,0}\{\sigma_{\uparrow \uparrow},\sigma_+,\sigma_{\downarrow \downarrow}\}$ for $m_j=1,0$, and $-1$ states respectively.
It turns out that out that the pseudoscalar meson decay constant calculated from a $\psi_{0,0,0}$ wavefunction, and the vector meson decay constant calculated from $\psi_{1,0,0}$ have exactly the same expressions as a function of $\kappa$ and $m_f$ as we have for the $J/\psi$ [also see Eqs.~\eqref{eq:fV_mj0_basis}, ~\eqref{eq:fV_mj1_basis}, and ~\eqref{eq:fP_basis} and the associated discussions]. 
Since these LFWF's are expected to be close to $\eta_c$ and $\psi'$,  it is interesting to plot also the curves corresponding to the experimental values $f_{\psi'}$ and $f_{\eta_c}$ in the same $\kappa, m_f$ plane. These curves are also shown in  Fig.~\ref{fig:C_kappa_mf_base_d}. 
One immediately observes that, having fixed $\kappa$ and  $m_f$ using the $J/\psi$ decay, the decay widths of the $\psi_{0,0,0}\sigma_-$ and $ \psi_{1,0,0}$ states do not match the experimental values for $\eta_c$ and $\psi'$. 
Therefore, we must allow both of these mesons to have some contribution from other basis states to fit their decay widths. 

\subsection[etac]{The LFWF of $\eta_c$ as a $0^{-+}$ state}
In the previous section, we have constructed the LFWF for $J/\psi$ as $1^{--}$ state in Eq.~\eqref{eq:Jpsi_WF_base1}. We find the values of $\kappa$ and $m_f$ by fitting the calculated $J/\psi$ decay constant to the experimental value, and we will use the same values in constructing other states.
The pseudoscalar meson $\eta_c$ with quantum number $0^{-+}$ is the ground state in the charmonium system. It is a 1S state in the nonrelativistic limit. In this work, we build the $\eta_c$ LFWF as a relativistic bound state, admitting a predominant LF-1S component with admixtures of LF-2S and LF-1P components,
\begin{align}\label{eq:C_etac_SSP}
\begin{split}
  \psi_{\eta_c} = &
  C_{\eta_c,1S}
  \psi_{\text{LF}-1S,0-+}
  + C_{\eta_c,2S}
  \psi_{\text{LF}-2S,0-+}\\
 & + C_{\eta_c,1P}
  \psi_{\text{LF}-1P,0-+}
  \;.
\end{split}
\end{align}
The LF-1P component has a pure relativistic origin. It is forbidden in the NR limit by the nonrelativistic parity relation $\mathsf P =(-1)^{\ell+1}=1$ ($\ell=1$ for the P wave), but allowed in our case as long as the charge conjugation and mirror parity symmetries are satisfied.

We determine the values of the basis coefficients by matching the diphoton decay width $\Gamma_{\eta_c\to \gamma\gamma}$ to the experimental value. 
Using the framework developed in Sec.~\ref{sec:decay_Pgg},
we impose the constraint that the diphoton decay width calculated through the transition form factor $ F_{\eta_c\gamma}(0,0)|_{J^+}$ and $ F_{\eta_c\gamma}(0,0)|_{J^\perp}$ match the PDG value simultaneously, as illustrated in Eq.~\eqref{eq:FPgg0_constraint}.
For comparison, we also use the decay constant formula, Eq.~\eqref{eq:fP}, to determine the basis coefficients by fitting to the diphoton decay width. In this case, only the spin singlet component is involved, and the LF-1P component is absent, i.e., $ C_{\eta_c,1P}=0$.
The results are presented in Table.~\ref{tab:C_i_etac_SSP}.
The obtained $\eta_c$ LFWF, for which the basis coefficients are listed in the third column, has a large LF-1S component and small LF-2S and LF-1P components.
This set of basis coefficients are very close to the one obtained via the decay constant fit. However, there is an essential difference between the two; as we have discussed in Sec.~\ref{sec:decay_Pgg}, the LF-1P component is necessary for preserving the rotational symmetry detected via the diphoton decay. 
In comparison, the BLFQ LFWF has a smaller percentage for the dominant LF-1S component, leaving probability available for higher excited modes.

\begin{figure}[tbp]
  \centering 
  \subfigure[\  ]{
  \includegraphics[width=0.32\textwidth]{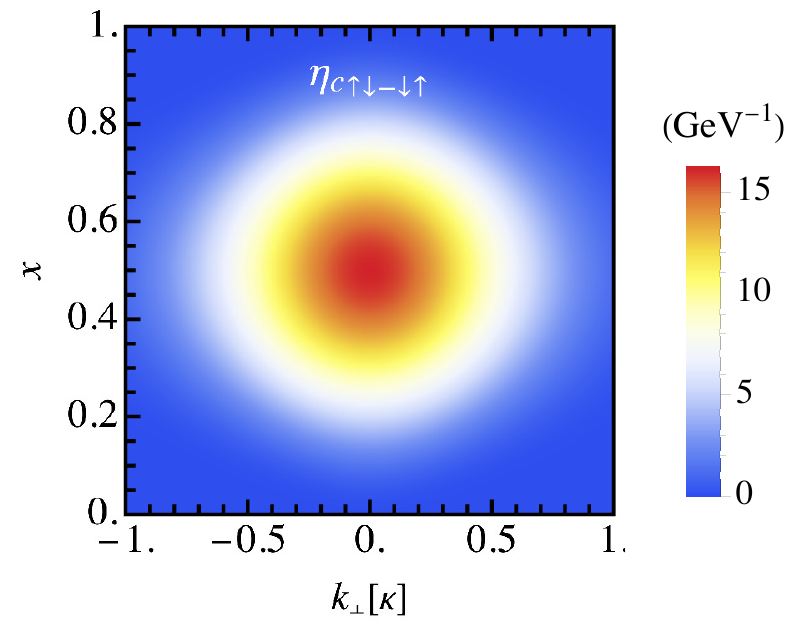} 
  }
  \subfigure[\ ]
  {\label{fig:WF_etac_SP_uu}
  \includegraphics[width=0.333\textwidth]{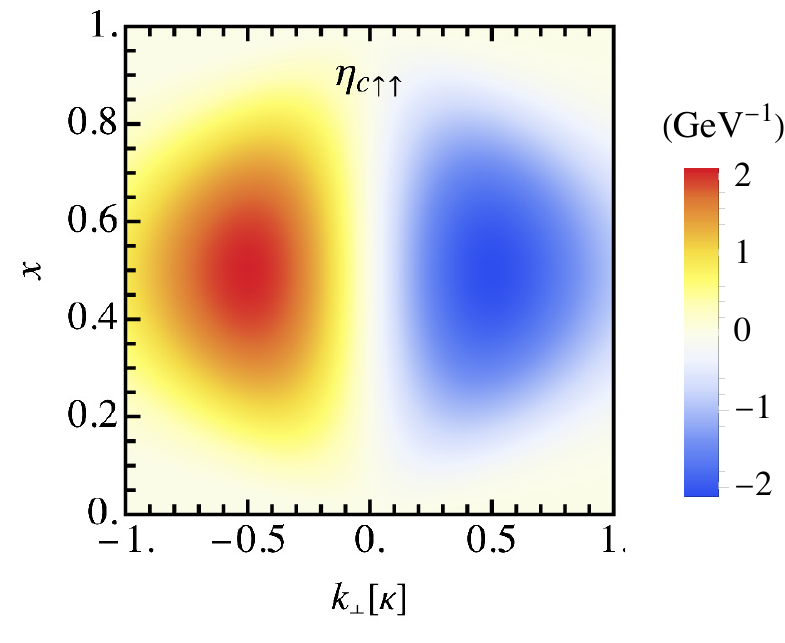}
  }
  \caption{Plot of the $\eta_c$ wavefunction according to Eq.~\eqref{eq:C_etac_SSP} as a function of $x$ and $\vec k_\perp$ at $\theta_k=0,\pi$, with the basis parameter $\kappa =1.34~\GeV$,
  (a) ${\psi}_{\uparrow\downarrow-\downarrow\uparrow/\eta_c}(\vec k_\perp, x)$, (b)${\psi}_{\uparrow\uparrow/\eta_c}(\vec k_\perp, x)={\psi}^*_{\downarrow\downarrow/\eta_c}(\vec k_\perp, x)$.
  }
  \label{fig:WF_etac_SP}
\end{figure}

\begin{figure}[htp]
  \centering 
  \subfigure[\  ]{
  \includegraphics[width=0.32\textwidth]{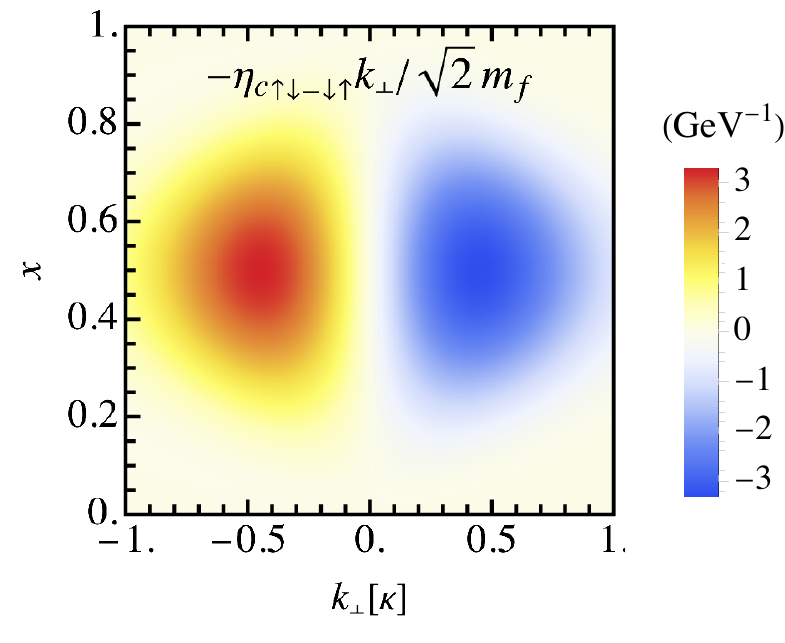} 
  }
  \subfigure[\ ]
  {
  \includegraphics[width=0.34\textwidth]{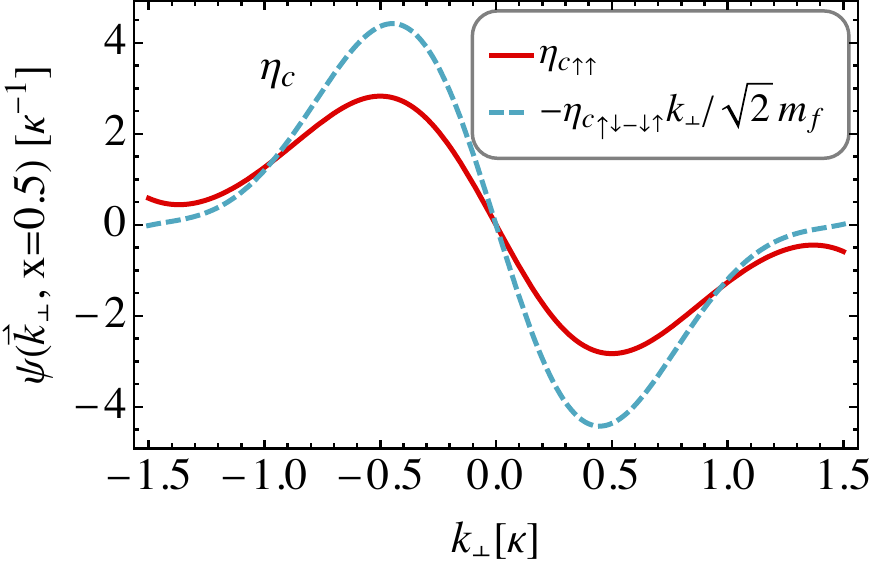}
  }
  \caption{Comparisons of different spin components of $\eta_c$. The wavefunctions are defined in Eq.~\eqref{eq:C_etac_SSP}, with the basis parameter $\kappa =1.34~\GeV$ and $m_f =1.27~\GeV$.
  (a) The spin component $\psi_{\uparrow\uparrow}$ generated from $\psi_{-}$ by the CLFD relation in Eq.~\eqref{eq:P_WF_CLFD_phi0_phi1} [contrast with Fig.~\ref{fig:WF_etac_SP_uu}]; (b) comparison of the $\psi_{\uparrow\uparrow}$ component in this work and CLFD generated component at $x=0.5$.
  }
  \label{fig:WF_etac_SP_CLFD}
\end{figure}

We present a plot of the spatial wavefunction of $\eta_c$  in Fig.~\ref{fig:WF_etac_SP}. 
In addition, we compare the relation of the two helicity components to the condition from CLFD in Eq.~\eqref{eq:P_WF_CLFD_phi0_phi1} in Fig. \ref{fig:WF_etac_SP_CLFD}. 
We see that the actual $\sigma_{\uparrow\uparrow}$ component in our LFWF and the one obtained from the $\sigma_-$ component via the CLFD condition have the same overall form and roughly the same magnitude, but are not exactly equal.

\begin{table*}[t]
  \caption{\label{tab:C_i_etac_SSP} 
  The wavefunction of $\eta_c$ defined in Eq.~\eqref{eq:C_etac_SSP}. The values of basis coefficients are obtained by fitting the diphoton decay width $\Gamma_{\eta_c \to \gamma\gamma}$ to the PDG value (see Table.~\ref{tab:cc_mass_f}) at $\kappa=\hat\kappa, m_f = \hat m_f $. Uncertainties come from those of the parameters and the PDG decay width. 
 In the second column, the values of basis coefficients are obtained by fitting the diphoton decay width via the decay constant formula according to Eq.~\eqref{eq:fP}. There, the LF-1P component is taken away.
 In the third column, the values of basis coefficients are obtained by fitting the diphoton decay width via the transition form factor $F_{\eta_c\gamma}(0,0)$ according to Eqs.~\eqref{eq:FPg_00_plus} and \eqref{eq:FPg_00_perp}.
 The corresponding LFWF is what we provide from this work.
 The percentage of each basis state is calculated by taking the square of the corresponding basis coefficient.
 The BLFQ values are extracted from the LFWF that is solved from the Hamiltonian formalism~\cite{Li:2017mlw}. 
 The meson mass is estimated by evaluating the expectation value of $H$ according to Eq.~\eqref{eq:H}.
  }
  \centering
  \begin{ruledtabular}
  \begin{tabular}{lcccc}
     & value by $f_{\eta_c}$   & value by $F_{\eta_c\gamma}(0,0)$ & percentage (\%) 
     & ~~~BLFQ percentage (\%)~\cite{Li:2017mlw}, normalized\\
  \hline
  $C_{\eta_c,1S}$ & 0.987(7)
  & 0.926(2) &85.8 
  &64.3, 77.0\\
  $C_{\eta_c,2S}$ &0.159(39) & 0.157(71)  &2.5
  &13.9, 16.6\\
  $C_{\eta_c,1P}$ &-  & 0.342(49) & 11.7
  &5.4, 6.4\\
  $\sqrt{\braket{H_0}} (\GeV)$ & $3.47(2)$   & $3.00(3)$ & & \\
  $\sqrt{\braket{H_{BLFQ}}} (\GeV)$ & $3.53(2)$   & $3.02(5)$& & 
  \end{tabular}
  \end{ruledtabular}
\end{table*}

\subsection[psi (2S)]{The LFWF of $\psi'$ as a $1^{--}$ state }
As the first excited vector state, $\psi'$ should receive a dominant contribution from the LF-2S state. However, mixing with other states is also required to obtain the measured decay constant. Thus, we allow admixture of the the LF-1S state. Hence, we construct $\psi'$ as a combination of LF-1S and LF-2S states, 
\begin{subequations} \label{eq:C_psi2s_2R_SD} 
    \begin{align}
        \psi^{(m_j=0)}_{\psi'} = &
          C^{(m_j=0)}_{\psi',1S}
          \psi_{\text{LF}-1S,1--}^{(m_j=0)}
         + C^{(m_j=0)}_{\psi',2S}
          \psi_{\text{LF}-2S,1--}^{(m_j=0)}
          \;, 
        \\
          \psi^{(m_j=1)}_{\psi'} =   &
              C^{(m_j=1)}_{\psi',1S}
          \psi_{\text{LF}-1S,1--}^{(m_j=1)}
         +  C^{(m_j=1)}_{\psi',2S}
          \psi_{\text{LF}-2S,1--}^{(m_j=1)}
          \;,
        \\
          \psi^{(m_j=-1)}_{\psi'} =   &
              C^{(m_j=-1)}_{\psi',1S}
          \psi_{\text{LF}-1S,1--}^{(m_j=-1)}
         +  C^{(m_j=-1)}_{\psi',2S}
          \psi_{\text{LF}-2S,1--}^{(m_j=-1)}
          \;.
    \end{align} 
\end{subequations}
Note that we do not require exact orthogonality between our $J/\psi$ and $\psi'$ states, which would require the admixture of additional basis states and the corresponding parameters.
The basis coefficients are determined by fitting the decay constant  [i.e., imposing Eq. (36)], at the values $\kappa=\hat\kappa, m_f = \hat m_f $ determined from the $J/\psi$. 
We also estimate the meson mass by calculating the expectation value of $H$ defined in Eq.~\eqref{eq:H}. 
The results are presented in Table.~\ref{tab:C_i_2basis_psi2s_SD}.

\begin{figure}
  \centering 
  \subfigure[\ ${\psi}^{(m_j=0)}_{\uparrow\downarrow+\downarrow\uparrow/\psi'}(\vec k_\perp, x)$ ]{
  \includegraphics[width=0.333\textwidth]{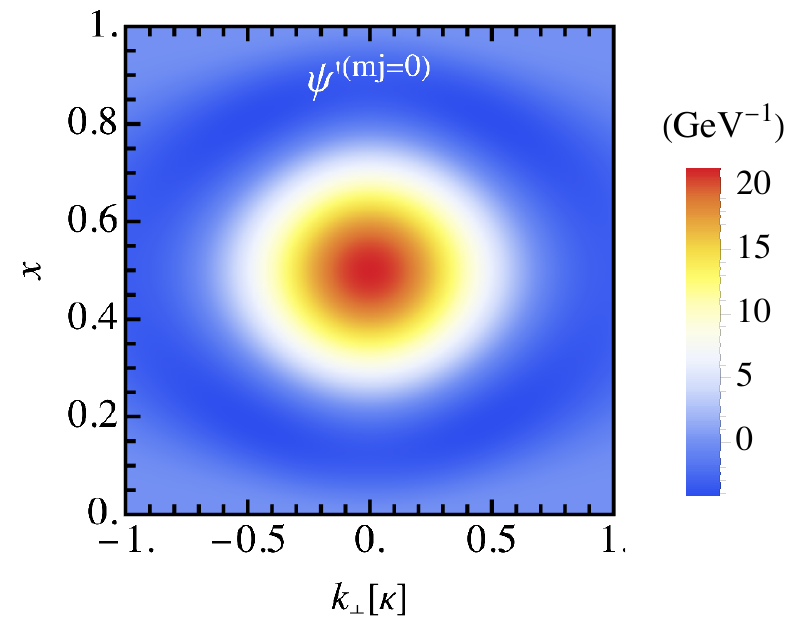} 
  }
  \subfigure[\ ${\psi}^{(m_j=1)}_{\uparrow\uparrow/\psi'}(\vec k_\perp, x)={\psi}^{(m_j=-1)}_{\downarrow\downarrow/\psi'}(\vec k_\perp, x)$]
  {
  \includegraphics[width=0.333\textwidth]{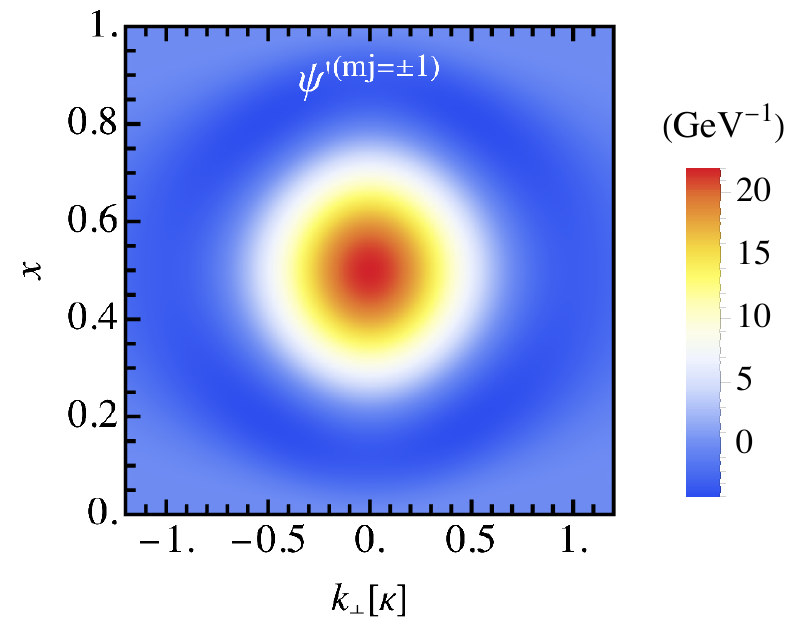}
  }
  \caption{Plot of the ${\psi'}$ wavefunction according to Eq.~\eqref{eq:C_psi2s_2R_SD} as a function of $x$ and $\vec k_\perp$ at $\theta_k=0,\pi$, with the basis parameter $\kappa =1.34~\GeV$.
  (a) ${\psi}^{(m_j=0)}_{\uparrow\downarrow+\downarrow\uparrow/\psi'}(\vec k_\perp, x)$, (b)${\psi}^{(m_j=1)}_{\uparrow\uparrow/\psi'}(\vec k_\perp, x)={\psi}^{(m_j=-1)}_{\downarrow\downarrow/\psi'}(\vec k_\perp, x)$.
  }
  \label{fig:WF_psi2s}
\end{figure}

\begin{figure}
  \centering 
  \subfigure[\ ${\psi}(\vec k_\perp=\vec 0_\perp, x)$ ]{
  \includegraphics[width=0.36\textwidth]{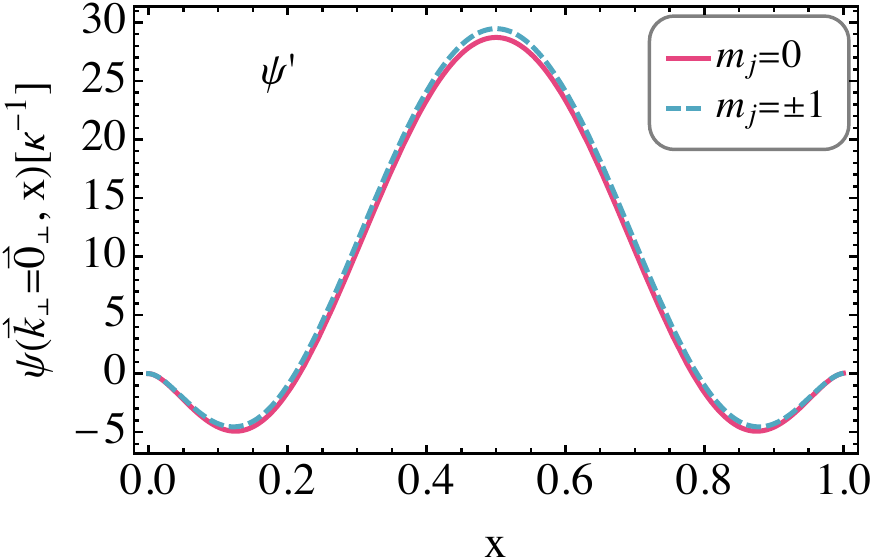} 
  }
  \qquad
  \subfigure[\ ${\psi}(\vec k_\perp, x=0.5)$]
  {
  \includegraphics[width=0.36\textwidth]{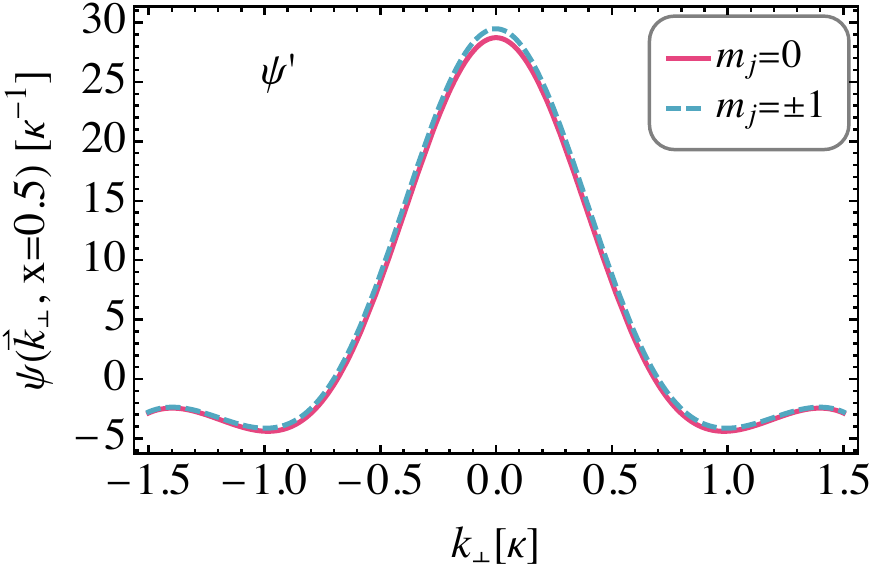}
  }
  \caption{Comparisons of different polarized states of ${\psi'}$. The wavefunctions are defined in Eq.~\eqref{eq:C_psi2s_2R_SD}, with the basis parameter $\kappa =1.34~\GeV$ and $m_f =1.27~\GeV$.
  (a) dependence of the wavefunction on $x$ at $\vec k_\perp=\vec 0_\perp$ (b) dependence of the wavefunction on $\vec k_\perp (\theta_k=0,\pi)$ at $x=0.5$.
  }
  \label{fig:WF_psi2s_slice}
\end{figure}

The LFWFs of $\psi'$ are presented in Fig.~\ref{fig:WF_psi2s}. 
In addition, we compare different $m_j$ states of $\psi'$ in the longitudinal and the transverse dimension separately in Fig.~\ref{fig:WF_psi2s_slice}.
The ${\psi'}^{(m_j=0)}$ state and the ${\psi'}^{(m_j=\pm 1)}$ state largely resemble each other in both the $x$ and $\vec k_\perp$ directions. This is expected by looking at their spatial decomposition, each being a predominantly ($> 90\%$, see Table.~\ref{tab:C_i_2basis_psi2s_SD}) LF-2S state.
The ${\psi'}^{(m_j=\pm 1)}$ state has a slightly higher peak at $\{x=1/2, \vec k_\perp=\vec 0_\perp\}$ compared to the ${\psi'}^{(m_j=0)}$ state, due to the fact that the ${\psi'}^{(m_j=\pm 1)}$ state has a somewhat larger LF-1S component ($9.2\%$ compared to $5.7\%$, see Table.~\ref{tab:C_i_2basis_psi2s_SD}).

\begin{table}[t]
  \caption{\label{tab:C_i_2basis_psi2s_SD} 
  The wavefunction of $\psi'$ defined in Eq.~\eqref{eq:C_psi2s_2R_SD}. The values of basis coefficients are obtained by fitting the decay constant to the PDG value, at $\kappa=\hat\kappa, m_f = \hat m_f $. Uncertainties come from those of the parameters and the PDG decay constants. The percentage of each basis state is calculated by taking the square of the corresponding basis coefficient.
  The BLFQ values are extracted from the LFWF that is solved from the Hamiltonian formalism~\cite{Li:2017mlw}. 
  The meson mass is estimated by evaluating the expectation value of $H$ according to Eq.~\eqref{eq:H}.
  }
  \centering
  \begin{ruledtabular}
  \begin{tabular}{lccc}
    & value & percentage (\%) & 
    \begin{tabular}{c}
         BLFQ percentage\cite{Li:2017mlw}, \\
          normalized (\%) 
    \end{tabular}
    \\
  \hline
  $C^{(m_j=0)}_{\psi',1S}$ & 0.240(5) &5.7 &18, 22.8 \\
  $C^{(m_j=0)}_{\psi',2S}$ &0.971(2) & 94.3 &61.0,77.2 \\
  $\sqrt{\braket{H_0}}(\GeV)$ & $4.42(0)$  & & \\
  $\sqrt{\braket{H_{BLFQ}}}(\GeV)$ & $4.01(0)$  & & \\
  \hline
  $C^{(m_j=\pm 1)}_{\psi',1S}$ & 0.302(18)  & 9.2 & 15.2, 19.5 \\
  $C^{(m_j=\pm 1)}_{\psi',2S}$ & 0.953(6)  & 90.8 &62.7, 80.5 \\
  $\sqrt{\braket{H_0}}(\GeV)$ & $4.39(1)$  & & \\
  $\sqrt{\braket{H_{BLFQ}}}(\GeV)$ & $4.00(1)$  & & 
  \end{tabular}
    \end{ruledtabular}
\end{table}

\subsection[psi(3770)]{The LFWF of $\psi(3770)$ as a $1^{--}$ state}
\begin{figure*}[ht]
  \centering 
  \subfigure[\ ${\psi}^{(m_j=0)}_{\uparrow\uparrow-\downarrow\downarrow/\psi(3770)}(\vec k_\perp, x)$]
  {
  \includegraphics[width=0.3\textwidth]{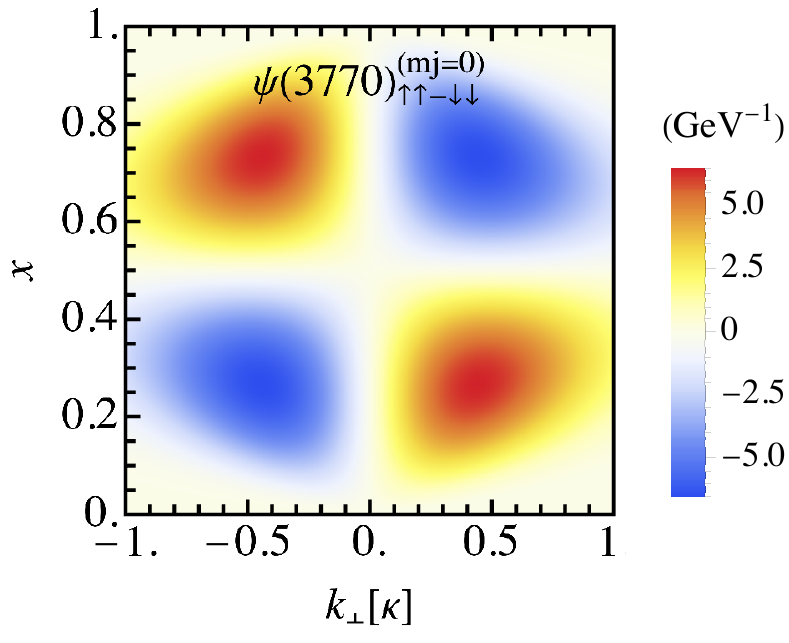}
  }  
  \subfigure[\ ${\psi}^{(m_j=0)}_{\uparrow\downarrow+\downarrow\uparrow/\psi(3770)}(\vec k_\perp, x)$ ]{
  \includegraphics[width=0.3\textwidth]{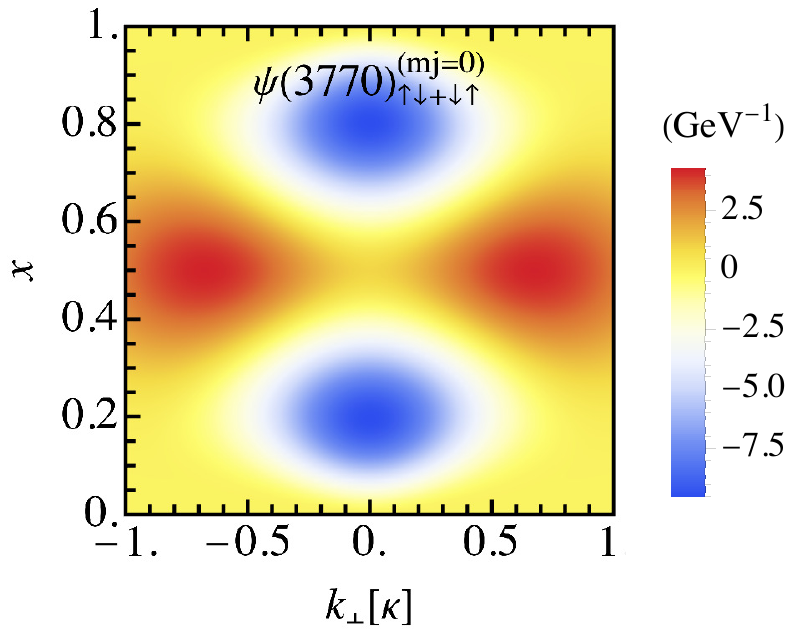} 
  }
\\
  \subfigure[\ ${\psi}^{(m_j=1)}_{\downarrow\downarrow/\psi(3770)}(\vec k_\perp, x)={\psi}^{*(m_j=-1)}_{\uparrow\uparrow/\psi(3770)}(\vec k_\perp, x)$]
  {
  \includegraphics[width=0.3\textwidth]{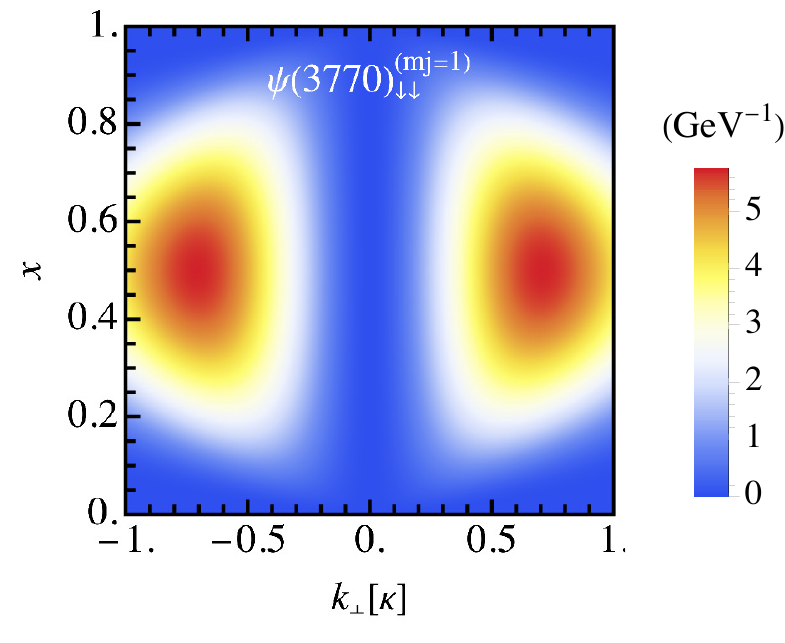}
  }
  \subfigure[\ ${\psi}^{(m_j=1)}_{\uparrow\downarrow+\downarrow\uparrow/\psi(3770)}(\vec k_\perp, x)=-{\psi}^{*(m_j=-1)}_{\uparrow\downarrow+\downarrow\uparrow/\psi(3770)}(\vec k_\perp, x)$ ]{
  \includegraphics[width=0.3\textwidth]{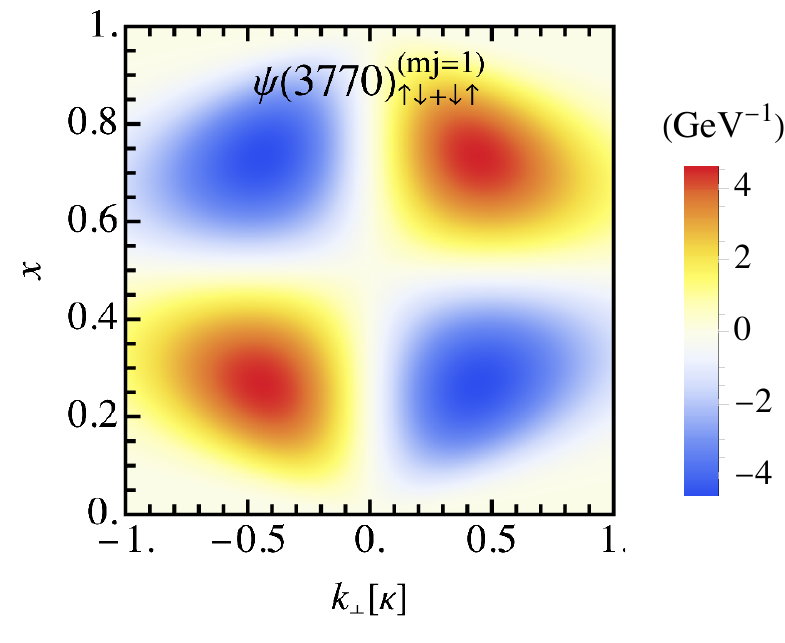} 
  }
  \subfigure[\ ${\psi}^{(m_j=1)}_{\uparrow\uparrow/\psi(3770)}(\vec k_\perp, x)={\psi}^{(m_j=-1)}_{\downarrow\downarrow/\psi(3770)}(\vec k_\perp, x)$]
  {
  \includegraphics[width=0.3\textwidth]{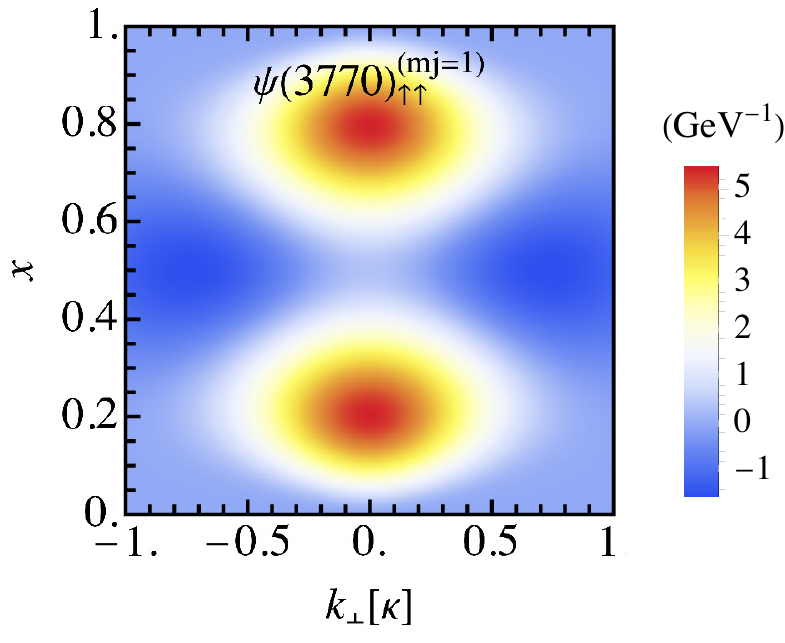}
  }
  \caption{Plot of the ${\psi(3770)}$ wavefunction according to Eq.~\eqref{eq:C_psi3770_1S2S1D} as a function of $x$ and $\vec k_\perp$ at $\theta_k=0,\pi$, with the basis parameter $\kappa =1.34~\GeV$.}
  \label{fig:WF_psi3770_1S2S1D}
\end{figure*}

The vector meson $\psi(3770)$ is recognized as primarily a 1D wave, with admixtures such as 1S and 2S waves, in reference to potential models and BLFQ calculation.
Here, we design the $\psi(3770)$ state as a linear combination of the LF-1D, LF-1S, and LF-2S states, 
\begin{subequations}\label{eq:C_psi3770_1S2S1D}
    \begin{align} \label{eq:C_psi3770_1S2S1D_mj0} 
       \begin{split}
    \psi^{(m_j=0)}_{\psi(3770)}
        =& C^{(m_j=0)}_{\psi(3770),1S}
      \psi_{\text{LF}-1S,1--}^{(m_j=0)}\\
       &+  C^{(m_j=0)}_{\psi(3770),2S}
      \psi_{\text{LF}-2S,1--}^{(m_j=0)}
    \\
       &+
     C^{(m_j=0)}_{\psi(3770),1D}
      \psi_{\text{LF}-1D,1--}^{(m_j=0)}
       \;, 
       \end{split}
    \end{align} 
    \begin{align} \label{eq:C_psi3770_1S2S1D_mj1} 
      \begin{split}
        \psi^{(m_j=1)}_{\psi(3770)} 
        =  & 
        C^{(m_j=1)}_{\psi(3770),1S}
        \psi_{\text{LF}-1S,1--}^{(m_j=1)}\\
      &+C^{(m_j=1)}_{\psi(3770),2S}
      \psi_{\text{LF}-2S,1--}^{(m_j=1)}
        \\
        &+
       C^{(m_j=1)}_{\psi(3770),1D}
        \psi_{\text{LF}-1D,1--}^{(m_j=1)}
        \;,
       \end{split}
    \end{align}  
    \begin{align} \label{eq:C_psi3770_1S2S1D_mjm1}
      \begin{split}
        \psi^{(m_j=-1)}_{\psi(3770)} = & 
        C^{(m_j=-1)}_{\psi(3770),1S}
        \psi_{\text{LF}-1S,1--}^{(m_j=-1)}\\
      &+C^{(m_j=-1)}_{\psi(3770),2S}
      \psi_{\text{LF}-2S,1--}^{(m_j=-1)}
        \\
         &+
        C^{(m_j=-1)}_{\psi(3770),1D}
         \psi_{\text{LF}-1D,1--}^{(m_j=-1)}
        \;.
       \end{split}
    \end{align}  
\end{subequations}
We impose two constraints: (1) the LF-1D component is dominant, (2) $\psi(3770)$ is orthogonal to $\psi'$.
With these considerations, we solve the basis coefficients of $\psi(3770)$ by fitting its decay constant to the PDG value [i.e., imposing Eq. (36)], at $\kappa=\hat\kappa, m_f = \hat m_f $. 
We also estimate the meson mass by calculating the expectation value of $H$ defined in Eq.~\eqref{eq:H}. 
The results are presented in Table.~\ref{tab:C_i_2basis_psi3770_1S2S1D}.
We present the different spin components of the $\psi(3770)$ state in Fig.~\ref{fig:WF_psi3770_1S2S1D}. Those plots display the features one would expect from 1D and 2S states.

\begin{table}[htp!]
  \caption{\label{tab:C_i_2basis_psi3770_1S2S1D} 
  The wavefunction of $\psi(3770)$ defined in Eq.~\eqref{eq:C_psi3770_1S2S1D}. The values of basis coefficients are obtained by fitting the decay constant to the PDG value, at $\kappa=\hat\kappa, m_f = \hat m_f $. Uncertainties come from those of the parameters and those of the PDG decay constants. The percentage of each basis state is calculated by taking the square of the corresponding basis coefficient.
  The BLFQ values are extracted from the LFWF that is solved from the Hamiltonian formalism~\cite{Li:2017mlw}. 
  The meson mass is estimated by evaluating the expectation value of $H$ according to Eq.~\eqref{eq:H}.
  }
  \centering
  \begin{ruledtabular}
       \begin{tabular}{lccc}
    & value & percentage (\%) 
    & 
        \begin{tabular}{c}
         BLFQ percentage\cite{Li:2017mlw}, \\
          normalized (\%) 
    \end{tabular}
    \\
  \hline
  $C^{(m_j=0)}_{\psi(3770),1S}$ & 0.015(10) & 0.023 & 0.04, 0.04 \\
  $C^{(m_j=0)}_{\psi(3770),2S}$ & -0.004(3) & 0.002 & 0.20, 0.23 \\
  $C^{(m_j=0)}_{\psi(3770),1D}$ & 1.000(0) & 99.975 & 87.59, 99.73 \\
  $\sqrt{\braket{H_0}} (\GeV)$ & 4.52(0) & & \\
  $\sqrt{\braket{H_{BLFQ}}} (\GeV)$ & 4.15(0) & & \\
  \hline
  $C^{(m_j=\pm 1)}_{\psi(3770),1S}$ &0.092(4)  & 0.84  & 0.007, 0.01 \\
  $C^{(m_j=\pm 1)}_{\psi(3770),2S}$ &-0.029(1)  & 0.08  & 0.09, 0.10\\
  $C^{(m_j=\pm 1)}_{\psi(3770),1D}$ & 0.995(1) & 99.08 & 88.00, 99.89 \\
  $\sqrt{\braket{H_0}} (\GeV)$ & 4.41(0)  & & \\
  $\sqrt{\braket{H_{BLFQ}}} (\GeV)$ & 3.99(0) & & 
  \end{tabular}
\end{ruledtabular}
\end{table}

\section{Meson states and observables}\label{sec:observables}
In this section, we calculate several observables for the charmonium states using the LFWFs we have constructed. 
For the $\eta_c$ observables, we adopt the amplitudes presented in the third column of Table \ref{tab:C_i_etac_SSP}.
We first estimate the masses of those states, and then calculate their charge radii and parton distribution functions. 
We use the $J/\psi$ wavefunction to calculate exclusive meson production in DIS and ultraperipheral heavy-ion collisions and compare with other model calculations.
We use the $\eta_c$ wavefunction to calculate the diphoton transition form factor and compare it with the experimental measurement.

\subsection{An estimated mass spectrum}
\begin{figure}[th!]
  \centering
  \includegraphics[width=0.4\textwidth]{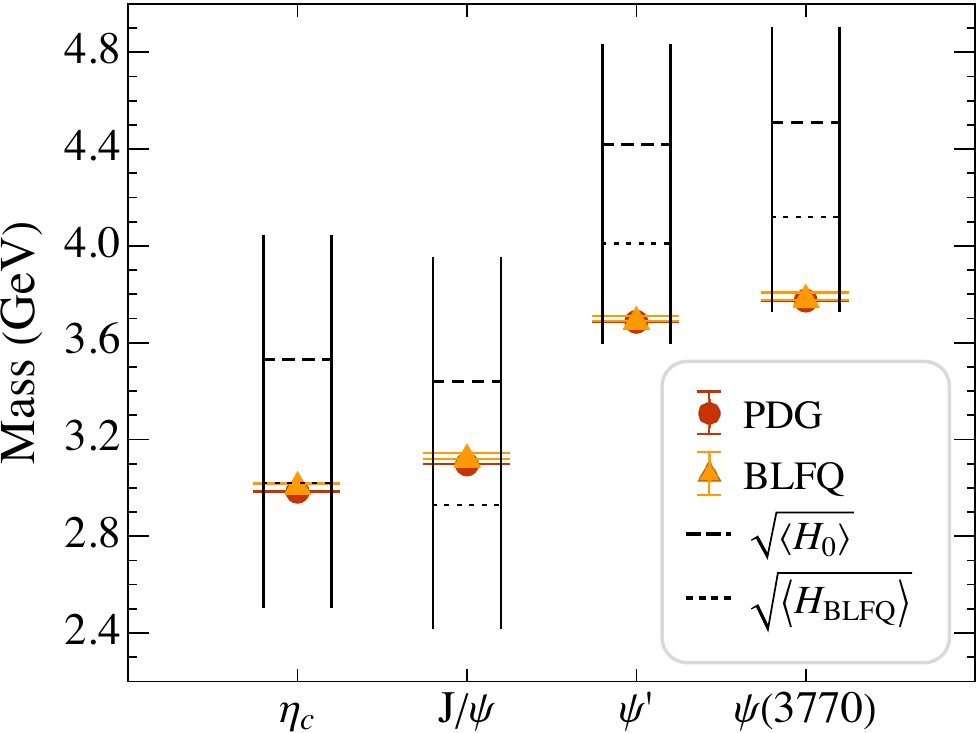}
  \caption{ The masses for the constructed states in this work are estimated by evaluating the expectation value of $H_0$ and $H_{BLFQ}$ as presented in Tables~\ref{tab:C_i_etac_SSP},~\ref{tab:C_i_2basis_psi2s_SD} and \ref{tab:C_i_2basis_psi3770_1S2S1D}.
  The dashed line is the value of $\sqrt{\braket{H_0}}$, and the dotted line is the value of $\sqrt{\braket{H_{BLFQ}}}$, both calculated using the $m_j=0$ state for vector mesons.
  The vertical solid lines extend from $\sqrt{\braket{H_{BLFQ}}}-\delta$ to $\sqrt{\braket{H_0}}+\delta$ ($\delta\equiv\left|\sqrt{\braket{H_0}}-\sqrt{\braket{H_{BLFQ}}}\right|$), indicating an open range for the estimated masses of the constructed states.
  The PDG values in the red dots are experimental measurements from Ref.~\cite{PDG2020}.
  The BLFQ values in the yellow triangles are solved from the Hamiltonian approach in Ref.~\cite{Li:2017mlw}.
  }  
  \label{fig:mass}
\end{figure}

When designing the LFWFs in Sec.~\ref{sec:LFWF}, we estimate the masses for the constructed states by evaluating their resulting expectation values for the operators $H_0$ and $H_{BLFQ}$.
We now use those values to construct an estimated mass spectrum. 
The results are shown in Fig.~\ref{fig:mass}, with a comparison to the PDG values in Ref.~\cite{PDG2020} and the BLFQ values in Ref.~\cite{Li:2017mlw}.
For each meson state, we take its values for $\sqrt{\braket{H_0}}$ and $\sqrt{\braket{H_{BLFQ}}}$ calculated from the $m_j=0$ state, and draw them in dashed and dotted lines respectively.
For all the four charmonium states, $\sqrt{\braket{H_0}}>\sqrt{\braket{H_{BLFQ}}}$, showing that the one-gluon-exchange operator reduces the mass expectation values.
We draw vertical solid lines that extend from $\sqrt{\braket{H_{BLFQ}}}-(\sqrt{\braket{H_0}}-\sqrt{\braket{H_{BLFQ}}})$ to $\sqrt{\braket{H_0}}+(\sqrt{\braket{H_0}}-\sqrt{\braket{H_{BLFQ}}})$ to indicate an open range for the estimated masses of the constructed states.
The estimated masses are in the vicinity of the PDG and the BLFQ values.

\subsection{Charge radii}
The charge radius of the meson bound state is defined in terms of the slope of the charge form factor at zero momentum transfer.
It provides important insight on the spatial structure of the system, 
\begin{align}
  \langle r^2_h \rangle = -6 \frac{\partial }{\partial Q^2} G_0(Q^2)\Big|_{Q\to 0} \;.
\end{align}
With the constructed LFWFs, the form factors can be obtained from the Drell-Yan-West formula within the Drell-Yan frame ${P'}^+=P^+$,
\begin{align}
\begin{split}
  I_{m_J,m'_J}(Q^2) 
= &\langle \psi^{(m_j')}_{h}(P')|J^+|\psi^{(m_j)}_{h}(P)\rangle / (2P^+) \\
 =& \sum_{s,\bar s} \int_0^1 \frac{\diff x}{2x(1-x)} \int
\frac{\diff^2 k_\perp}{(2\pi)^3} \\
&\psi^{(m_j')*}_{s\bar s/h}(\vec k_\perp+(1-x)\vec q_\perp, x) 
\psi^{(m_j)}_{s\bar s/h}(\vec k_\perp, x)
\;,
\end{split}
\end{align}
where $q = P'-P$, and $Q^2 = -q^2 = \bm q^2_\perp$. 
For the pseudoscalar $\eta_c$, it directly produces the charge form factor $G_0(Q^2) = I_{0,0}(Q^2)$. For vector mesons, we adopt the prescription of Grach and Kondratyuk \cite{Grach:1983hd},
\begin{align}
  G_0 = \frac{1}{3}\big[ (3-2\eta) I_{1,1} + 2\sqrt{2\eta}I_{1,0} + I_{1,-1}
\big]\;, 
\end{align}
where $\eta = Q^2/(4M_h^2)$, $M_h$ is the mass of the hadron. 
Table~\ref{tab:rms_radii} lists the r.m.s. radii of the states studied in this work.
The values calculated from the BLFQ wavefunctions~\cite{Li:2017mlw} are also listed for comparison.
 From our results, the radius of $\eta_c$ is about 46\% larger than the radius of the $J/\psi$. 
 This is because the $J/\psi$ is a LF-1S state by design, whereas $\eta_c$ admits small LF-2S and LF-1P components. Both the LF-2S and LF-1P states have a larger radius than the LF-1S state.
 
\begin{table}
  \centering 
 \caption{ \label{tab:rms_radii}
 The mean squared radii (in $\text{fm}^2$) of $\eta_c$, $J/\psi$, $\psi'$ and $\psi(3770)$ calculated from their constructed LFWFs.
 }
\begin{ruledtabular}
\begin{tabular}{l cccc}
  ($\text{fm}^2$)
  & $~~~\langle r^2_{\eta_c} \rangle~~~~~~$ 
  & $~~~\langle r^2_{J/\psi} \rangle~~~~~~$ 
  & $~~~\langle r^2_{\psi'} \rangle~~~~~~$ 
  & $~~~\langle r^2_{\psi(3770)} \rangle$ 
 \\
 \hline
 this work                     
  & 0.098 
  & 0.046
  & 0.154 
  & 0.138
   \\
   BLFQ~\cite{Li:2017mlw}                    
    & 0.029(1)
    & 0.0402(2)
    & 0.13(0)
    & 0.13(0)
\end{tabular}
  \end{ruledtabular}
\end{table}

\subsection{Parton distributions}
The quark Parton Distribution Functions (PDFs) represent the probability of finding a quark carrying momentum fraction $x$ in the hadron state.
They are essential ingredients in describing the hadron structure for hard scattering processes such as deep inelastic scattering from a hadron target and the hadron-hadron Drell-Yan process~\cite{Mulders:1995dh,Bacchetta:2001rb}. 
In the light-front formalism, the PDFs of a hadron state can be evaluated by integrating out the transverse momentum of the wavefunction overlaps.
For the pseudoscalar meson $\eta_c$, the quark's PDF is defined as
\begin{align}\label{eq:q}
q(x) = \sum_{s, \bar s}
\int
\frac{\diff^2k_\perp}{2x(1-x)(2\pi)^3} 
\left| \psi_{s\bar s/h}(\vec k_\perp, x) \right|^2 \;.
\end{align}
This PDF is normalized to unity in our model in the sense that there is one valence quark,
\begin{align}
  \int_0^1  \diff x \,q(x) =1\;.
\end{align}

For vector mesons, the PDFs are defined in terms of the quark-quark correlation function probed in the spin-one target lepton-hadron scattering~\cite{Soper:1976jc,Soper:1979fq,Manohar:1990kr,Mulders:1995dh,Bacchetta:2001rb,Kaur:2020emh}.
The PDFs appear as the parametrization coefficients in front of the Dirac $\gamma$-matrices, and there are four time-reversal even distributions at the leading twist. 
We write those PDFs in terms of the light-front helicity matrix elements, defined as
\begin{multline}
  A^{m_j',m_j}_{s',s}(x) = \\
  \sum_{\bar s}
  \int
  \frac{\diff^2k_\perp}{2x(1-x)(2\pi)^3} 
  \psi^{(m_j')*}_{s'\bar s/h}(\vec k_\perp, x) 
  \psi^{(m_j)}_{s\bar s/h}(\vec k_\perp, x) 
  \;,
\end{multline}
where the initial (final) state helicity of the hadron is $m_j$ ($m_j'$) and that of the quark is $s$ ($s'$).
The unpolarized PDF $f_1(x)$ is expressed as
\begin{align}\label{eq:f1}
  \begin{split}
    f_1(x) = &
    \frac{1}{3}
    \sum_{s}
    \bigg[
    A^{0,0}_{s,s}(x) 
    +A^{1,1}_{s,s}(x) 
    +A^{-1,-1}_{s,s}(x) 
    \bigg]
    \;.      
  \end{split}
\end{align}
It represents the unpolarized quark distributions in the unpolarized spin-one hadron.
Similar to the pseudoscalar PDF $q(x)$ in Eq.~\eqref{eq:q}, $f_1(x)$ is also normalized to unity in our model,
\begin{align}
  \int_0^1  \diff x f_1(x) =1\;.
\end{align}
The tensor polarized PDF $f_{1LL}(x)$ represents the difference of unpolarized quark distributions in the transversely polarized spin-one hadron with spin projection $m_j=0$ and $m_j=\pm 1$, and is sensitive to the quark's orbital angular momentum~\cite{Hoodbhoy:1988am,Kumano:2010vz,Ninomiya:2017ggn}. 
It is expressed as
\begin{align}\label{eq:f1LL}
  \begin{split}
    f_{1LL}(x) = &
    \sum_{s}
    \bigg[
    A^{0,0}_{s,s}(x) 
    -\frac{1}{2} A^{1,1}_{s,s}(x) 
    -\frac{1}{2} A^{-1,-1}_{s,s}(x) 
    \bigg]
    \;.      
  \end{split}
\end{align}

The longitudinally polarized PDF $h_1(x)$ is expressed as
  \begin{align}\label{eq:h1}
    \begin{split}
      h_1(x) = &
      \frac{1}{2\sqrt{2}}
      \bigg[
      A^{1,0}_{\uparrow,\downarrow}(x) 
      + A^{0,1}_{\downarrow,\uparrow}(x) \\
      &
      + A^{0,-1}_{\uparrow,\downarrow}(x) 
      + A^{-1,0}_{\downarrow,\uparrow}(x) 
      \bigg]
      \;.      
    \end{split}
    \end{align}
It describes the distribution of the longitudinally polarized quark in the longitudinally polarized meson.

The transversely polarized PDF $g_1(x)$ is expressed as
\begin{align}\label{eq:g1}
  \begin{split}
    g_1(x) = &
    \frac{1}{2}
    \bigg[
    A^{1,1}_{\uparrow,\uparrow}(x) 
    - A^{1,1}_{\downarrow,\downarrow}(x) \\
    &
    - A^{-1,-1}_{\uparrow,\uparrow}(x) 
    + A^{-1,-1}_{\downarrow,\downarrow}(x)
    \bigg]
    \;.      
  \end{split}
  \end{align}
It describes the distribution of the transversely polarized quark in the transversely polarized meson.
Note that there are different conventions in naming the polarized PDFs $h_1(x)$ and $g_1(x)$. For example, in Ref.~\cite{Tangerman:1994eh}, the former is referred to as the transversely polarized PDF whereas the latter as the longitudinally polarized PDF, which is linked with calling the $\lambda=1$ nucleon a longitudinally polarized state.

\begin{figure*}
 \centering 
 \includegraphics[width=0.34\textwidth]{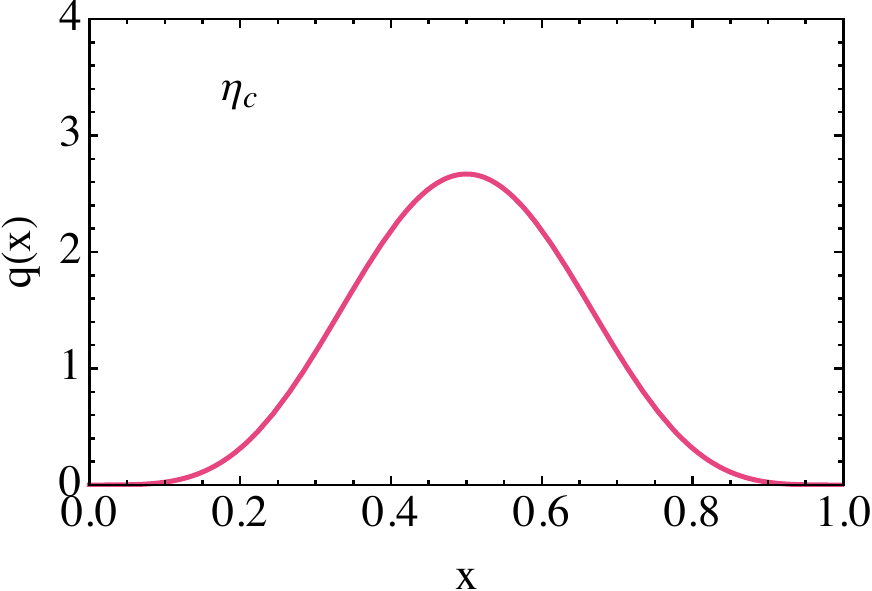} \quad 
 \includegraphics[width=0.34\textwidth]{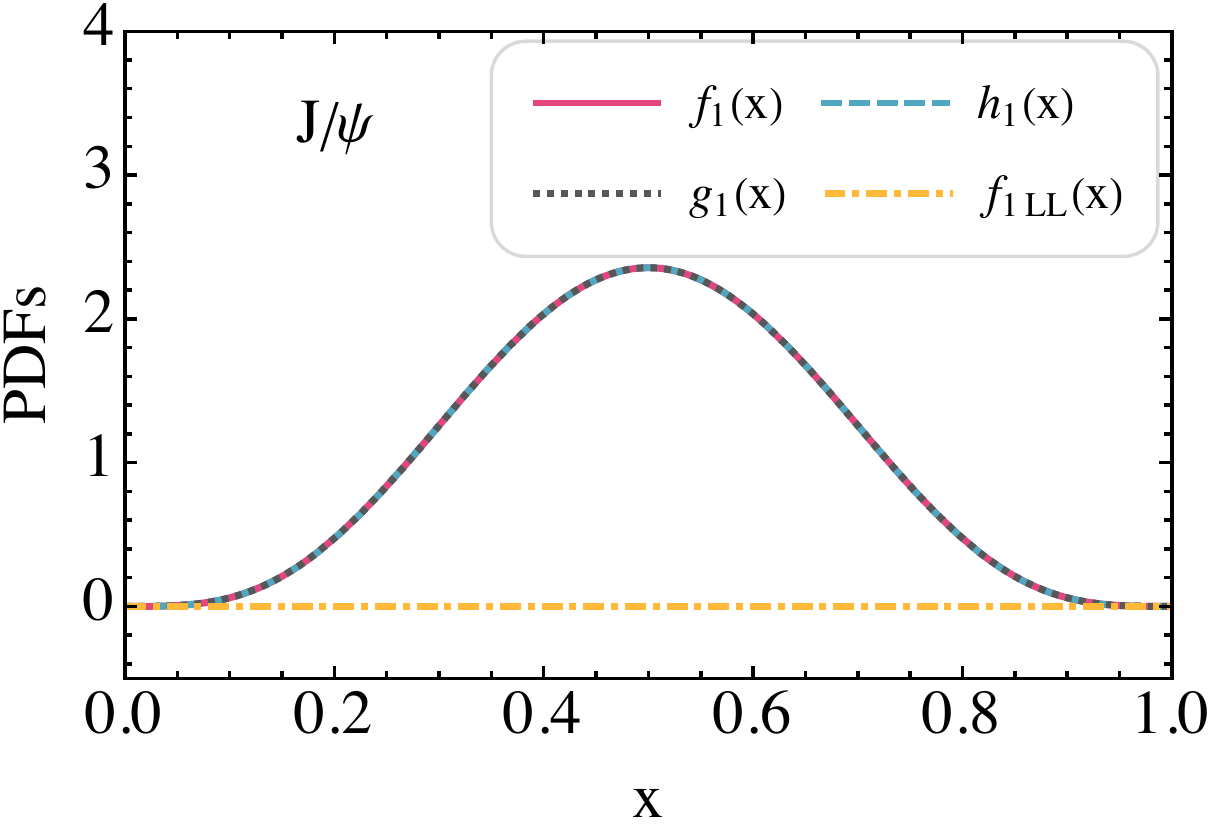}
 \includegraphics[width=0.34\textwidth]{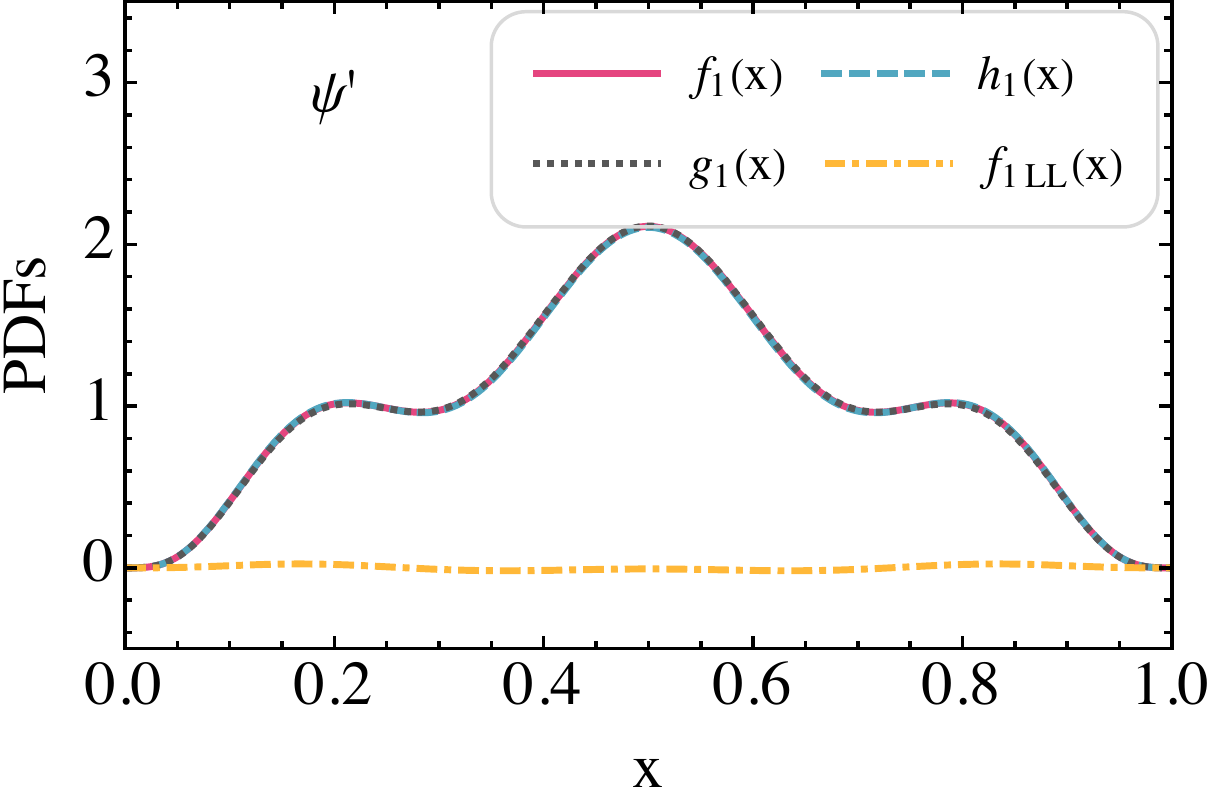} \quad 
 \includegraphics[width=0.34\textwidth]{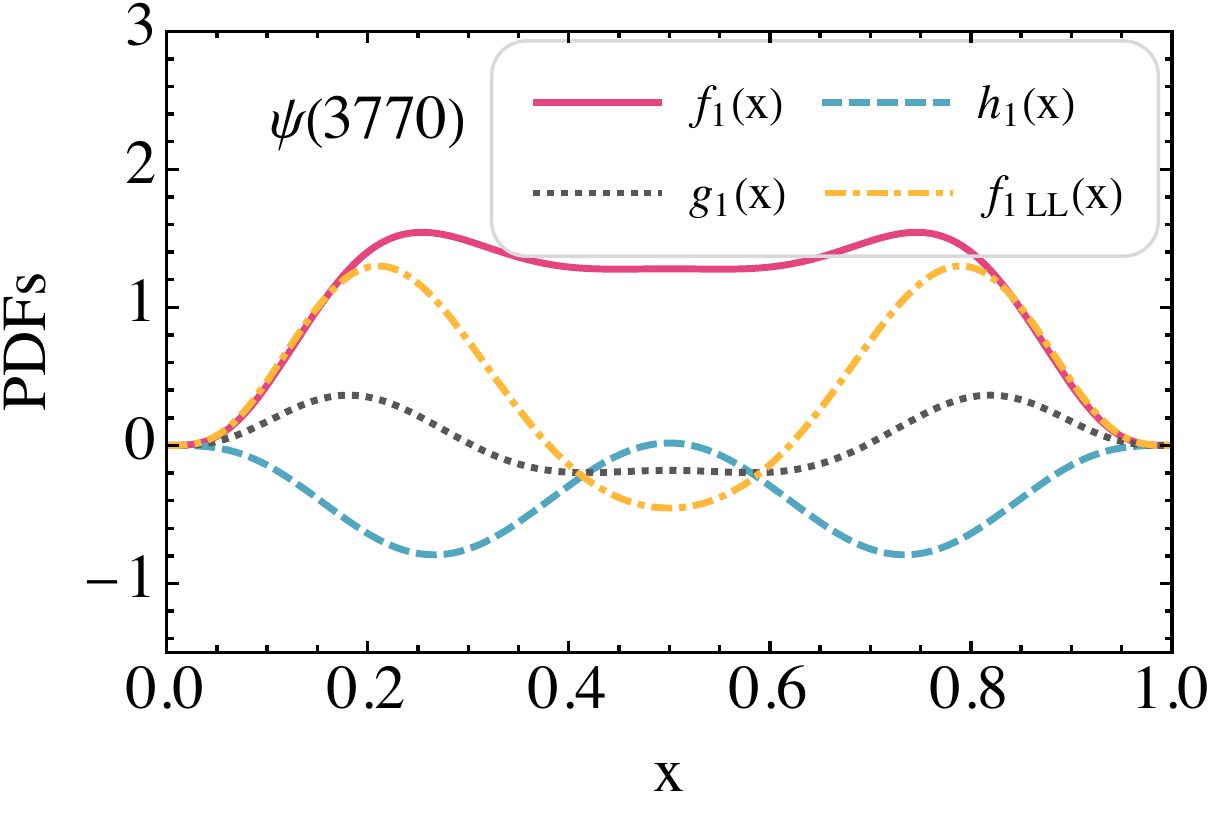}
 \caption{PDFs of $\eta_c$, $J/\psi$, $\psi'$ and $\psi(3770)$, calculated from the constructed LFWFs according to Eq.~\eqref{eq:q} and Eqs.~\eqref{eq:f1}-\eqref{eq:f1LL}. }
 \label{fig:pdf}
\end{figure*}

Figure~\ref{fig:pdf} shows the PDFs of $\eta_c$, $J/\psi$, $\psi'$ and $\psi(3770)$. 
The PDF of $\eta_c$ is peaked at $x=1/2$, reflecting its structure as a predominantly LF-1S wave.
For $J/\psi$, the three PDFs $h_1(x)$, $g_1(x)$, and $f_1(x)$ are identical, and the tensor polarized PDF $f_{1LL}(x)$ is 0.
This is because the designed $J/\psi$ is a pure LF-1S wave with spin-triplet configuration, so its spatial dependence of each polarized state is the same.
In the case of $\psi'$, the three PDFs $h_1(x)$, $g_1(x)$, and $f_1(x)$ are only slightly different, and they reflect excitations in the longitudinal direction from the large LF-2S component.
The $\psi'$'s tensor polarized PDF $f_{1LL}(x)$ deviates slightly from 0, indicating the resemblance between the transversely and longitudinally polarized states.
The PDFs of $\psi(3770)$ admit several interesting features.
The unpolarized PDF $f_1(x)$ of $\psi(3770)$ has an extensive flat region, indicating its large angular excitation. A recent study on $\rho$ meson PDF also reveal a large flat region using the light-front holographic wavefunction~\cite{Kaur:2020emh}.
Compared to $J/\psi$ and $\psi'$, the polarized PDF $h_1(x)$ and $g_1(x)$ of $\psi(3770)$ are each different from $f_1(x)$. In addition, the tensor polarized PDF $f_{1LL}(x)$ of $\psi(3770)$ has a much larger amplitude, indicating the difference between the spatial dependences of different polarized states.

\subsection{Vector meson production}
In this section, we study exclusive charmonium production in diffractive deep inelastic scattering and ultra-peripheral heavy ion collisions within the dipole picture. 
We employ the LFWF of $J/\psi$ obtained in Sec.~\ref{sec:LFWF}. 
We also make calculations of charmonium production using the BLFQ LFWF \cite{Li:2015zda} and boosted-Gaussian LFWF for comparisons. 

For exclusive heavy quarkonium production in DIS, the amplitude in the dipole model can be calculated as~\cite{Kowalski:2006hc}  
\begin{multline}
 \label{eq:newampvecm}
  \mathcal{A}^{\gamma^* p\rightarrow h p}_{T,L}
  (x_B,Q,t) = i 
  \int \diff^2 r_\perp
  \int_0^1 \frac{\diff x }{4\pi}
  \int \diff^2 b_\perp\\
(\psi_h^{*}\psi_\gamma)_{T,L} (\vec r_\perp,x,Q) 
  e^{-i[\vec b_\perp-(1-x)\vec r_\perp]\cdot \vec \Delta_\perp}
  \frac {\diff\sigma_{q\bar q}}{\diff^2 b_\perp} (x_B,\vec r_\perp) \; ,    
\end{multline} 
where $T$ and $L$ specify the transverse and longitudinal polarization of the virtual photon (with virtuality $Q^2$) and the produced 
quarkonium, and $t= - |\vec \Delta_\perp|^2$ is the momentum transfer squared. On the right-hand side, the transverse size of the color
dipole is denoted by $\vec r_\perp$, the LF longitudinal momentum fraction of the quark is denoted by $x$, the impact parameter of the dipole relative to the proton is denoted by $\vec b_\perp$ and $x_B$ is the Bjorken variable. 
Here, $\psi_\gamma$ and $\psi_h$ are the LFWFs of the virtual photon and the exclusively produced quarkonium, respectively (see the explicit expression of the photon LFWF in Appendix~\ref{app:photon}).
The cross section is related to the amplitude via 
\begin{align}
\frac{\diff \sigma^{\gamma^* p\rightarrow h p}_{T,L}}{\diff t} (x_B,Q) = \frac{1}{16 \pi} 
\left| \mathcal{A}^{\gamma^* p\rightarrow
h p}_{T,L}(x_B,Q,t)  \right|^2 \; .
\end{align}
Furthermore, we implement two phenomenological corrections in the calculation of the cross section: the contribution from the real part of the scattering amplitude \cite{Kowalski:2006hc}, and the skewedness correction \cite{Shuvaev:1999ce}, which takes into account the fact that two gluons interacting with the dipole are carrying slightly different momentum fractions (consult Ref.~\cite{Chen:2016dlk} for the details of the implementations).
\begin{figure}[htp!]
  \centering
    \subfigure[ \label{OLapL} Longitudinal polarization ($m_j=0$)]{
  \includegraphics[width=0.4\textwidth]{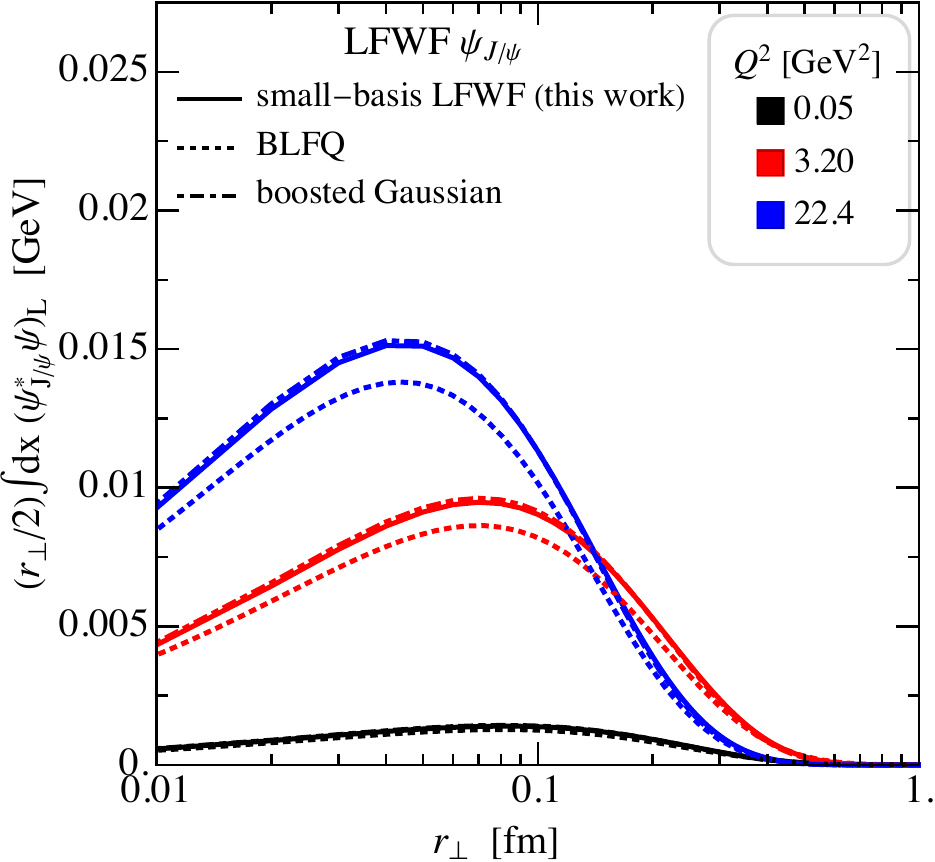}
  }
    \subfigure[ \label{OLapT}Transverse polarization ($m_j=\pm 1$)]{
  \includegraphics[width=0.4\textwidth]{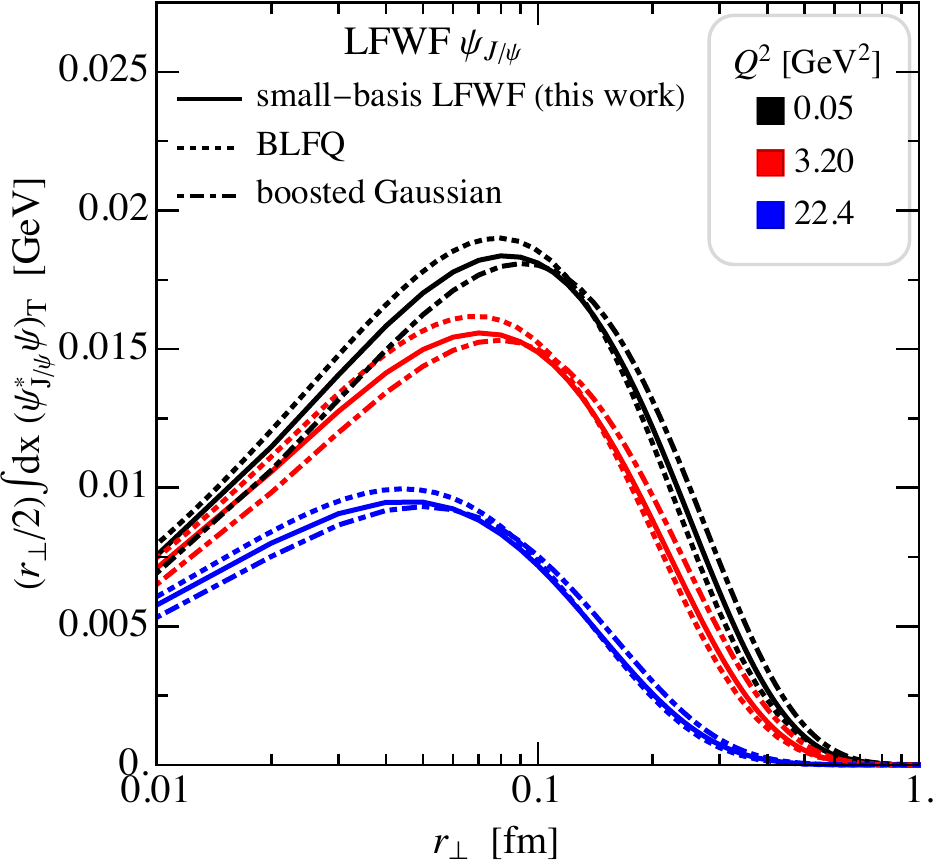}
  }
  \caption{
  The $J/\psi$-photon overlap functions [Eq. \eqref{eq:overlap}] using three different $J/\psi$ LFWFs at $Q^2$ values representative of the data, for (a) the longitudinal polarization and (b) the transverse polarization.
  Solid lines are obtained using the small-basis LFWF (this work), dotted lines the BLFQ LFWF \cite{Li:2015zda}, and dot-dashed lines the boosted Gaussian LFWF~\cite{Armesto:2014sma}.
  }
  \label{fig:overlap}
\end{figure}

We take the leading order perturbative calculations to obtain the photon's wavefunction.
With the $J/\psi$ LFWF from Eq.~\eqref{eq:Jpsi_WF_base1}, the overlap functions are, written explicitly,
\begin{subequations}
    \begin{align}
      (\psi_{J/\psi}^*\psi_\gamma)_{L} &=\mathcal Q_f e \, \frac{\sqrt{2N_c}}{\pi}\,
      Qx(1-x)\,K_0(\epsilon r_\perp)\,
      \phi(r_\perp,x),
        \label{eq:overl}
        \\
      (\psi_{J/\psi}^*\psi_\gamma)_{T} &= \mathcal Q_f e\, \frac{\sqrt{2N_c}}{2\pi} m_f K_0(\epsilon r_\perp)\phi(r_\perp,x),
      \label{eq:overt}
  \end{align} 
\end{subequations}
with 
\begin{align}
    \begin{split}
        \phi(r_\perp,x) =& \kappa \sqrt{4(2\alpha+1)} \frac{ \sqrt{\Gamma(2\alpha+1)}}{\Gamma(\alpha+1)}\\ &\exp\left(-\frac{\kappa^2 x(1-x) r_\perp^2}{2}\right) [x(1-x)]^{\frac{\alpha+1}{2}}\;,
    \end{split}
\end{align}
where $e=\sqrt{4\pi\alpha_{\mathrm{em}}}$, $\mathcal Q_f=\mathcal Q_c=2/3$ is the charge fraction carried by the quark, $K_0$ the modified Bessel function of the second kind, $\epsilon^2 \equiv x(1-x)Q^2+m_f^2$ and $N_c=3$ the number of colours. 
In Fig.~\ref{fig:overlap}, we show the overlap between the photon and the $J/\psi$ wavefunctions integrated over $x$ at different photon virtualities.
To be precise, we plot the quantity
\begin{equation}\label{eq:overlap}
  2\pi r_\perp\int_0^1\!
  \frac{\diff x }{4\pi}\;(\psi_{J/\psi}^*\psi_\gamma)_{T,L}.
\end{equation}
We take three different versions of the $J/\psi$ LFWF, the one designed in this work  (denoted as ``$1$-basis" BLFQ in the figure), 
the BLFQ LFWF \cite{Li:2015zda}, and the boosted Gaussian with $m_c=1.27 \GeV$~\cite{Armesto:2014sma}. 
The LFWF in this work is built from the same basis functions used in BLFQ. 
However, the BLFQ LFWF is solved by diagonalizing the light-front Hamiltonian in a much larger basis space and fitting to the meson mass spectrum, whereas our LFWF and the boosted Gaussian are obtained by fitting to the decay widths.
Our LFWF exhibits similarity with the boosted Gaussian by having a simple analytical form, but the structures are different.
From Fig.~\ref{fig:overlap}, we see that the $J/\psi$-photon overlaps calculated with the three $J/\psi$ LFWFs are of the same magnitude and similar overall shape at each $Q^2$. In the case of the longitudinally polarized states, as in Fig.~\ref{OLapL}, the overlaps from this work and the boosted Gaussian are very close and slightly higher than BLFQ.   
For the transversely polarized states, as in Fig.~\ref{OLapT}, the result from this work is roughly in between that from BLFQ and boosted Gaussian.

We use the bCGC dipole model for the dipole cross section,
\begin{align}\label{eq:bcgc}
\begin{split}
   &\frac {\diff\sigma_{q\bar q}}{\diff^2\vec b_\perp}
   =  2 \mathcal{N}(r_\perp Q_s,x_B) \\
  & =  2 
  \begin{cases}
    \mathcal{N}_0 \left(\frac{r_\perp Q_s}{2}\right)^{2(\gamma_s + \frac{1}{\kappa_s \lambda_s \ln (1/x_B)} \ln \frac{2}{r_\perp Q_s})}\;,  
    & \quad r_\perp Q_s\le 2\\
    1-\mathrm{e}^{-\mathcal{A} \ln^2(\mathcal{B} r_\perp Q_s)} \;,
    & \quad r_\perp Q_s>2
  \end{cases}\;.
\end{split}
\end{align}
Here, $Q_s\equiv Q_s(x_B)=(x_{B,0}/x_B)^{\lambda_s/2} Q_0$ and $Q_0=1$~GeV; $\gamma_s$, $\kappa_s$, $\lambda_s$ are parameters to be determined by inclusive DIS data \cite{Abramowicz:2015mha}; $\mathcal{A}$ and $\mathcal{B}$ should be evaluated by continuity conditions at $r_\perp Q_s = 2$. We use one of the parametrizations in Ref.~\cite{Rezaeian:2013tka} for this investigation, which we provide in Table~\ref{tab:bCGC}. 
\begin{table}
  \centering
  \caption{Parameters of the bCGC model in Eq.~\eqref{eq:bcgc} determined from fits to combined HERA data \cite{Rezaeian:2013tka}.}
  \label{tab:bCGC}
\begin{ruledtabular}
  \begin{tabular}{ccccccccc}
    \begin{tabular}{c}
          $B_\text{CGC}$ \\
        ($\GeV^{-2}$) 
    \end{tabular}
      & 
        \begin{tabular}{c}
          $m_c$ \\
      ($\GeV$) 
    \end{tabular}
      & $\gamma_s$ & $\mathcal{N}_0$ & $x_{B,0}$ & $\lambda_s$ & $\chi^2/\text{d.o.f.}$ \\ \hline
     $5.5$ & $1.27$ & $0.6599$ & $0.3358$ & $0.00105$ & $0.2063$ &
     $1.241$ 
  \end{tabular}
  \end{ruledtabular}
\end{table}

\begin{figure*}[phtb!]
 \centering 
  \subfigure[]{\label{fig:hera_Q2}
\includegraphics[width=.3\textwidth]
{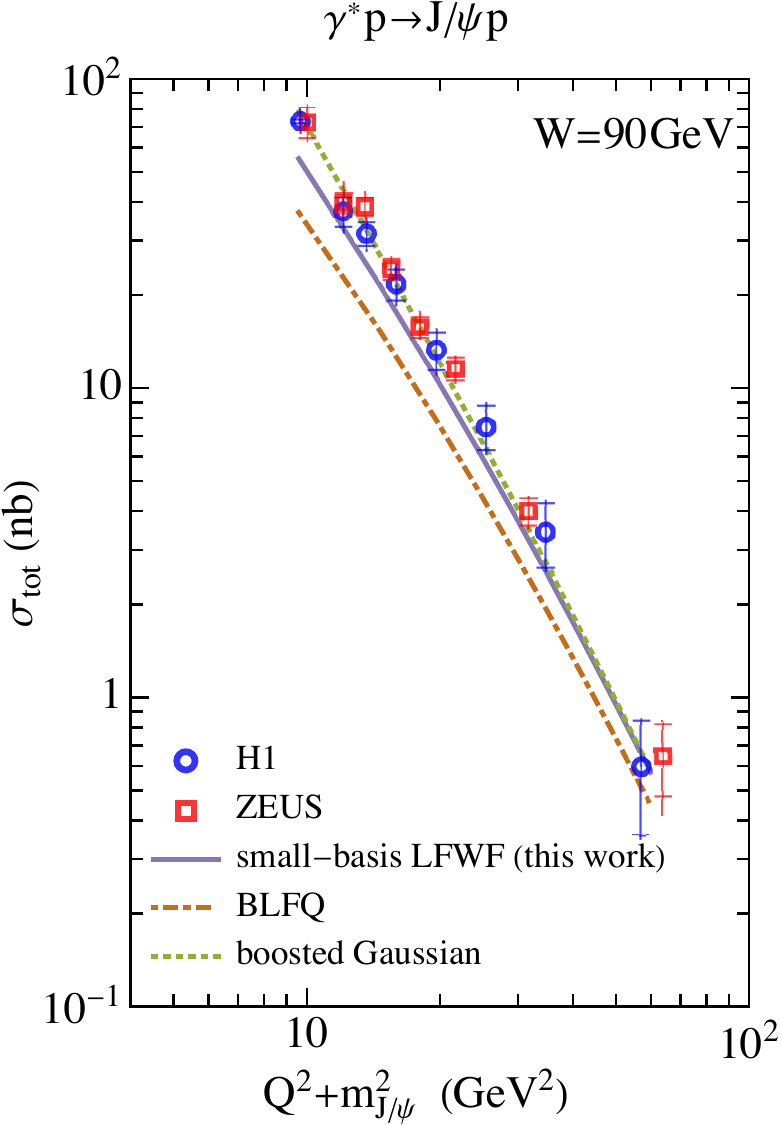}
}
\subfigure[]{\label{fig:hera_W}
\raisebox{0.02\height}
{\includegraphics[width=.295\textwidth]{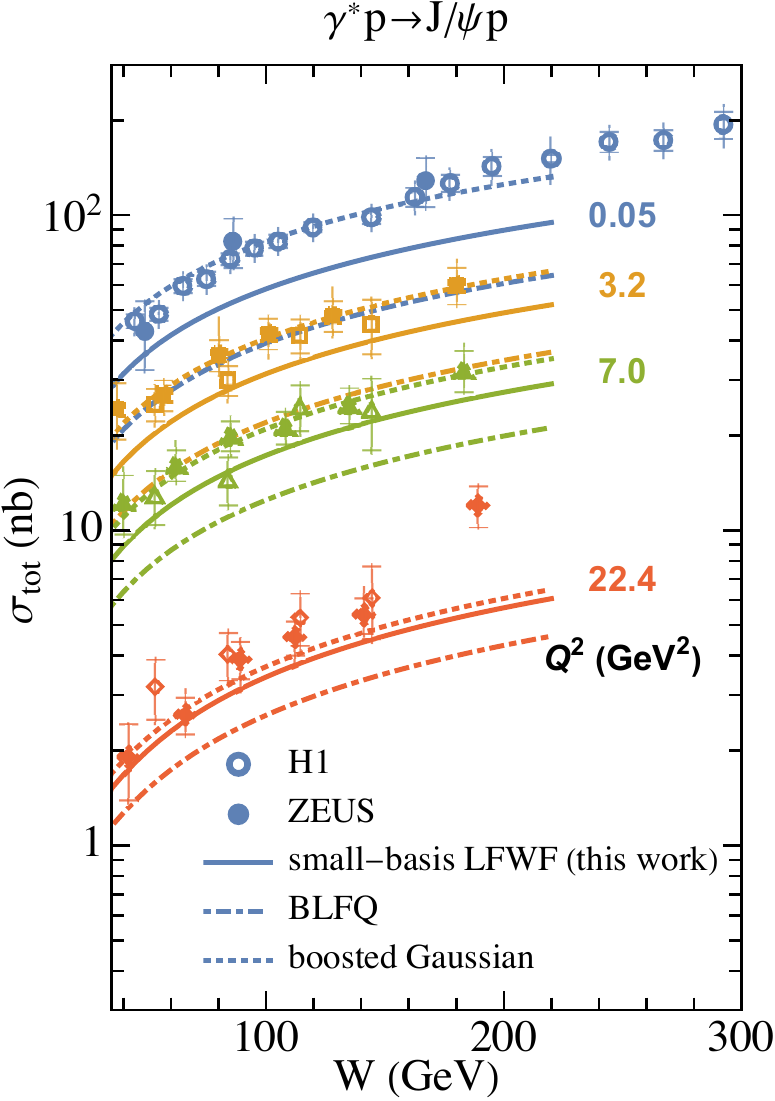}}
}
\subfigure[]{\label{fig:hera_t}
\includegraphics[width=.3\textwidth]{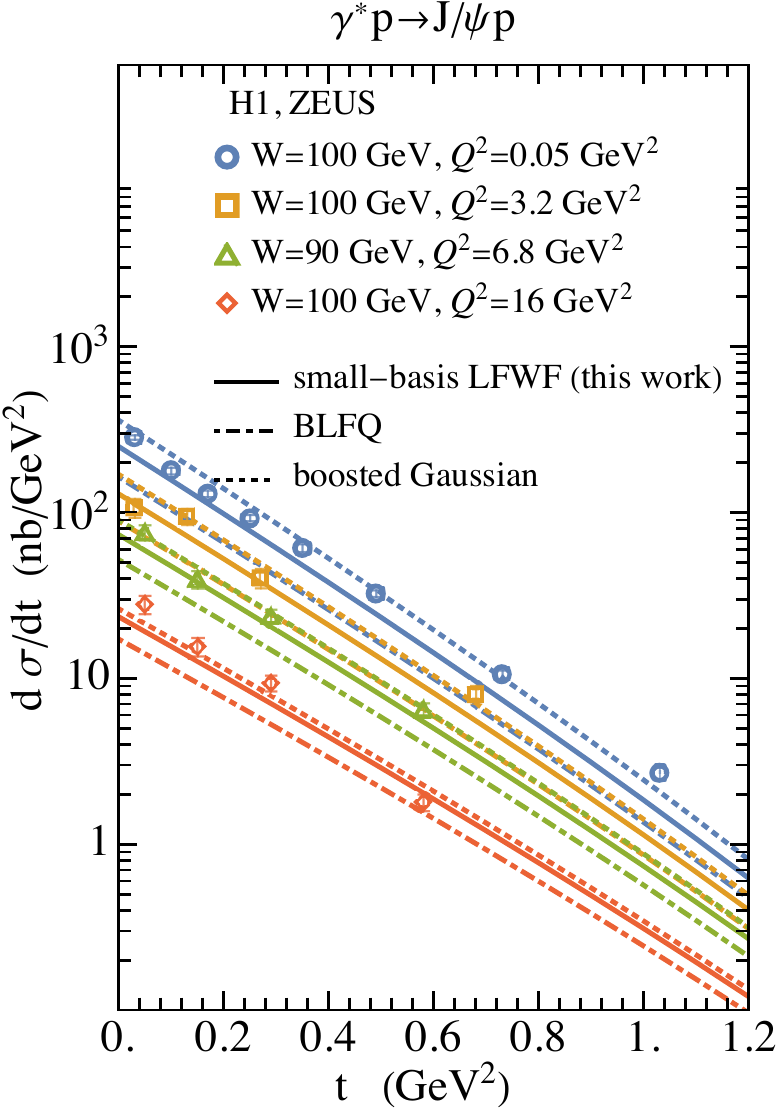}
}
\caption{Predictions of the small-basis LFWF (solid curves), the BLFQ LFWF \cite{Li:2015zda} (dotted curves) and the boosted Gaussian LFWF \cite{Armesto:2014sma} (dot-dashed curves) compared to the HERA experimental data \cite{Chekanov:2004mw,Aktas:2005xu}. The inner bars indicate the statistical uncertainties; the outer bars are the statistical and systematic uncertainties added in quadrature. 
(a) Total $J/\psi$ cross section for different value of $(Q^2+M_V^2)$ at $W=90$~GeV. 
(b) Total $J/\psi$ cross section for different values of $Q^2$ and $W$.
(c) The $J/\psi$ differential cross section $\diff \sigma/ \diff t$ as a function of $t$. 
}
\label{fig:hera}
\end{figure*}

We then calculate the $J/\psi$ production in the kinematic range of the HERA experiment \cite{Chekanov:2004mw,Aktas:2005xu}.
Various cross sections obtained as a function of the kinematic variables $Q^2$, $W$, and $t$ reasonably agree with experimental data. 
As an illustration, we present some representative results in Fig.~\ref{fig:hera}, together with calculations using BLFQ and boosted Gaussian wavefunctions for comparison. 
In all three panels, the solid curves are calculated with the $J/\psi$'s LFWF designed in this work, the dotted curves are calculated with BLFQ vector meson LFWF, and the dot-dashed curves are calculated with the boosted Gaussian LFWF of Ref.~\cite{Armesto:2014sma}, respectively. The bCGC parametrization listed in Table~\ref{tab:bCGC} for dipole cross section was used for all wavefunctions.

Figure~\ref{fig:hera_Q2} shows the total $J/\psi$ cross section as function of $(Q^2+m_{J/\psi}^2)$ for photon-proton c.m. energy $W=90 \GeV$. 
In Fig.~\ref{fig:hera_W}, we show the total $J/\psi$ cross section as function of $W$ at various values of $Q^2$.
The differential cross section $\diff \sigma/\diff t$ is shown in Fig.~\ref{fig:hera_t} as function of the momentum transfer $t$. 
Qualitatively, all three wavefunctions provide reasonable descriptions of the $J/\psi$ cross section data at HERA. 
The calculations using the small basis LFWF give very similar results to those using the boosted Gaussian LFWF. 
The BLFQ LFWF calculation generally underestimates the $J/\psi$ production at HERA, especially in the small $Q^2$ regime. 
However, the $J/\psi$ cross section at small $Q^2$ may have a stronger dependence on the dipole cross section model and the photon wavefunction~\cite{Chen:2016dlk}.

\begin{figure}
 \centering 
     \includegraphics[width=.43\textwidth]{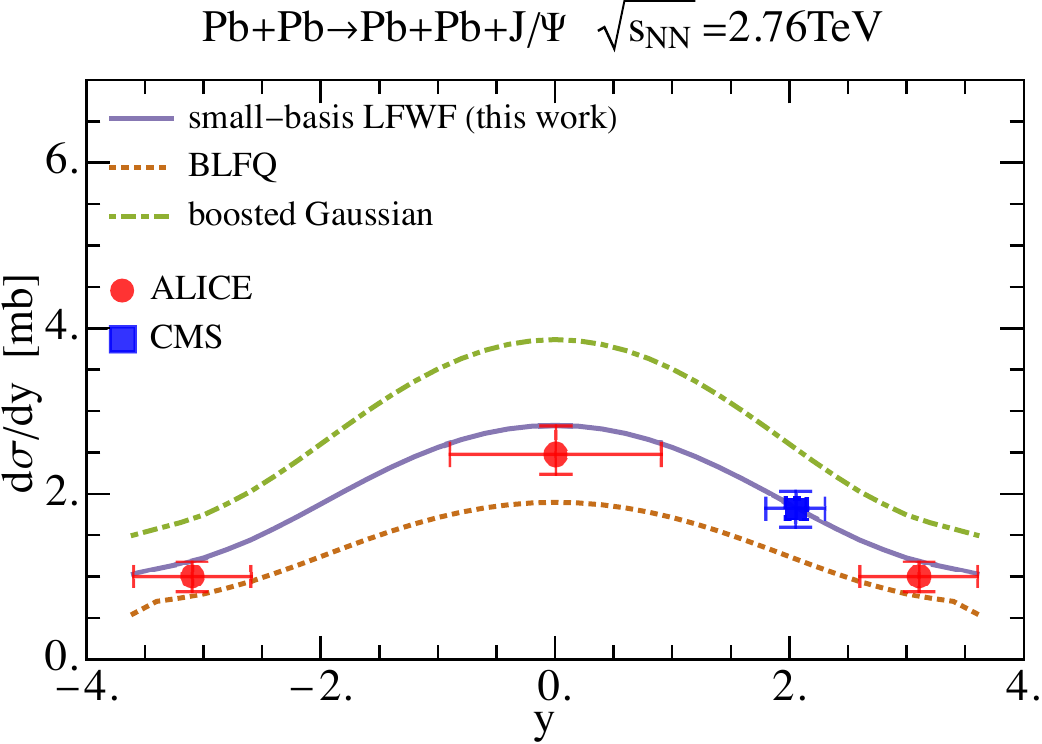}
    \caption{Predictions of the small-basis LFWF (solid curves), the BLFQ LFWF \cite{Li:2015zda} (dotted curves) and the boosted Gaussian LFWF \cite{Armesto:2014sma} (dot-dashed curves) for the coherent production of $J/\psi$ production in Pb-Pb ultra-peripheral collision at $\sqrt{s_{NN}}=2.76$~TeV, compared with the measurements by the ALICE collaboration \cite{Abbas:2013oua, Adam:2015sia} and CMS collaboration \cite{Khachatryan:2016qhq} at LHC. Error bars show statistical uncertainties only.
    }
    \label{fig:lhc}
\end{figure}

We also apply the proposed $J/\psi$ LFWF to calculate the coherent production of $J/\psi$ at LHC at mid-rapidity in Fig.~\ref{fig:lhc}, using the same procedure we adopted in Refs.~\cite{Chen:2016dlk,Chen:2018vdw}. 
Here, the solid curve, the dotted curve, and the dot-dashed curve show the predictions of the small-basis LFWF, the BLFQ LFWF \cite{Li:2015zda}, and the boosted Gaussian LFWF, respectively, for the coherent production of $J/\psi$ in Pb-Pb ultra-peripheral collision at $\sqrt{s_{NN}}=2.76$~TeV, compared to the measurements of the ALICE \cite{Abbas:2013oua, Adam:2015sia} and CMS collaborations \cite{Khachatryan:2016qhq} at the LHC. 
The bCGC parametrization listed in Table~\ref{tab:bCGC} for the dipole cross section was used for all wavefunctions. 
The prediction of the small-basis LFWF is within the statistical uncertainty of the experimental data. 
The prediction of the boosted Gaussian LFWF slightly overshoots the data, and that of the BLFQ LFWF underestimates the experimental data.

Based on the above discussion, we arrive at the conclusion that the small-basis LFWF for $J/\psi$ designed in this work can make quantitatively reasonable predictions for diffractive charmonium production in both $ep$ collisions and ultra-peripheral collisions.

The $\psi'$ cross section calculated with our designed LFWF for $\psi'$ is, on the other hand, far below the experimental data. 
In contrast to the ground state $J/\psi$, the $\psi'$ wavefunction has a node in the $r_\perp$ direction, so in calculating the scattering amplitude in the dipole model, there is a cancellation between the negative and the positive regions. 
The value of the $\psi'$ cross section is therefore very sensitive to the location of the node relative to the typical transverse separation of the virtual $c\bar{c}$ pair in the dipole model.
This cancellation turns out to be very dramatic with our designed $\psi'$ LFWF, resulting in a greatly suppressed cross section.
For example, due to the sensitivity to the location of this node, a 5\% increase (decrease) in $\kappa$ (with no other changes) results in a factor of $\sim 3$ increase (a factor of $\sim 5$ decrease) in the $\psi'$ production in Pb-Pb ultra-peripheral collision at $\sqrt{s_{NN}}=5.02$~\TeV. 
On the contrary, the $J/\Psi$ production is insensitive to such a change of $\kappa$.
We do not present the $\psi'$ production in this paper, and we hope to return to this aspect of the $\psi'$ LFWF in future work.

\subsection[Pseudoscalar-photon transition form factor]{ The \texorpdfstring{$\gamma^*\gamma\to\eta_c$}{gamma+ gamma etac} transition form factor}

The $\eta_c$ meson is produced at $e^+e^-$ colliders in the process $e^+e^-\to e^+e^-\eta_c$ via the diphoton production mechanism~\cite{BaBar:2010siw}.
In this section, we study the $\gamma\gamma^*\to\eta_c$ transition with the LFWF of $\eta_c$ obtained in Sec.~\ref{sec:LFWF}. 
We have defined the diphoton transition form factor $F_{\mathcal{P}\gamma}(Q_1^2,Q_2^2)$ for the process $\mathcal{P} (P)\to\gamma^*(q_1)+\gamma^*(q_2)$ in Sec.\ref{sec:decay_Pgg}, and used its value at $Q_1^2=Q_2^2=0$ to determine the basis coefficients in the $\eta_c$ LFWF. 
We now use the obtained $\eta_c$ LFWF to calculate the transition form factor for the case of one photon on-shell and the other being spacelike.

The transition form factor can be extracted with either the $J^+$ or the $J^\perp$ current, and the results should be the same by Lorentz invariance. 
Therefore, using both currents for the calculation would help check the rotational symmetry embedded in the LFWF.
We take the Drell-Yan frame, such that the vertex photon has zero longitudinal momentum, i.e., $q^+_2=0$, which is the preferred frame for LFWF in the valence sector~\cite{Li:2019kpr}.
In the LFWF representation, the transition form factor extracted from the $J^+$ current reads
\begin{align}\label{eq:FPgg_plus}
  \begin{split}
    F_{\mathcal{P}}&(Q_1^2,Q_2^2)|_{J^+}\\
  =&2 \mathcal{Q}^2_f\sqrt{N_c}
  \int_0^\infty\frac{ k_\perp \diff k_\perp}{(2\pi)^2} 
  \int_{0}^1\frac{\diff x}{\sqrt{2x(1-x)}}
    \\
   &\times 
   \bigg\{
   \frac{1}{\sqrt{A^2-B^2}}
   \bigg[
    \frac{A-\sqrt{A^2-B^2}}{B Q_2 }[
      \phi_{0/\mathcal P}(k_\perp, x)k_\perp\\
      &+\sqrt{2}m_f\phi_{1/\mathcal P}(k_\perp, x)
    ]
    -(1-x) \phi_{0/\mathcal P}(k_\perp, x)  \bigg]\\
     &-\frac{1}{\sqrt{\bar A^2-\bar B^2}}
    \bigg[
      \frac{\bar A-\sqrt{\bar A^2-\bar B^2}}{\bar B Q_2 }[
        \phi_{0/\mathcal P}(k_\perp, x)k_\perp\\
        &+\sqrt{2}m_f\phi_{1/\mathcal P}(k_\perp, x)
      ]
      +x \phi_{0/\mathcal P}(k_\perp, x)
       \bigg]
       \bigg\}
     \;,
  \end{split}
\end{align}
and the expression from the $J^\perp$ current
\begin{align}\label{eq:FPgg_perp}
  \begin{split}
    F_{\mathcal{P}}&(Q_1^2,Q_2^2)|_{J^\perp}\\
    =&-2\mathcal{Q}^2_f \sqrt{N_c}
    \int_0^\infty\frac{ k_\perp \diff k_\perp}{(2\pi)^2} \\
    &
    \bigg\{
      \int_{0}^1\frac{\diff x}{\sqrt{2x(1-x)}}
    \frac{1-x}{\sqrt{A^2-B^2}}
    \phi_{0/\mathcal P}(k_\perp, x)\\
    &+
    \int_{0}^1\frac{\diff x}{\sqrt{2x(1-x)}}
     \frac{x}{\sqrt{\bar A^2-\bar B^2}}
      \phi_{0/\mathcal P}(k_\perp, x)
    \bigg\}
    \;.
  \end{split}
\end{align}
In both expressions, $A= k_\perp^2+(1-x)^2Q_2^2+m_f^2+x(1-x)Q_1^2$, $B= 2(1-x)k_\perp Q_2$, $\bar A= k_\perp^2+x^2Q_2^2+m_f^2+x(1-x)Q_1^2$ and $\bar B= -2xk_\perp Q_2$. 
By Bose symmetry, the transition form factor should also be symmetric under the exchange of $Q_1^2$ and $Q_2^2$. This means that in the limit of one on-shell photon, we should have $F_{\eta_c\gamma}(Q^2)\equiv F_{\eta_c\gamma}(Q^2,0)=F_{\eta_c\gamma}(0,Q^2)$. 
However, such a symmetry is not explicit in the expressions~\eqref{eq:FPgg_plus} and~\eqref{eq:FPgg_perp}.
We examine this symmetry by taking both the two limits of $Q_1^2=0$ and $Q_2^2=0$. 

\begin{figure}
  \centering 
  \includegraphics[width=.4\textwidth]{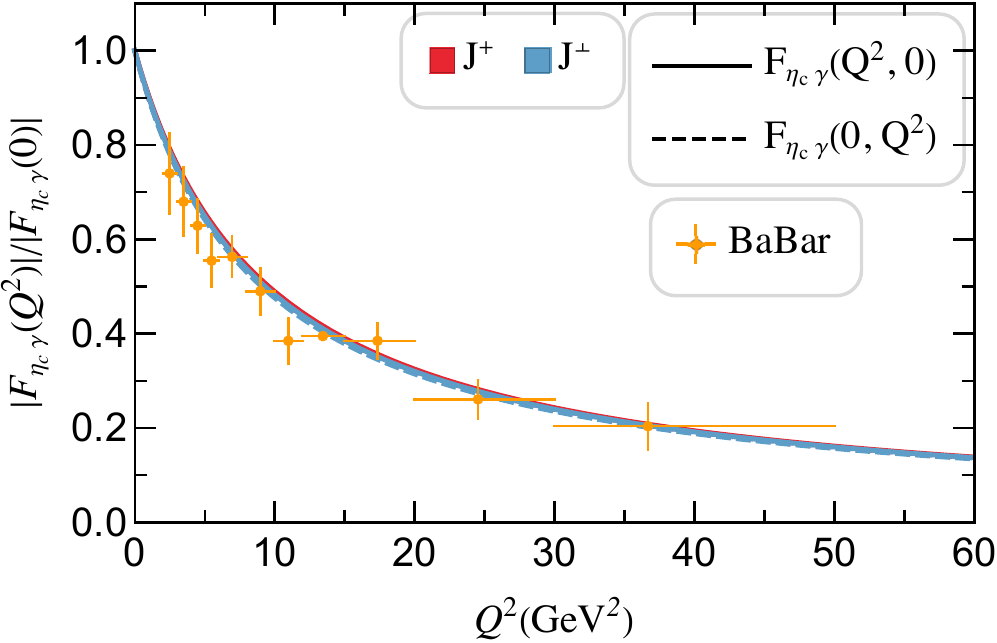}
  \caption{
  The normalized transition form factor of $\eta_c$ calculated according to Eqs.~\eqref{eq:FPgg_plus} and \eqref{eq:FPgg_perp}, in red and blue, respectively.
  We use the $\eta_c$ LFWF in Eq.~\eqref{eq:C_etac_SSP} (see values of basis coefficient in Table.~\ref{tab:C_i_etac_SSP}). 
  The solid line is obtained by taking the limit of $Q_2^2=0$ [$F_{\eta_c\gamma}(Q^2)= F_{\eta_c\gamma}(Q^2,0)$], and the dashed by taking $Q_1^2=0$ [$F_{\eta_c\gamma}(Q^2)= F_{\eta_c\gamma}(0,Q^2)$].
  The BaBar experimental data is taken from Ref.~\cite{BaBar:2010siw}.
  }
\label{fig:etacSP_TFF}
\end{figure}

We present the results for the $\eta_c$ diphoton transition form factor in the format of the normalized transition form factor $|F_{\eta_c\gamma}(Q^2)/F_{\eta_c\gamma}(0)|$ in Fig.~\ref{fig:etacSP_TFF}. 
We found that the results calculated with the two different current components, and by taking the two limits of $Q_i^2=0 (i=1,2)$ agree. This indicates that the designed $\eta_c$ LFWF quite closely preserves both the rotational symmetry and the Bose symmetry for this observable.
We also compare our results with the experimental data from BaBar, finding a reasonable agreement.

\section{Summary}\label{sec:summary}
We proposed a method to build the LFWFs of meson bound states on a small-sized basis function representation. 
In this work, the basis functions are the eigenfunctions of an effective Hamiltonian developed from light-front holography. 
However, the basis coefficients and parameters of the basis functions are obtained using constraints on the wavefunction directly, not through the Hamiltonian.
We use physical constraints, including the orthonormalization relation, symmetries, insights from nonrelativistic state identification, and decay widths from experimental measurements, to determine the parameters within the basis function and the basis coefficients.
The resulting LFWFs inherit the physical interpretation of relativistic bound states from the phenomenological Hamiltonians of LFH and BLFQ, while admitting simple-functional forms that are feasible in calculating observables. 
We have adopted a ``by design'' approach where we choose by hand a set of sufficiently many phenomenologically most important constraints to achieve a unique determination of the parameters. 

With this formalism, we designed the LFWFs for $\eta_c, J/\psi$, $\psi'$, and $\psi(3770)$. 
First, we make the assumption that the $J/\psi$ state is the ground vector state of the charmonium system and its spatial wavefunction is a LF-1S state. 
The two adjustable parameters in the basis function are determined by the $J/\psi$ decay constant, which is established by its experimental dilepton decay width. 
We then construct the pseudoscalar state $\eta_c$ and the two excited vector states $\psi'$ and $\psi(3770)$ as superpositions of the ground and excited light-front basis states with proper spin structure assignments.
The basis coefficients are determined by their diphoton/dilepton decay widths, as well as other theoretical considerations, such as spatial symmetries and the orthogonality conditions. 
This step-by-step approach could be extended by including further constraints and could be developed towards a simultaneous global analysis in the future.

Using the charmonium LFWFs by design, we calculate several physical quantities that are accessible by experimental measurements.
For instance, we calculate the charge radii and parton distribution functions; we also estimated the masses of the charmonium states we designed by evaluating the expectation value of an approximated Hamiltonian guided by the full BLFQ formalism. 
Our predictions for masses, charge radii, and parton distribution functions using the constructed LFWFs are in reasonable agreement with the experimental measurements and are quantitatively consistent with other established methods such as Dyson-Schwinger Equations and Lattice calculations. 
Furthermore, we calculate the $J/\psi$ production in DIS at HERA and UPC at LHC using the constructed LFWF in the dipole model, the theoretical prediction is consistent with experiment data within uncertainties, and are comparable to calculations using the boosted Gaussian and BLFQ LFWFs. 
We calculate the $\eta_c$ diphoton transition form factor using the obtained LFWF and find a reasonable agreement with experimental data. 

With this work, we provided light-front wavefunctions of mesons in a simple-functional form while retaining physical interpretation and matching to a variety of selected experimental observables. 
We anticipate these analytical LFWFs can be used for making predictions of various physical processes that involve the meson states, e.g., exclusive processes at the EIC~\cite{Accardi:2012qut}.

\section*{Acknowledgements}
M. Li and T. Lappi are supported by the Academy of Finland, project 321840 and under the European Union’s Horizon 2020 research and innovation programme by the European Research Council (ERC, grant agreements No. ERC-2015-CoG-681707 and ERC-2018-AdG-835105) and by the STRONG-2020 project (grant agreement No 824093).
Y. Li, G. Chen and J. P. Vary are supported in part by the US Department of Energy (DOE) under Grant No. DE-FG02-87ER40371. The content of this article does not reflect the official opinion of the European Union and responsibility for the information and views expressed therein lies entirely with the authors. 
M.~Li acknowledges financial support from Xunta de Galicia (Centro singular de investigación de Galicia accreditation 2019-2022), European Union ERDF, the “María de Maeztu” Units of Excellence program, the Spanish Research State Agency, and European Research Council project ERC-2018-ADG-835105 YoctoLHC.

\appendix
\section{Conventions}\label{app:conventions}
We follow the conventions of Refs. \cite{Li:2017mlw, Li:2018uif}. Here, we provide a concise summary.
\subsection{Light-Front coordinates}\label{app:LF_cor}
The light-front coordinates are defined as \( x^\mu= (x^+, x^-, x^1, x^2) \), where \(x^+=x^0 + x^3\) is the light-front time,  \(x^-=x^0- x^3\) is the longitudinal coordinate, and \(\vec{x}_\perp=(x^1, x^2)\) are the transverse coordinates. 
We also write the transverse components with subscript $x$ ($y$) in place of $1$ ($2$), for example, \(\vec{r}_\perp=(r^x,r^y)\).
The covariant vectors are obtained by $x_\mu=g_{\mu\nu}x^\nu$, with the metric tensors $g_{\mu\nu}$ and $g^{\mu\nu}$. 
The nonzero components of the metric tensors are,
\begin{align}
    \begin{split}
    & g^{+-}=g^{-+}=2, \qquad 
    g_{+-}=g_{-+}=\frac{1}{2}, \\
    & g^{ii}=g_{ii}=-1~(i=1,2) \;. 
    \end{split}
\end{align}
The inner product of two 4-vectors is therefore
\(
a\cdot b = a^\mu b_\mu=\frac{1}{2}(a^+b^- + a^- b^+)-\vec a_\perp\cdot \vec b_\perp
\).
For a transverse vector $\vec k_\perp=(k^x, k^y)$, we will write its complex forms as 
\begin{align}\label{eq:RL}
    \begin{split}
      k^R\equiv & k^x +i k^y=k_\perp e^{i \theta_k}\;,\\ k^L\equiv & k^x -i k^y=k_\perp e^{-i\theta_k}\;.
    \end{split}
\end{align}
 where $k_\perp=|\vec k_\perp |$ and $\theta_k=\arg k^R$.
\subsection[The gamma matrices]{The \texorpdfstring{$\gamma$}{gamma} matrices}\label{app:gamma}
The Dirac matrices are four unitary traceless $4 \times 4$ matrices:
\begin{align}
\begin{split}
     & \gamma^0=\beta=
  \begin{pmatrix}
    0&-i\\
    i&0
  \end{pmatrix},
  \quad
  \gamma^i=
  \begin{pmatrix}
    -i\hat{\sigma}^i&0\\
    0&i\hat{\sigma}^i
  \end{pmatrix}\;,\\
  &\gamma^+=
  \begin{pmatrix}
    0 & 0\\
    2i & 0
  \end{pmatrix},
  \quad
  \gamma^-=
  \begin{pmatrix}
    0&-2i\\
    0&0
  \end{pmatrix}
  \;,
\end{split}
\end{align}
in which the Pauli matrices are
\begin{align}
  \hat{\sigma}^1=\sigma^2=
  \begin{pmatrix}
    0&-i\\
    i&0
  \end{pmatrix},
   \quad
   \hat{\sigma}^2=-\sigma^1=
   \begin{pmatrix}
     0&-1\\
     -1&0
   \end{pmatrix}
   \;.
\end{align}
It is also convenient to define $\gamma^R\equiv\gamma^1+i\gamma^2$ and $\gamma^L\equiv\gamma^1+i\gamma^2$. The chiral matrix is $\gamma^5=i\gamma^0\gamma^1\gamma^2\gamma^3$, in which $\gamma^3=\gamma^+-\gamma^0$.
\subsection{Polarization vectors}\label{app:polarization_vectors}
The polarization vector for a vector boson with momentum $k^\mu$, mass $m$, and helicity $\lambda$ is
\begin{align}
  \begin{split}
  \epsilon^\mu_{\lambda=0}(k)
  =&(\frac{k^+}{m},\frac{{\vec{k}_\perp}^2-m^2}{mk^+},\frac{\vec{k}_\perp}{m})\;,\\
  \epsilon^\mu_{\lambda=\pm 1}(k)
  =&(0,\frac{2\bm{\epsilon}^\perp_\lambda\cdot \vec{k}_\perp}{k^+},\bm{\epsilon}^\perp_\lambda)\;,   
  \end{split}
\end{align}
where $\bm{\epsilon}^\perp_\pm=(1,\pm i)/\sqrt{2}$.
The polarization vector for a photon with momentum $q^\mu$, virtuality $Q^2=-q^2>0$, and helicity $\lambda$ is
\begin{align}
  \begin{split}
    \epsilon^\mu_{\lambda=0}(q)
  =&(\frac{q^+}{Q},\frac{{\vec{q}_\perp}^2+Q^2}{Q q^+},\frac{\vec{q}_\perp}{Q})\;,\\
  \epsilon^\mu_{\lambda=\pm 1}(q)
  =&(0,\frac{2\bm{\epsilon}^\perp_\lambda\cdot \vec{q}_\perp}{q^+},\bm{\epsilon}^\perp_\lambda)\;. 
  \end{split}
\end{align}

\section{Transformation coefficients of the three-dimensional harmonic oscillator (3D HO)}\label{app:3DHO}
In this Appendix, we calculate the transformation coefficients of the three-dimensional harmonic oscillators (3D HOs) from the spherical coordinate to the cylindrical coordinate. 
The meson states, especially heavy quarkonia, are sometimes identified as 3D-HO states in the spherical coordinate, e.g., 1S wave, 1P wave, etc.
On the other hand, the light-front basis functions we take, as in Eq.~\eqref{eq:LFWF_full}, are very similar to the 3D HOs in the cylindrical coordinate. 
Therefore, we take the transformation between the two coordinates as a guidance to help us construct and identify 3D-HO states in the chosen light-front basis functions.

A vector in the Cartesian coordinates is given by $\bm{\eta} = (\eta_x, \eta_y, \eta_z)= \eta_x \bm e_x + \eta_y \bm e_y + \eta_z \bm e_z$. Then in the cylindrical coordinates, $\bm{\eta} = (\eta_\rho, \eta_z, \theta)$, and in the spherical coordinates, $\bm{\eta} = (\eta, \theta, \phi)$, where
\[\eta_\rho = \sqrt{\eta_y^2 + \eta_x^2}\;,\]
\[\phi = \tan^{-1}\left(\frac{\eta_y}{\eta_x}\right)\;,\]
\[\theta = \tan^{-1}\left(\frac{\eta_\rho}{\eta_z}\right)\;.\]

The 3D-HO state in a spherical representation is written as $\ket{n,\ell, m}$, where $n$ is the radial quantum number, $\ell$ and $m$ are the orbital angular momentum and its $z$ component, and the total energy of the state is $N=2n+\ell$,
\begin{align}\label{eq:spec_state}
  \begin{split}
    \Phi_{n,\ell, m} (\bm{\eta})\equiv&\braket{n,\ell, m|\eta,\theta,\phi}\\
   =&(-1)^n
    \sqrt{\frac{4\pi 2^\ell (n+\ell)!}{n!(2n+2\ell+1)!}}
    \eta^{2n+\ell} Y_{\ell, m}(\theta,\phi)
    \;.
  \end{split}
\end{align}
Note that the quantum numbers ($n$, $\ell$, $m$) here are different from the ($n$, $m$, $l$) in the light-front basis functions as in Eq.~\eqref{eq:LFWF_full}. Note especially that the ``$l$" in the latter set is the light-front longitudinal quantum number.

The spherical harmonics read
\begin{multline}\label{eq:Ylms}
  Y_{\ell, m}(\theta,\phi)
  =(-1)^{m}\sqrt{\frac{(2\ell+1)(\ell-m)!}{4\pi (\ell+m)!}}P^{m}_{\ell}(\cos\theta)e^{i m\phi}
  \;,
\end{multline}
with the phase convention
\begin{align}\label{eq:Ylms_minus}
  Y_{\ell,-m}(\theta,\phi)=(-1)^{m}
  Y_{\ell,-m}^*(\theta,\phi)
  \;,
\end{align}
where $P^{m}_{\ell}$ are the associated Legendre polynomials without the Condon–Shortley phase $(-1)^{m}$ (to avoid counting the phase twice).

The 3D-HO state in a cylindrical representation is written as $\ket{n_\rho,m,n_z}$, where $n_\rho$ is the quantum number for the $\rho$ coordinate, and the total energy of the state is $N=2n_\rho+n_z+|m|$,
\begin{align}\label{eq:3DHO_cyl}
  \begin{split}
   \tilde\Phi_{n_\rho,m,n_z} (\bm{\eta})\equiv&\braket{n_\rho,m,n_z|\eta_\rho, \eta_z, \theta}\\
   &=
    (-1)^{n_\rho}
    \sqrt{\frac{2\pi}{2^{2n_\rho+m}n_\rho! (n_\rho+m)!}}
    \frac{1}{\sqrt{n_z!}}
    \\
   &\eta_z^{n_z}
    \eta_\rho^{2n_\rho+m}
    \frac{1}{\sqrt{2\pi}}
    e^{i m\phi}
    \;.
  \end{split}
\end{align}

The relation of the two representations is 
\begin{align}\label{eq:3DHO_transf}
  \begin{split}
   &\ket{n,\ell,m} \\
  & =\sum_{n_\rho, n_z}
  \delta_{2n+\ell,2n_\rho+n_z+|m|}
  \braket{n_\rho,m,n_z|n,\ell,m}
  \ket{n_\rho,m,n_z}\;,   
  \end{split}
\end{align}
and the transformation coefficient is derived in Ref.~\cite{HO_transformation}.
Written explicitly,
\begin{widetext}
\begin{align}\label{eq:3DHO_transf_expression}
  \begin{split}
    \braket{n_\rho,m,n_z|n,\ell,m}
    =&\delta_{2n+\ell,2n_\rho+n_z+|m|}
    (-1)^{n+m+n_\rho}\\
    &\times
    \left[
      2^{2n_\rho-\ell+m}
      \frac{(2\ell+1)(\ell-m)!(n+\ell)!(n_\rho+m)!(2n-2n_\rho+\ell-m)!}{(\ell+m)!n!(2n+2\ell+1)!n_\rho!}
    \right]^{1/2}\\
    &\sum_{s=s_{\min}}^{s_{\max}}
    (-1)^s\frac{(2\ell-2s)!(n+s)!}{s!(\ell-s)!(\ell-2s-m)!(n-n_\rho+s)!}
    \;,
  \end{split}
\end{align}
\end{widetext}
where
\begin{align}
  s_{\min}=\begin{cases}
  0, & n\ge n_\rho\\
  n_\rho -n ,& n< n_\rho
  \end{cases}
  \;,
\end{align}
and 
\begin{align}
  s_{\max}=\begin{cases}
  \dfrac{\ell-m}{2}, & \ell-m \text{ is even}\\
  \dfrac{\ell-m-1}{2} ,& \ell-m \text{ is odd}
  \end{cases}
  \;.
\end{align}

The transformation coefficients for selected states are calculated according to Eqs.~\eqref{eq:3DHO_transf} and ~\eqref{eq:3DHO_transf_expression} and listed in Tables.~\ref{tab:3DHO_S},~\ref{tab:3DHO_P} and ~\ref{tab:3DHO_D}.
\begin{table}[ht]
  \caption{\label{tab:3DHO_S} 
  The transformation of S waves (states with $\ell=0$) from the spherical coordinate to the cylindrical coordinate.
  }
  \centering
  \begin{ruledtabular}
  \begin{tabular}{lll}
   State & $\ket{n,\ell,m}$ &$\ket{n_\rho, m, n_z}$ \\
    \hline
  1S & $\ket{0,0,0}$ & $\ket{0,0,0}$ \\
  2S & $\ket{1,0,0}$ & $\sqrt{\frac{2}{3}}\ket{1,0,0}- \sqrt{\frac{1}{3}}\ket{0,0,2}$ \\
  3S & $\ket{2,0,0}$ & $\sqrt{\frac{8}{15}}\ket{2,0,0}+ \sqrt{\frac{1}{5}}\ket{0,0,4} -\sqrt{\frac{1}{15}}\ket{1,0,2}$ 
  \end{tabular}
  \end{ruledtabular}
\end{table}

\begin{table}[ht]
  \caption{\label{tab:3DHO_P} 
  The transformation of the 1P wave (states with $n=0$ and $\ell=1$) from the spherical coordinate to the cylindrical coordinate.
  }
  \centering
  \begin{ruledtabular}
  \begin{tabular}{lll}
   State & $\ket{n,\ell,m}$ &$\ket{n_\rho, m, n_z}$ \\
  \hline
  \multirow{3}{*}{1P} & $\ket{0,1,0}$ & $\ket{0,0,1}$ \\
   & $\ket{0,1,1}$ & $-\ket{0,1,0}$ \\
   & $\ket{0,1,-1}$ & $-\ket{0,-1,0}$ 
  \end{tabular}
  \end{ruledtabular}
\end{table}

\begin{table}[ht]
  \caption{\label{tab:3DHO_D} 
  The transformation of the 1D wave (states with $n=0$ and $\ell=2$) from the spherical coordinate to the cylindrical coordinate.
  }
  \begin{ruledtabular}
  \begin{tabular}{lll}
   State & $\ket{n,\ell,m}$ &$\ket{n_\rho, m, n_z}$ \\
  \hline
  \multirow{4}{*}{1D} & $\ket{0,2,0}$ & $\sqrt{\frac{2}{3}}\ket{0,0,2}+ \sqrt{\frac{1}{3}}\ket{1,0,0}$ \\
   & $\ket{0,2,1}$ & $-\ket{0,1,1}$ \\
  & $\ket{0,2,-1}$ & $-\ket{0,-1,1}$ \\
  & $\ket{0,2,2}$ & $\ket{0,2,0}$ \\
   & $\ket{0,2,-2}$ & $\ket{0,-2,0}$ 
  \end{tabular}
  \end{ruledtabular}
\end{table}

\section{Nonrelativistic limit of the light-front spectroscopic states}\label{app:NR}
In this Appendix, we write out the nonrelativistic (NR) limit of the spatial part of the light-front spectroscopic states, $\psi_{\text{LF}-W}$, and compare them to the spherical harmonic oscillators. 
In the NR limit of $|\vec k| \ll m_f$ and $x\to 1/2+ k_z/(2m_f)$, the light-front basis function $ \psi_{nml}$ reduces to
\begin{align}
  \begin{split}
    \psi_{nml, NR}(\vec k)
  =&
  C_{nml}R(k)
  {\bigg(\frac{2k\sin\theta}{\kappa 
  }\bigg)}^{|m|}
  L_n^{|m|}
  \left(
  \frac{2k^2\sin^2\theta}{\kappa^2}
  \right)\\
  &
  e^{im\phi}  
  P_l^{(\alpha,\alpha)}\left(\frac{k\cos\theta}{m_f}\right) 
  \;,
  \end{split}
\end{align} 
where $C_{nml}$ is a positive constant depending on $n,m$, and $l$,
\begin{align}
  \begin{split}
    C_{nml}=&
  \kappa^{-1}\sqrt{\frac{4\pi n!}{(n+|m|)!}}
  \sqrt{4\pi(2l+2\alpha+1)}\\
  &\sqrt{\frac{\Gamma(l+1)\Gamma(l+2\alpha+1)}{\Gamma(l+\alpha+1)^2}}2^{-\alpha}
  \;,
  \end{split}
\end{align} 
and $R(k)$ is the Gaussian part, 
\begin{align}
R(k)=\exp\left(-
\frac{2k_\perp^2}{\kappa^2}
\right)\;.
\end{align} 
Here the spherical polar angles $\theta$ and $\phi$ represent the colatitude and azimuthal angle of $\vec k$, respectively.

To compare the NR limit of the basis function with the spherical harmonic oscillators, we expand them in the small momentum region.
The solutions of the Schr\"{o}dinger equation with a spherically symmetric harmonic oscillator potential are in the form of $\Psi_{n \ell m}(r,\theta_r,\phi_r)=R_{n \ell}(r)Y_{\ell m}(\theta_r,\phi_r)$, where $R_{n \ell}(r)$ is the radial function.
In the small momentum limit of $|\vec k|\to 0$, the spherical harmonic oscillator in the momentum space has the following dependence on $\vec k$,
 \begin{align}\label{eq:spectroscopic_NR}
  \lim_{k\to 0} \tilde\Psi_{n \ell m}(k,\theta,\phi)
  = f(n,\ell) k^\ell Y_{\ell m}(\theta,\phi)  
  \;,
\end{align}
where $f(n,\ell)$ is some function of the $n$ and $\ell$ inherited from the radial wavefunction.
In the spectroscopic notation, $\ell=0,1,2$ correspond to the S, P, and D wave, respectively, and $n=1,2,\ldots$ labels the energy level in the ascending order. 
For each LF spectroscopic state in the NR limit, we identify its $\ell$ quantum number by comparing its dependence on $k$ to the right hand side of Eq.~\eqref{eq:spectroscopic_NR}, and $n$ by its energy level.

The NR limit of the LF-1S state is that of the ground basis state $\psi_{000}$
\begin{align}\label{eq:LF_1S_NR}
  \begin{split}
  \psi_{\text{LF}-1S,NR} = 
    \lim_{k\to 0}\psi_{000,NR}(\vec k)
    = C_{000}
    = C k^0 Y_{0,0}(\theta,\phi)
  \;.
  \end{split}
\end{align} 
Here and in the following, we use $C$ (and $C'$, $C'', \ldots$) to indicate a positive constant.
Thus the NR limit of the LF-1S state is indeed a 1S state, by seeing the radial number $n=0$ and $Y_{0,0}$ as the S wave.

To find the NR limits of the LF-2S state and the LF-1D0 state, we first look at their components $\psi_{100}$ and $\psi_{002}$, 
\begin{align}
    \psi_{100,NR}(\vec k)
    =&
    C_{100}R(k)
  \left(1-
  \frac{4k^2}{\kappa^2}\sin^2\theta
  \right)
  \;,\\
    \psi_{002,NR}(\vec k)
  =&
  C_{002}R(k)
  \frac{\alpha+2}{4}
    \left[-1+(2\alpha+3)
   \frac{k^2\cos^2\theta}{m_f^2}
    \right]
  \;.
\end{align} 
The ratio of the coefficient of $\sin^2\theta$ in $ \psi_{100,NR}$ and that of $\cos^2\theta$ in $ \psi_{002,NR}$ in the $\alpha\to\infty$ (i.e. $m_f\to\infty$ at fixed $\kappa$) limit is
\begin{align}
  \begin{split}
   &\lim_{\alpha\to\infty} -\frac{2\alpha C_{100}}{(\alpha+2)(2\alpha+3)C_{002}}\\
   =&\lim_{\alpha\to\infty}
   -\frac{4\alpha }{(\alpha+2)(2\alpha+3)}
   \sqrt{
   \frac{ 
     (2\alpha+1)
     (\alpha+2)^2(\alpha+1)^2
     }
   {
     (2\alpha+5)2(2\alpha+2)(2\alpha+1)
   }
   }
   \\
   =&-\frac{1}{\sqrt{2}}
  \;.
  \end{split}
\end{align} 
The ratio of the $\theta$ independent term in $ \psi_{100,NR}$ and that in $ \psi_{002,NR}$ in the $\alpha\to\infty$  limit is
\begin{align}
  \begin{split}
   \lim_{\alpha\to\infty} -\frac{4 C_{100}}{(\alpha+2)C_{002}}
   =-\sqrt{2}
  \;.
  \end{split}
\end{align} 
In the LF-2S state defined in Eq.~\eqref{eq:LF_2S}, the angular dependence in the coefficient of $k^2$ vanishes by seeing that $\sin^2\theta+\cos^2\theta=1$. At $k\to 0$ limit, the constant term outweighs the $k^2$ term, so the function is considered as proportional to $k^0 Y_{0,0}$, 
\begin{align}\label{eq:LF_2S_NR}
  \begin{split}
    \psi_{\text{LF}-2S,NR} = &
   \lim_{k\to 0}
   \left(
   \sqrt{\frac{2}{3}}
    \psi_{ 1,0,0,NR}  
  -    \sqrt{\frac{1}{3}}
  \psi_{ 0,0,2,NR}  
  \right)\\
  =&C k^0
   Y_{0,0}(\theta,\phi)
  \;.  
  \end{split}
\end{align}
Thus the NR limit of the LF-2S state is a 2S state, by seeing the radial number $n=1$ and $Y_{0,0}$ as the S wave.
On the contrary, in the LF-1D0 state defined in Eq.~\eqref{eq:LF_1D0}, the ratio  between $\psi_{ 1,0,0}$ and $\psi_{ 0,0,2}$ is $\sqrt{2}$. 
The constant term cancels, and the angular dependence in the coefficient of $k^2$ becomes $2\cos^2\theta-\sin^2\theta\propto Y_{2,0}(\theta,\phi)$,
\begin{align}\label{eq:LF_1D0_NR}
  \begin{split}
  \psi_{\text{LF}-1D0,NR} = &
     \lim_{k\to 0}
   \left(
    \sqrt{\frac{1}{3}}
    \psi_{ 1,0,0,NR}  
  +    \sqrt{\frac{2}{3}}
  \psi_{ 0,0,2,NR}  
    \right)\\
  =& C 
  k^2(-\sin^2\theta+2\cos^2\theta)\\
  =& C'
  k^2 Y_{2,0}(\theta,\phi)
  \;.
  \end{split}
\end{align}
Thus the NR limit of the LF-1D0 state is the $m=0$ component of the 1D wave.

The NR limit of LF-1D1 state and LF-1D-1 states are 
\begin{align}\label{eq:LF_1Dpm1_NR}
  \begin{split}
  \psi_{\text{LF}-1D1,NR} = &
   \lim_{k\to 0}
    -\psi_{0,1, 1,NR}(\vec k)\\
    =& \lim_{k\to 0}
    - C_{0,1,1}R(k)
\frac{2(\alpha+1)k^2}{\kappa m_f}
  e^{i\phi}  
   \cos\theta \sin\theta\\
   = & C k^2
Y_{2,1}(\theta,\phi)\;,\\
  \psi_{\text{LF}-1D-1,NR} = &
   \lim_{k\to 0}
-\psi_{0,-1, 1,NR}(\vec k)\\
=&\lim_{k\to 0}
-
C_{0,-1,1}R(k)
\frac{2(\alpha+1)k^2}{\kappa m_f}
e^{i\phi}  
\cos\theta \sin\theta\\
=&-
C k^2
Y_{2,-1}(\theta,\phi)
  \;.
  \end{split}
\end{align} 
Note that there is a `-' sign on the NR limit of the LF-1D-1 state. It follows that the NR limit of the LF-1D1 state is the $m=1$ component of the 1D wave, and that of the LF-1D-1 state is negative of the $m=-1$ component of the 1D wave.

The NR limit of LF-1D2 state and LF-1D-2 states are
\begin{align}\label{eq:LF_1Dpm2_NR}
  \begin{split}
    \psi_{\text{LF}-1D2,NR} = &
   \lim_{k\to 0}
    \psi_{0,2, 0,NR}(\vec k)\\
    =&\lim_{k\to 0}
    C_{0,2, 0}R(k)
  \frac{4k^2}{\kappa^2}
  e^{i2\phi}  
  \sin^2 \theta\\
  =&C k^2
  Y_{2,2}(\theta,\phi)\;,\\
      \psi_{\text{LF}-1D-2,NR} = &
   \lim_{k\to 0}
  \psi_{0,-2, 0,NR}(\vec k)\\
  =&\lim_{k\to 0} C_{0,-2, 0}R(k)
\frac{4k^2}{\kappa^2}
e^{-i2\phi}  
\sin^2 \theta\\
=&
  C k^2
Y_{2,-2}(\theta,\phi)
  \;.
  \end{split}
\end{align} 
Therefore the NR limit of the LF-1D$\pm2$ states are the $m=\pm2$ components of the 1D state, respectively. 

The NR limit of the LF-1P0 state is a 1P state with $m=0$,
\begin{align}\label{eq:LF_1P0_NR}
  \begin{split}
    \psi_{\text{LF}-1P0,NR} 
    = &
    \lim_{k \to 0}\psi_{001,NR}(\vec k)\\
    =&\lim_{k \to 0}
    C_{001}R(k)
  (\alpha+1)\frac{k}{m_q}\cos\theta\\
  = & C k
  Y_{1,0}(\theta,\phi)
  \;.
  \end{split}
\end{align} 
The NR limit of the LF-1P$\pm 1$ states are
\begin{align}\label{eq:LF_1Ppm1_NR}
  \begin{split}
    \psi_{\text{LF}-1P1,NR} 
    = &
    \lim_{k \to 0}-\psi_{0,1,0,NR}(\vec k)\\
    =&
    \lim_{k \to 0}-C_{0,1,0}R(k)
  \frac{2k}{\kappa 
  }
  e^{i\phi}  \sin\theta\\
  =& C k
Y_{1,1}(\theta,\phi)\;, \\
\psi_{\text{LF}-1P-1,NR} = &
    \lim_{k \to 0}-
  \psi_{0,- 1,0,NR}(\vec k)\\
  =&\lim_{k \to 0}
  -C_{0,- 1,0}R(k)
\frac{2k}{\kappa 
}
e^{- i\phi}  \sin\theta\\
=&-C k
Y_{1,-1}(\theta,\phi)
  \;.
  \end{split}
\end{align} 
Note that there is a `-' sign on the NR limit of the LF-1P-1 state. It follows that the NR limit of the LF-1P1 state is the $m=1$ component of the 1P state, and that of the LF-1P-1 state is negative of the $m=-1$ component of the 1P state.
\begin{table}[ht]
  \caption{\label{tab:psi_nml_NR}
  The NR limit of the LF spectroscopic states in comparison with the spherical harmonics states.
  }
  \centering
  \begin{ruledtabular}
  \begin{tabular}{l cc}
   $\psi_{\text{LF}-W}$
   & by design $\psi_{nml}$
   & NR limit\\
    \hline  
    $\psi_{\text{LF}-1S}$ & $\psi_{000}$ & $\Psi_{1S}$ \\
    $\psi_{\text{LF}-2S}$ & $    \sqrt{\frac{2}{3}}
    \psi_{ 1,0,0}  
  -    \sqrt{\frac{1}{3}}
  \psi_{ 0,0,2}  $ & $\Psi_{2S}$ \\
  $\psi_{\text{LF}-1P0}$ & $   
\psi_{ 0,0,1}  $ & $\Psi_{1P0}$ \\
$\psi_{\text{LF}-1P1}$ & $   
-\psi_{ 0,1,0}  $ & $\Psi_{1P1}$ \\
$\psi_{\text{LF}-1P-1}$ & $   
-\psi_{ 0,-1,0}  $ & $-\Psi_{1P-1}$ \\
  $\psi_{\text{LF}-1D0}$ & $    \sqrt{\frac{1}{3}}
  \psi_{ 1,0,0}  
+    \sqrt{\frac{2}{3}}
\psi_{ 0,0,2}  $ & $\Psi_{1D0}$ \\
$\psi_{\text{LF}-1D1}$ & 
$-\psi_{ 0,1,1}  $ & $\Psi_{1D1}$ \\
$\psi_{\text{LF}-1D-1}$ & $   
-\psi_{ 0,-1,1}  $ & $-\Psi_{1D-1}$ \\
$\psi_{\text{LF}-1D2}$ & $\psi_{ 0,2,0}  $ & $\Psi_{1D2}$ \\
$\psi_{\text{LF}-1D-2}$ & $ \psi_{ 0,-2,0}  $ & $\Psi_{1D-2}$ 
  \end{tabular}
\end{ruledtabular}
\end{table}
The above states are listed in Table.~\ref{tab:psi_nml_NR}.

\section{ Decay constant in the basis function representation}\label{app:fV_basis}
In this Appendix, we write out the decay constants in the basis function representation.
The decay constant of the vector meson in Eqs.~\eqref{eq:fV_mj0} and \eqref{eq:fV_mjpm1}, after integrating out the basis functions, reduce to,
\begin{align}\label{eq:fV_mj0_basis}
  \begin{split}
    &f_{\mathcal{V}}|_{m_j=0}\\
    =&
    \frac{2 \kappa \sqrt{3}}{\pi} \sum_{n, l} 
    (-1)^n \sqrt{\frac{2l+2\alpha+1}{l! \Gamma(l+2\alpha+1)}}\Gamma((\alpha+3)/2)
    \\
    &\sum_{k=0}^l \binom{l}{k} 
    (-1)^k
    \frac{\Gamma(l+k+2\alpha+1) \Gamma(k+\alpha/2+3/2)}{\Gamma(k+\alpha+1)\Gamma(k+\alpha+3)}
    \\
    &\times
    \psi_{\mathcal{V}}^{(m_j=0)}(n, m=0, l, \frac{1}{2}, -\frac{1}{2})
    \;,
  \end{split}
\end{align}
and
\begin{widetext}
\begin{align}\label{eq:fV_mj1_basis}
  \begin{split}
      f_{\mathcal{V}}|_{m_j=\pm 1}
      =&
      \frac{ \kappa \sqrt{3}}{\sqrt{2}\pi}\frac{m_f}{m_{\mathcal{V}}}
      \sum_{n, l} 
      (-1)^n \sqrt{\frac{2l+2\alpha+1}{l! \Gamma(l+2\alpha+1)}}\Gamma((\alpha+1)/2)
        \sum_{k=0}^l \binom{l}{k} 
      (-1)^k
      \frac{\Gamma(l+k+2\alpha+1) \Gamma(k+\alpha/2+1/2)}{\Gamma(k+\alpha+1)\Gamma(k+\alpha+1)}
      \\
      &\times
      \psi_{\mathcal{V}}^{(m_j=-1)}(n, m=0, l, -\frac{1}{2}, -\frac{1}{2})\\
      &+
      \frac{ \kappa^2 \sqrt{6}}{\pi m_{\mathcal{V}}}
      \sum_{n, l} 
      (-1)^n \sqrt{n+1}
      \sqrt{\frac{2l+2\alpha+1}{l! \Gamma(l+2\alpha+1)}}\Gamma((\alpha+2)/2)
      \sum_{k=0}^l \binom{l}{k} 
      (-1)^k
      \frac{\Gamma(l+k+2\alpha+1) \Gamma(k+\alpha/2+2)}{\Gamma(k+\alpha+1)\Gamma(k+\alpha+3)}
      \\
      &\times
      \psi_{\mathcal{V}}^{(m_j=-1)}(n, m=-1, l, \frac{1}{2}, -\frac{1}{2})
      \\
      &-
      \frac{ \kappa^2 \sqrt{6}}{\pi m_{\mathcal{V}}}
      \sum_{n, l} 
      (-1)^n \sqrt{n+1}
      \sqrt{\frac{2l+2\alpha+1}{l! \Gamma(l+2\alpha+1)}}\Gamma(\alpha/2+2)
      \sum_{k=0}^l \binom{l}{k} 
      (-1)^k
      \frac{\Gamma(l+k+2\alpha+1) \Gamma(k+\alpha/2+1)}{\Gamma(k+\alpha+1)\Gamma(k+\alpha+3)}
      \\
      &\times
      \psi_{\mathcal{V}}^{(m_j=-1)}(n, m=-1, l, -\frac{1}{2}, \frac{1}{2})
      \;.
    \end{split}
\end{align}
\end{widetext}

The decay constant of the pseudoscalar meson in Eq.~\eqref{eq:fP}, after integrating out the basis functions, reduce to,
\begin{align}\label{eq:fP_basis}
  \begin{split}
  f_{\mathcal{P}}=&
\frac{2 \kappa \sqrt{3}}{\pi} \sum_{n, l} 
(-1)^n \sqrt{\frac{2l+2\alpha+1}{l! \Gamma(l+2\alpha+1)}}\Gamma((\alpha+3)/2)
\\
&\sum_{k=0}^l \binom{l}{k} 
(-1)^k
\frac{\Gamma(l+k+2\alpha+1) \Gamma(k+\alpha/2+3/2)}{\Gamma(k+\alpha+1)\Gamma(k+\alpha+3)}
\\
&\times
\psi_{\mathcal{P}}(n, m=0, l, \frac{1}{2}, -\frac{1}{2})
\;.
  \end{split}
\end{align}
By comparing Eqs.~\eqref{eq:fV_mj0_basis} and ~\eqref{eq:fP_basis}, we can see that the decay constant of a pseudoscalar meson with a $\psi_{n,0,0}\sigma_-$ wavefunction and that of a vector meson in the $m_j=0$ state with a $\psi_{n',0,0}\sigma_+$ wavefunction have the same functional form (taking the absolute value) in terms of $\kappa$ and $m_f$.
Similarly, from Eq.~\eqref{eq:fV_mj1_basis}, we can see that the decay constant of a vector meson in the $m_j=\pm 1$ state with a $\psi_{n,0,0}\sigma_{\uparrow\uparrow}(\sigma_{\downarrow\downarrow})$ wavefunction has the same functional form in terms of $\kappa$ and $m_f$ for different values of $n$.

\section{The photon LFWF}\label{app:photon}
We derive the photon wavefunction in the framework of light-cone perturbation theory to lowest order~\cite{Dosch:1996ss,Lepage:1980fj, Kowalski:2006hc},
 \begin{subequations}\label{eq:photon_WF_r}
  \begin{multline}
       \psi_{s \bar s/\gamma}^{(m_j=0)}(\vec r_\perp,x,Q)
    =\\\mathcal{Q}_f e\sqrt{N_c} \delta_{s,-\bar s} 2Q x(1-x)\frac{K_0(\epsilon r_\perp)}{2\pi}
    \;,
  \end{multline}
    \begin{multline}
      \psi_{s \bar s/\gamma}^{(m_j=1)}(\vec r_\perp,x,Q)
      =\sqrt{2}\mathcal{Q}_f e\sqrt{N_c}
      \{
      -i e^{i\theta_r}  
      [
        x\delta_{s,+}\delta_{\bar s,-}\\
        -(1-x)\delta_{s,-}\delta_{\bar s,+}
      ]\partial_r
      +m_f\delta_{s,+}\delta_{\bar s,+}
      \}\frac{K_0(\epsilon r_\perp)}{2\pi}
      \;,
  \end{multline}
    \begin{multline}
      \psi_{s \bar s/\gamma}^{(m_j=-1)}(\vec r_\perp,x,Q)
      =
      -\sqrt{2}\mathcal{Q}_f e\sqrt{N_c}
      \{
      i e^{-i\theta_r}  
      [
        x\delta_{s,-}\delta_{\bar s,+}\\
        -(1-x)\delta_{s,+}\delta_{\bar s,-}
      ]\partial_r
      +m_f\delta_{s,-}\delta_{\bar s,-}
      \}\frac{K_0(\epsilon r_\perp)}{2\pi}
      \;,
  \end{multline}
  \end{subequations}
  where $r_\perp=|\vec r_\perp|$ and $\theta_r=\arg (r^x + ir^y)$, $\epsilon^2\equiv x(1-x)Q^2+m_f^2$, $\mathcal Q_f e$ is the quark charge, and $\partial_r K_0(\epsilon r_\perp)=-\epsilon K_1(\epsilon r_\perp)$.
  Note that there is a sign difference in the second equation compared to that in Ref.~\cite{Kowalski:2006hc} by the convention we use for the spinors. 
The wavefunction in the momentum space is obtained by the Fourier transformation as in Eq.~\eqref{eq:FT_kt_rt}.
The photon wavefunction in the momentum space, written explicitly, is
\begin{subequations}\label{eq:photon_WF_p}
  \begin{multline}
      \psi_{s \bar s/\gamma}^{(m_j=0)}(\vec k_\perp,x,Q)
      =\\
      \mathcal{Q}_f e\sqrt{N_c}\delta_{s,-\bar s} \sqrt{x(1-x)}
      \frac{2Qx(1-x)}{\epsilon^2+k_\perp^2}
      \;, 
  \end{multline}
  \begin{multline}
      \psi_{s \bar s/\gamma}^{(m_j=1)}(\vec k_\perp,x,Q)=
      \mathcal{Q}_f e\sqrt{2N_c}\frac{\sqrt{x(1-x)}}{\epsilon^2+k_\perp^2}\\
      [
        xk^R\delta_{s,+}\delta_{\bar s,-}
        -(1-x)k^R\delta_{s,-}\delta_{\bar s,+}
        +m_f\delta_{s,+}\delta_{\bar s,+}
      ]
      \;,
  \end{multline}
  \begin{multline}
      \psi_{s \bar s/\gamma}^{(m_j=-1)}(\vec k_\perp,x,Q)
      =-\mathcal{Q}_f e\sqrt{2N_c}
      \frac{\sqrt{x(1-x)}}{\epsilon^2+k_\perp^2}\\
      [
        -xk^L\delta_{s,-}\delta_{\bar s,+}
        +(1-x)k^L\delta_{s,+}\delta_{\bar s,-}
        +m_f\delta_{s,-}\delta_{\bar s,-}
      ]
      \;.
    \end{multline}
\end{subequations}

\bibliography{LFWF}
\end{document}